\documentclass[a4paper,11pt]{article}
\usepackage{jheppub}
\usepackage[T1]{fontenc}
\usepackage{lmodern}

\usepackage{amstext}
\usepackage{babel}
\usepackage{verbatim}
\usepackage{amsfonts}
\usepackage{mathrsfs}
\usepackage{amsmath}
\usepackage{amssymb}
\usepackage{bm}
\usepackage{extarrows}
\usepackage{slashed}
\usepackage{isodateo}
\usepackage{xcolor}
\usepackage{graphicx}
\usepackage[low-sup]{subdepth}
\usepackage{float}
\usepackage{isodateo}
\usepackage{tensor}
\usepackage{multirow}
\usepackage{enumitem}
\usepackage{framed}
\usepackage{scalerel}
\usepackage{stackengine,wasysym}
\usepackage{tikz-feynhand}
\usepackage{hyperref}
\usepackage{stackengine}
\usepackage{subcaption}
\hypersetup{
colorlinks=true,
citecolor=blue,
linkcolor=blue,
urlcolor=blue,
pdfauthor={},
pdftitle={},
pdfsubject={}}
\makeatletter
\def\simgt{\mathrel{\lower2.5pt\vbox{\lineskip=0pt\baselineskip=0pt
\hbox{$>$}\hbox{$\sim$}}}}
\def\simlt{\mathrel{\lower2.5pt\vbox{\lineskip=0pt\baselineskip=0pt
\hbox{$<$}\hbox{$\sim$}}}}
\numberwithin{equation}{section}
\definecolor{shadecolor}{gray}{0.925}
\definecolor{}{gray}{0.925}
\definecolor{lblu}{RGB}{15,125,255}
\definecolor{lgre}{RGB}{40,170,40}
\newcommand{\red}{\color{red}}
\newcommand{\blu}{\color{blue}}

\newcommand{\org}{\color{orange}}
\newcommand{\lgray}{\color{gray}}
\newcommand{\lblu}{\color{lblu}}

\newcommand{\eqrefe}{eq.\,\eqref}
\newcommand{\beqs}{\begin{subequations}}
\newcommand{\eeqs}{\end{subequations}}
\newcommand{\bla}{\big\langle}  
\newcommand{\bra}{\big\rangle}
\renewcommand{\rm}{\mathrm}
\newcommand\xrowht[2][0]{\addstackgap[.5\dimexpr#2\relax]{\vphantom{#1}}}
\graphicspath{{fig/}}       
\DeclareGraphicsExtensions{.pdf}
\newcommand{\letterfig}[4]{%
\begin{minipage}{#4}
\includegraphics[scale=#3]{#1/#2.pdf}
\end{minipage}
}
\newcommand{\widesim}[1][1.5]{
\mathrel{{\scalebox{#1}[1]{$\thicksim$}}}
}
\def\leqq{\leqslant}
\def\geqq{\geqslant}
\def\({\left(}
\def\){\right)}
\def\[{\left[\,}
\def\]{\,\right]}

\def\nn{\nonumber}
\def\pd{\partial}
\def\pp{\prime}
\def\to{\rightarrow}
\def\summ{\scalebox{0.77}{$\sum$}}
\def\dlog{\text{d}\hspace*{-.5mm}\log}
\def\vs{\vspace*{1mm}}
\def\hs{\hspace*{0.3mm}}
\def\hsm{\hspace*{-0.3mm}}
\def\td{\text{d}}

\def\D{\mathcal{D}}
\def\EE{\mathcal{E}}
\def\F{\mathsf{F}}
\def\bF{\textbf{F}}
\def\FF{\mathcal{F}}

\def\FFt{\widetilde{\mathcal{F}}}
\def\ii{\text{i}}
\def\bI{\textbf{I}}
\def\II{\mathcal{I}}
\def\JJ{\mathcal{J}}
\def\bk{\textbf{k}}
\def\L{\text{L}}
\def\La{\mathcal{L}}
\def\PP{\mathcal{P}}
\def\P{\mathsf{P}}
\def\QQ{\mathcal{Q}}
\def\Q{\mathsf{Q}}
\def\R{\text{R}}
\def\mS{\mathcal{S}}
\def\SS{\mathbf{S}}
\def\NN{\mathcal{N}}
\def\VV{\mathcal{V}}
\def\YY{\mathbf{Y}}
\def\mY{\mathcal{Y}}
\def\tmY{\widetilde{\mathcal{Y}}}
\def\mZ{\mathcal{Z}}
\def\al{\alpha}
\def\be{\beta}
\def\ga{\gamma}
\def\Ga{\Gamma}
\def\ab{\alpha\beta}

\def\ep{\epsilon}
\def\vep{\varepsilon}
\def\lam{\lambda}
\def\bOme{\overline{\Omega}}
\def\uu{\text{u}}
\def\ctr{\text{ctr}}
\def\bub{\text{bub}}
\def\tad{\text{tad}}
\def\sun{\text{sun}}
\def\dbl{\text{dbl}}
\def\lop{\text{lop}}
\def\grd{\text{grd}}
\def\tree{\text{tree}}

\allowdisplaybreaks
\title{A Note on Kinematic Flow and Differential Equations for Two-Site One-Loop Graph in FRW Spacetime}

\author[]{Yanfeng Hang}
\emailAdd{yfhang@northwestern.edu}

\author[]{and~Cong Shen}
\emailAdd{CongShen2028@u.northwestern.edu}

\affiliation[]{Department of Physics and Astronomy,\\ 
Northwestern University, Evanston, IL 60208, USA}

\abstract{In this work, we systematically study the differential systems governing loop-level wavefunction coefficients of conformally-coupled scalar field theory within a general power-law FRW cosmology.\ By utilizing the twisted cohomology, hyperplane arrangements, and IBP techniques, we derive the canonical differential equations for two-site one-loop bubble and tadpole systems, revealing distinct structural differences.\
We present new insights into the one-loop tadpole system, uncovering that its integral family can include multiple parent functions due to distinct pairs of relative hyperplanes associated with each function, unlike the single parent function appearing in the one-loop bubble case.\ Moreover, we demonstrate that the tadpole correlator selectively probes only a subset of the cohomology space, despite the hyperplane arrangement suggesting a higher-dimensional structure.\ 
Another novel contribution of this work is the extension of kinematic flow framework to the loop-level scenarios for the first time.\ Using a graphical approach based on family trees generated by marked tubing graphs, which encode singularity structures, we efficiently construct the differential equations and uncover the hierarchical relationships among the associated master integrals.\ 
Additionally, we provide a preliminary discussion on generalization to two-site higher-loop configurations. We propose a general decomposition formula for the canonical form of a two-site diagram with arbitrary loops, breaking it into unshifted and shifted components associated with the fundamental tree-level and bubble-like structures, and establish a block-wise decomposition rule for the matrix $A$ in the corresponding differential system.\
These advancements provide a unified framework for two-site loop-level correlators and lay the groundwork for future study of more complex multi-site loop systems.

\vspace*{1cm}
\noindent
JHEP 09 (2025) 209~ arXiv:\,2410.17192 [hep-th]
}

\begin{document}
\maketitle
\flushbottom

\section{Introduction}
\label{sec:1} 

The study of cosmological dynamics has emerged as a central focus in modern theoretical physics, especially in understanding the early universe and the mechanisms that shaped its evolution.\ 
Among the many toy models, the investigation of wavefunction coefficients for conformally-coupled scalar theory in a power-law Friedmann-Robertson-Walker (FRW) universe \cite{Arkani-Hamed:2017fdk,Arkani-Hamed:2018bjr} has proven instrumental in advancing our understanding of cosmological processes.\ 
These wavefunction coefficients contain key information about the inflationary dynamics and the generation of primordial fluctuations, serving as a bridge between the theoretical models and observational cosmology.

In recent years, various innovative methods have been developed to compute and study cosmological correlators, unveiling new mathematical structures and offering deeper insights into the nature of the early universe.\ Notable approaches include:  
the Boolean-lattice construction for two-site wavefunction coefficients at arbitrary loop order \cite{Hang:2026men};\ 
the framework of cosmological polytope \cite{Arkani-Hamed:2017fdk,Arkani-Hamed:2018bjr,Benincasa:2018ssx,Benincasa:2019vqr}, which geometrically encodes the combinatorial and analytic properties of correlators;\ 
the techniques based on the (relative dual) twisted cohomology, intersection theory and algebraic geometry \cite{De:2023xue, Benincasa:2024ptf}, which comprehensively study the differentiation system of FRW wavefunction coefficients at both tree and loop level;
the kinematic flow method \cite{Arkani-Hamed:2023bsv,Arkani-Hamed:2023kig}, which organizes the canonical differential equations of tree-level FRW wavefunction coefficients based on the hierarchical structure of their singularities represented by tubing graphs.

In this work, we focus on the loop-level wavefunction coefficients of conformally-coupled scalar field theory with non-conformal polynomial interactions in a general power-law FRW universe.\ 
We utilize the twisted cohomology \cite{aomoto1975vanishing,Aomoto:2011ggg,Mastrolia:2018uzb} alongside the
integration-by-parts (IBP) techniques to derive the canonical differential equations (DEs) \cite{Henn:2013pwa,Henn:2014qga} associated with the wavefunction coefficients of two-site one-loop diagrams.\
Our study investigates both one-loop bubble and tadpole systems within the context of hyperplane arrangements, highlighting novel features in the loop-level case.\ 
A major finding is that the one-loop bubble system resembles a combination of three tree-like subsystems, where the parent function of each subsystem can be merged into a single parent function through a linear relation.\
Unlike the bubble case, the one-loop tadpole system can involve multiple parent functions.\ 
This is a result of distinct pairs of relative hyperplanes (which are parallel to the twist hyperplanes) associated with each parent function, which introduces a different structure compared to the same pair appearing in bubble case.\
And based on the analysis of one-loop case, we also develop a unified framework for analyzing two-site diagram with arbitrary loops.\ 
We propose a general formula for the canonical form associated with two-site $L$\hs-loop correlator, decomposing it into a combination of unshifted and shifted forms, with the shifts induced combinatorially from the presence of tadpole loops.\
This approach enables any two-site $L$\hs-loop diagram to be expressed in terms of the most fundamental tree and bubble-like diagrams, along with their shifted counterparts.\
Another important point is the application and extension of the kinematic flow framework to loop level.\ 
This powerful methodology enables the systematic prediction of canonical differential equations by representing the singularities of the system as marked tubing graphs \cite{Arkani-Hamed:2023bsv,Arkani-Hamed:2023kig,carr2006coxeter,devadoss2020colorful,balduf2024tubings}.\
These graphs encode the hierarchical relationships among the integrals, forming family trees that organize the system’s structure according to some simple rules.\
By utilizing this framework, we streamline the derivation of differential equations from the family trees constructed by the tubing graphs.\
Furthermore, based on the analysis of hyperplane arrangements, the kinematic flow framework can be effectively generalized for any two-site loop order.\
All of these findings are presented for the first time in this work.

This paper is organized as follows:\  
In section\,\ref{sec:2}, we set up the model of conformally-coupled scalars in the general power-law FRW background and provide a brief review of the cosmological wavefunction coefficients.\
In section\,\ref{sec:3}, we systematically analyze the two-site one-loop FRW wavefunction coefficients within bubble and tadpole types using the twisted cohomology and hyperplane arrangements.\  
We perform a detailed study of the canonical forms associated with the integral family, investigating their structure within the bounded chambers (triangle basis), and derive the corresponding canonical differential equations.\  
This derivation relies on matching residues at different codimension-2 boundaries using IBP techniques.\  
Further, we show that in the tadpole system, its integral family contains two distinct parent functions, in contrast with the bubble system, where multiple parent functions can be merged into a single function.\
In section\,\ref{sec:4}, we present the extension of kinematic flow framework from tree-level to loop-level computations for the first time.\ Singularities in the differential system are represented as the marked tubing graphs, establishing a direct correspondence between these graphical structures and the canonical differential equations.\ 
Following the simple rules, the differential equations can be directly constructed from the family trees generated by the evolution of those marked tubes.\
In section\,\ref{sec:5}, we provide four examples of wavefunction coefficients for two-site two-loop bubble and tadpole diagrams, exploring the configurations of their hyperplane arrangements based on the analysis of one-loop cases.\ We also examine the feasibility of extending the kinematic flow method to scenarios beyond one-loop.\
Besides, we provide the decomposition rules for two-site diagram with arbitrary loops, and the block decomposition relation for matrix $A$ associated with the corresponding differential system.\
Finally, in section\,\ref{sec:6}, we summarize our results and discuss potential directions for future research.\ Appendices\,\ref{app:A}--\ref{app:C} provide additional details that are missing from the main text.

\noindent
\textbf{Note:} As this paper was being finalized, we have found that the ref.\,\cite{He:2024olr} also investigated the differential equations for cosmological loop integrands but employed a different methodology.\ 
While our approach differs from theirs, we find agreement with their results and view their findings as complementary to our work.

\section{Toy Model and Cosmological Wavefunction Coefficients}
\label{sec:2}

We consider a toy model involving a scalar field with a non-conformal polynomial self-interaction in a general Friedmann-Robertson-Walker (FRW) background.\ 
The relevant action in a $(d+1)$-dimensional spacetime takes the form:
\begin{equation}
\label{eq:S-FRW}
S_{\rm{FRW}} \,=\, -\int\td^d x \int^{0}_{-\infty}\td\eta\hs \sqrt{-g}\, \Bigg[\,\frac{1}{2}\hs g^{\ab} \pd_{\al}^{}\phi\hs \pd_{\be}^{}\phi +\frac{1}{2}\xi R\hs \phi^2 + \sum_{n\geqq3} \frac{\,\lam_n\,}{n!}\hs \phi^n \Bigg]\,,
\end{equation}
where $\eta$ denotes the conformal time, $R$ refers to the Ricci scalar and $\lam_n$ is the coupling constant.\
In addition, the parameter $\xi$ satisfies the following conditions: when $\xi = \frac{d-1}{4d}$, the theory describes a conformally coupled scalar, which is the focus of our analysis. When $\xi = 0$, the scalar is minimally coupled, a case that is not considered in this work.

The FRW metric is defined as follows:
\begin{equation}
\label{eq:FRW-metric}
\td s^2 = g_{\ab}^{} \hs\td x^{\al} \td x^{\be} = [a(\eta)]^2\big(\!-\td\eta^2 + \delta_{ij}\td x^i \td x^j\,\big)\,,
\end{equation}
where $a(\eta)$ represents the scale
factor and the indices $(i,j)\!=\!1,\dots,d$\, span the spatial dimensions.\
Further, by substituting FRW metric \eqref{eq:FRW-metric} into the action \eqref{eq:S-FRW} and rescaling the scalar field $\phi\to[a(\eta)]^{(1-d)/2}\phi$\,, the FRW action \eqref{eq:S-FRW} will reduce to the action describing massless scalar fields effectively in a $(d+1)$-dimensional Minkowski (flat) spacetime:
\begin{equation}
\label{eq:S-Flat}
S_{\rm{Mink}}^{\hs\rm{eff}} \,=\, -\int \td^d x \int_{-\infty}^0 \td\eta \Bigg[\,\frac{1}{2} (\pd \phi)^2 + \sum_{n\geqq3} \frac{\lam_n(\eta)}{n!}\hs \phi^n \Bigg]\,,
\end{equation}
where the time-dependent coupling $\lam_n(\eta)$ is defined as follows:
\begin{equation}
\label{eq:lam-eta}
\lam_n(\eta) \,\equiv\,\lam_n[a(\eta)]^{2-(n-2)(d-1)/2}
= \lam_n \eta^{-(1+\vep)(2-(n-2)(d-1)/2)} \,.
\end{equation}
Note that the final equality in \eqrefe{eq:lam-eta} holds when the scale factor takes the power-law form \eqref{eq:scalefactor-power-law}, as will be discussed later.

Next, we provide a brief review of the cosmological wavefunction coefficients.\ 
These coefficients describe the amplitude for different configurations of fields in the early universe and encode crucial information about cosmological fluctuations.\ 
It can be derived from the path integral formalism, where the wavefunction $\Psi[\phi]$ is obtained by summing over all possible field configurations in a given cosmological background.\
Specifically, we can express the cosmological wavefunction $\Psi[\phi]$ as a path integral over the field configurations with boundary conditions:
\begin{equation}
\Psi[\phi] = \int_{\varphi[-\infty(1-\ii\ep)]=0}^{\hs\varphi(0)=\phi} \D\varphi \, e^{\ii S[\varphi]},
\end{equation}
where $S[\varphi]$ is the action of the field $\varphi$ as given in \eqrefe{eq:S-Flat} and the integral is performed over all field configurations.\ 
And, the $\ii\ep$-prescription is crucial for suppressing the negative-frequency component and ensuring that the initial state corresponds to a pure, minimal-energy (Bunch-Davies) vacuum at the early time $\eta\to-\infty$.\
Further, $|\Psi[\phi]|^2$ provides the physical meaning of the probability density for spatial field configurations.\ 
Thus, the equal-time $m$-point correlation function for $\phi$ can be written as:
\begin{equation}
\bla\phi(\textbf{x}_1^{})\cdots\phi(\textbf{x}_m^{})\bra \,=\, \int \D\phi\, \phi(\textbf{x}_1^{})\cdots\phi(\textbf{x}_m^{}) |\Psi[\phi]|^2 \,.
\end{equation}
The wavefunction can also be expanded in terms of its coefficients, which are related to cosmological correlation functions. For example, expanding $\Psi[\phi]$ perturbatively gives:
\begin{equation}
\Psi[\phi] \,=\, \exp{\Bigg\{\!-\!\sum_{m>1} \frac{1}{m!}\int\prod_{i=1}^m\Bigg[\frac{\td^d\bk_i}{(2\pi)^{d}}\,\phi_{\bk_i}^{}\Bigg]
\psi_m^{}(\bk_1,\ldots,\bk_m)(2\pi)^d\delta^{(d)}\bigg(\sum_{j=1}^{m}\bk_j^{}\bigg)\Bigg\}}\,,
\end{equation}
where we refer to $\psi_m^{}(\bk_1,\ldots,\bk_m)$ as the cosmological wavefunction coefficients, encoding the interactions between $m$-point fluctuations of the scalar field.

The path integral can be performed diagrammatically in the usual way with Feynman graphs.\ 
Each diagram corresponds to a specific contribution to the perturbative expansion of the wavefunction, which can be calculated by using the following Feynman rules:
%
(i).\,Draw the boundary surface at fixed time $\eta=0$ where the wavefunction will be computed and draw all diagrams ending on the boundary surface.\
(ii).\,Attach a vertex factor $\ii V_v$ to each bulk interaction, where $V_v$ is defined by the interaction terms in the action.\ For a scalar field with non-derivative polynomial interactions as given in \eqrefe{eq:S-Flat}, $V_v$ depends only on the coupling constant.\
(iii).\,Assign a bulk-to-boundary propagator, $K(E;\eta)$, to each external line.\ For a given vertex $v$, this propagator corresponds to the propagation of a field from a bulk vertex to the late-time boundary, and it takes the form:
\begin{equation}
\label{eq:bulk-boundary-prop}
K_v(E_{v}; \eta_{v}^{}) \,=\, e^{\ii E_{v} \eta_{v}^{}} \,,
\end{equation}
where $E_{v}=\sum_i|\bk_i$| denotes the sum of the energies of all $i$ external legs connected to the vertex $v$.\
(iv).\,Assign a bulk-to-bulk propagator, $G(E;\eta,\eta')$, to each internal line.
For an internal line $e$ connecting two vertices, $v_e$ and $v_e^\pp$, the bulk-to-bulk propagator $G_e(E_{e} ; \eta_{v_e}^{}, \eta_{v_e^\pp}^{})$ describes the propagation of fields between two vertices $v_e$ and $v_e^\pp$ located at different times and it takes the form:
\begin{align}
\label{eq:bulk-bulk-prop}
\!\!G_e\hsm(E_{e};\hsm\eta_{v_{\hsm e}}^{},\hsm\eta_{v_{\hsm e}^\pp}^{})\hsm=\hsm\frac{1}{2E_e}\hsm\Big[
e^{-\ii E_e\hsm(\eta_{v_{\hsm e}}^{}\!\hsm-\hs\eta_{v_{\hsm e}^\pp})}\theta(\eta_{v_{\hsm e}}^{}\!\!-\eta_{v_{\hsm e}^\pp}) 
\!+\!e^{\ii E_e\hsm(\eta_{v_{\hsm e}}^{}\!\hsm-\hs\eta_{v_{\hsm e}^\pp})}\theta(\eta_{v_{\hsm e}^\pp}\!-\eta_{v_{\hsm e}}^{})
\!-\!e^{\ii E_e\hsm(\eta_{v_{\hsm e}}^{}\!+\hs\eta_{v^\pp_{\hsm e}})}
\Big],
\end{align}
where $E_e=|\bk_e|$ represents the energy of an internal line $e$.
(v).\,Integrate over the time insertions of all bulk vertices and any loop momenta.\ Notice that in this analysis, we focus solely on the wavefunction coefficient at the integrand level, without performing integration over loop momenta.

Finally, we assume that the scale factor $a(\eta)$ appearing in the FRW metric \eqref{eq:FRW-metric} takes the following power-law form \cite{De:2023xue,Arkani-Hamed:2023kig}:
\vspace*{-1mm}
\begin{equation}
\label{eq:scalefactor-power-law}
a(\eta) \,=\, \eta^{-(1+\vep)}\,,
\vspace*{-1.2mm}
\end{equation}
where the values $\vep=0,-1,-2,-3$ correspond to the de Sitter (dS), flat, radiation- and matter-dominated universe, respectively.\
For broader physical relevance, we treat 
$\vep$ as a general real parameter, allowing non-integer values.\
Thus, in the case of a power-law form \eqref{eq:scalefactor-power-law}, we can analyze and compute the wavefunction coefficient.\
The corresponding Feynman rules need slight modifications, where the bulk-to-boundary propagator \eqref{eq:bulk-boundary-prop} and bulk-to-bulk propagator \eqref{eq:bulk-bulk-prop} remain unchanged from their flat-space forms.\ 
As for the vertices, the $n$-point time-dependent coupling \eqref{eq:lam-eta} can be expressed in the energy space via following Mellin transform:
\begin{equation}
\lam_n(\eta)=\lam_n\, \eta^{-\ga}
=\int_0^{+\infty(1-\ii\ep)} \td x \, \bar{\lam}_n(x)\hs e^{\ii x \eta} \,,
\end{equation}
where $\ga=(1+\vep)[2-(n-2)(d-1)/2]$ and the energy-dependent coupling constant $\bar{\lam}_n(x)$ is given by
\begin{equation}
\bar{\lam}_n(x)= \lam_n \frac{\,e^{-\ii\pi\ga/2}\,}{\Ga(\ga)}\hs x^{\ga-1}\,.
\end{equation}
Further, the $\ii\ep$\hs-prescription modifies the integration contour in the complex plane to ensure convergence of the integral at large $x$.

Therefore, for a given $N$-site $L$-loop graph characterized by the vertex set $\VV$ and edge set $\EE$, its FRW wavefunction coefficient can be obtained by appropriately shifting the flat-space expression over the external energies:
\begin{equation}
\label{eq:psi-FRW}
\psi^{\rm{FRW}}_{(N,L)}(X_v,Y_e)=\int_0^{+\infty} \prod_{v\in\VV} [\hs\td x_v \hs \bar{\lam}_n(x_v)] \hs \tilde{\psi}^{\rm{Mink}}_{(N,L)}(X_v+x_v, Y_e)\,,
\end{equation}
where $X_v$ represents the total energy flowing
from a vertex $v$ to the late-time boundary and $Y_e$ corresponds to the energy of each internal line $e$.\
In \eqrefe{eq:psi-FRW}, the integral involving the power parameter $\ga$ is valid for $\ga>0$ in $\bar{\lam}_n(x_v)$, while for $\ga\leqq0$, the integration over $x$ can be replaced by a differential operator acting on $\tilde{\psi}^{\rm{Mink}}_{(N,L)}$ \cite{Benincasa:2019vqr}.\
Further, the flat-space wavefunction coefficient integrand $\tilde{\psi}^{\rm{Mink}}_{(N,L)}$ is derived as follows:
\begin{equation}
\label{eq:psi-Flat}
\tilde{\psi}^{\rm{Mink}}_{(N,L)}(X_v,Y_e)=\ii^N\int_{-\infty}^0 \prod_{v\in\VV} \td \eta_v^{} K_v(X_v;\eta_v) \prod_{e\in\EE} G_e(Y_e;\eta_{v_e}^{}, \eta_{v^\pp_e}^{})\,,
\end{equation}
where the bulk-to-boundary propagator $K_v$ and bulk-to-bulk propagator $G_e$ are defined in eqs.\,\eqref{eq:bulk-boundary-prop}--\eqref{eq:bulk-bulk-prop}.\ 
Finally, for convenience, we will drop out the superscripts ``FRW'' and ``Mink'' and use $\psi_{(N,L)}$ and $\tilde{\psi}_{(N,L)}$ to denote the wavefunction coefficients in FRW and flat spacetimes in the following discussion.

\section{Two-Site One-Loop Wavefunction Coefficients and Canonical DEs}
\label{sec:3}

In this section, we investigate a detailed analysis of the two-site one-loop wavefunction coefficients of the bubble and tadpole diagrams within a $(d+1)$-dimensional power-law FRW universe.\ 
For the first time, we systematically find the corresponding integral basis associated with the loop-level wavefunction coefficients guided by the structure of the hyperplane arrangements.\
Furthermore, we also provide the comprehensive derivations for the canonical differential equations, employing integration-by-parts (IBP) techniques to capture their intricate dependencies.

\subsection{Two-Site One-Loop Bubble}
\label{sec:3.1}

We first consider the two-site one-loop bubble-type wavefunction coefficient whose Feynman diagram is shown in fig.\,\ref{fig:1}.\
Following the Feynman rules given in section\,\ref{sec:2}, we can compute the flat-space wavefunction coefficient as follows:
\begin{align}
\label{eq:psi-bub-flat}
\tilde{\psi}_{(2,1)}^\bub &= \ii^2\int_{-\infty}^{0}\td\eta_1 \int_{-\infty}^{0}  \td\eta_2 \, 
K_1(X_1;\eta_1) \hs K_2(X_2;\eta_2) \hs G_1(Y_1;\eta_1,\eta_2) \hs G_2(Y_2;\eta_2,\eta_1)
\nn\\
&=\frac{2(X_1\!+\!X_2\!+\! Y_1\!+\!Y_2)}{(X_1\!+\!X_2)(X_1\!+\!X_2\!+\!2Y_1)(X_1\!+\!X_2\!+\!2Y_2)(X_1\!+\!Y_1\!+\!Y_2)(X_2\!+\!Y_1\!+\!Y_2)}\,.
\end{align}
Then, to derive the two-site one-loop wavefunction coefficient in a general FRW universe, we can shift over the external energies of its corresponding flat-space form \eqref{eq:psi-bub-flat} via the following transformations: $(X_1,\,X_2)\to(X_1+x_1,\,X_2+x_2)$.\  
Further, we incorporate a twist term $(x_1x_2)^{\ga-1}$ to regularize the integrand.\ 
Specifically, we focus on the case $\ga=1+\vep$, where the twister parameter $\vep$ effectively controls and mitigates the singular behavior.\ Thus, the complete FRW wavefunction coefficient for two-site one-loop bubble diagram is derived in the following integration representation:
\begin{equation}
\label{eq:psi-bub-frw-1}
\psi_{(2,1)}^\bub = 4Y_1Y_2\, \frac{\lam_n^2 (-1)^{1+\vep}}{[\Ga(1\!+\!\vep)]^2} \int_{\mathbb{R}^2_+}\td x_1\!\wedge\!\td x_2 \hs (x_1x_2)^\vep \hs\tilde{\psi}_{(2,1)}^\bub\Big|{\substack{
\\[1.5mm]
X_1\to X_1+x_1 \\
X_2\to X_2+x_2}} \,,
\end{equation}
where we have also included an overall factor $\,4Y_1Y_2\,$ for later convenience.\
Moreover, from now on and in all subsequent examples, we will ignore the overall common factor $\lam_n^2 (-1)^{1+\vep}/[\Ga(1+\vep)]^2$ appearing in all FRW wavefunction coefficients.

\begin{figure}[t]
\centering
\tikzset{every picture/.style={line width=0.75pt}}
\begin{tikzpicture}[x=0.75pt,y=0.75pt,yscale=-1,xscale=1,scale=.8]
\draw  [line width=1.5]  (316.35,249.75) .. controls (316.35,230.69) and (331.81,215.23) .. (350.88,215.23) .. controls (369.94,215.23) and (385.4,230.69) .. (385.4,249.75) .. controls (385.4,268.82) and (369.94,284.28) .. (350.88,284.28) .. controls (331.81,284.28) and (316.35,268.82) .. (316.35,249.75) -- cycle ;
\draw  [fill={rgb, 255:red, 0; green, 0; blue, 0 }  ,fill opacity=1 ][line width=1.5]  (312.75,249.75) .. controls (312.75,247.77) and (314.36,246.16) .. (316.35,246.16) .. controls (318.34,246.16) and (319.95,247.77) .. (319.95,249.75) .. controls (319.95,251.74) and (318.34,253.35) .. (316.35,253.35) .. controls (314.36,253.35) and (312.75,251.74) .. (312.75,249.75) -- cycle ;
\draw  [fill={rgb, 255:red, 0; green, 0; blue, 0 }  ,fill opacity=1 ][line width=1.5]  (381.8,249.75) .. controls (381.8,247.77) and (383.41,246.16) .. (385.4,246.16) .. controls (387.39,246.16) and (389,247.77) .. (389,249.75) .. controls (389,251.74) and (387.39,253.35) .. (385.4,253.35) .. controls (383.41,253.35) and (381.8,251.74) .. (381.8,249.75) -- cycle ;
\draw [color={rgb, 255:red, 0; green, 0; blue, 0 }  ,draw opacity=1 ][line width=1.2]    (260,179.69) -- (316.35,249.75) ;
\draw [color={rgb, 255:red, 0; green, 0; blue, 0 }  ,draw opacity=1 ][line width=1.2]    (300.33,180.35) -- (316.35,249.75) ;
\draw [color={rgb, 255:red, 0; green, 0; blue, 0 }  ,draw opacity=1 ][line width=1.2]    (440,180.19) -- (385.85,250.09) ;
\draw [color={rgb, 255:red, 0; green, 0; blue, 0 }  ,draw opacity=1 ][line width=1.2]    (400.5,180.19) -- (385.85,250.09) ;
\draw [color=gray  ,draw opacity=1 ][line width=1.5]    (239,180.19) -- (461.67,180.19) ;
\draw (390,248) node [anchor=north west][inner sep=0.75pt]  [font=\small]  {$X_{2}$};
\draw (291,248) node [anchor=north west][inner sep=0.75pt]  [font=\small]  {$X_{1}$};
\draw (344,262) node [anchor=north west][inner sep=0.75pt]  [font=\small]  {$Y_{1}$};
\draw (344,195) node [anchor=north west][inner sep=0.75pt]  [font=\small]  {$Y_{2}$};
\draw (278.5,195) node [anchor=north west][inner sep=0.75pt]{$\cdots$};
\draw (398,195) node [anchor=north west][inner sep=0.75pt]{$\cdots$};
\end{tikzpicture}
\caption{Two-site one-loop bubble diagram, where $X_1, X_2$ are the sum of energies flowing from the vertices to the late-time boundary and $Y_1, Y_2$ represent the internal energies.}
\label{fig:1}
\end{figure}

\subsubsection{Hyperplane Arrangements}
\label{sec:3.1.1}

\begin{figure}[t]
\centering
\includegraphics[scale=.9]{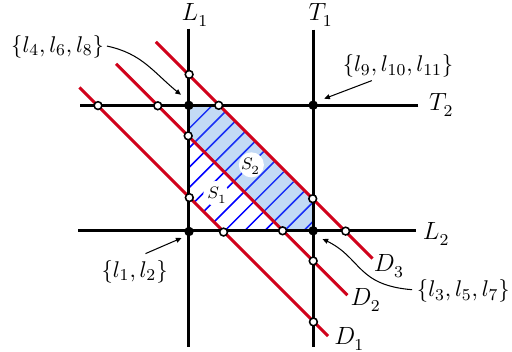}
\vspace*{-2mm}
\caption{Hyperplane arrangement for two-site one-loop bubble wavefunction coefficient.\
All black and white intersection points represent the codimension-2 boundaries.\ 
Among these, the four points $L_1\cap L_2,\, L_1\cap T_2,\, L_2\cap T_1$ and $T_1\cap T_2$ are specifically used to calculate the letters $\{l_i\}$ in our chosen basis.}
\label{fig:2}
\vspace*{-1mm}
\end{figure}

While the integrals in \eqrefe{eq:psi-bub-frw-1} can be computed directly, a more efficient and elegant approach leverages advanced mathematical tools.\ 
Specifically, we employ concepts from intersection theory and twisted cohomology \cite{aomoto1975vanishing,Aomoto:2011ggg,Mastrolia:2018uzb} (cf.\,appendix\,\ref{app:A}) in conjunction with the method of canonical differential equations \cite{Henn:2013pwa,Henn:2014qga} to evaluate \eqrefe{eq:psi-bub-frw-1}.\ 
This allows us to reformulate the two-site one-loop FRW wavefunction coefficient \eqref{eq:psi-bub-frw-1} as follows:
\begin{equation}
\label{eq:psi-bub-frw-2}
\psi_{(2,1)}^\bub
= \int_{\mathbb{R}^2_+} \td x_1\!\wedge\!\td x_2\hs (T_1T_2)^\vep\! \[\frac{4Y_1Y_2}{L_1 L_2 D_3}\! \(\frac{1}{D_1}+\frac{1}{D_2}\)\!\] \equiv \int (T_1T_2)^\vep \,\Omega^2_{\PP_\bub} \,.
\end{equation}
In \eqrefe{eq:psi-bub-frw-2}, we have defined:
\begin{align}
\label{eq:TLD-1}
T_1& = x_1 \,, \qquad T_2 = x_2 \,,  \qquad
\nn\\[.5mm]
L_1 &= x_1\!+\!X_1\!+\!Y_1\!+\!Y_2 \,, \hspace*{1.8cm}
L_2 = x_2\!+\!X_2\!+\!Y_1\!+\!Y_2 \,, \qquad
\nn\\[.5mm]
D_1 &= x_1\!+\!x_2\!+\!X_1\!+\!X_2\!+\!2Y_1 \,, \qquad
D_2 = x_1\!+\!x_2\!+\!X_1\!+\!X_2 \!+\!2Y_2\,,
\nn\\
D_3 &= x_1\!+\!x_2\!+\!X_1\!+\!X_2  \,,
\end{align}
where those divisors are expressed in the linear form, corresponding to seven independent hyperplanes determined by the equations: $\{T_i,L_j,D_k\}\!=\!0$.\ 
Their geometric arrangement in $(x_1, x_2)$-coordinate system is illustrated in fig.\,\ref{fig:2}.\ 
Further, inspecting \eqrefe{eq:psi-bub-frw-2}, the extra normalization factor $\,4Y_1Y_2\,$ is included in order to ensure that the canonical 2-form $\,\Omega^{2}_{\PP_\bub}\,$ has the residues of $\,\pm1\,$ on all codimension-2 boundaries \cite{Arkani-Hamed:2017tmz}.\
Moreover, the branch planes $\{T_1,T_2\}$ correspond to twisted singularities or ``mild'' singularities, regulated by the parameter $\vep$, while the hyperplanes $\{L_i,D_j\}$ are associated with the relative singularities or ``dangerous'' singularities of the integrand \cite{De:2023xue}.\
The overall analytic structure of these integrals is fundamentally governed by the intersections of the surfaces $\{T_i, L_j, D_k\}$, as these intersections dictate the behavior and singularities of the whole system.

As discussed in refs.\,\cite{aomoto1975vanishing,Aomoto:2011ggg,Mastrolia:2018uzb}, the number of independent master integrals is equal to the number of chambers bounded by the hyperplanes \eqref{eq:TLD-1}.\
A direct count in fig.\,\ref{fig:2} yields ten independent bounded chambers.\ However, we temporarily work with an overcomplete basis of twelve triangles, constructed from all distinct triples of intersecting hyperplanes.\ This redundancy is advantageous, as it makes tree-level factorization manifest and allows us to efficiently reuse existing results from the tree-level analysis.\ 
More precisely, our detailed analysis reveals that the differential system of fig.\,\ref{fig:2} arising from the two-site one-loop bubble
graph can be partitioned into three distinct subsystems.\ 
Each subsystem shares four common hyperplanes $\{T_1,T_2,L_1,L_2\}$, and contains one unique diagonally-placed hyperplane $\{D_1,D_2,D_3\}$.\
Accordingly, fig.\,\ref{fig:2} contains a total of twelve bounded triangles, each corresponding to a distinct canonical form from a triple of intersecting hyperplanes, consistent with the framework established in ref.\,\cite{Arkani-Hamed:2017tmz}.\ 
And based on this foundation, refs.\,\cite{Arkani-Hamed:2023bsv, Arkani-Hamed:2023kig} demonstrate that each triangle uniquely maps to a specific wavefunction coefficient.\
Therefore, within each subsystem of hyperplanes, there exists a hierarchical structure of functions comprising a ``parent'' function $\PP_i$ accompanied by three ``descendant'' functions: $\FF_i,\,\FFt_i$ and $\QQ_i$.\
These twelve functions collectively form three sets:
\begin{equation}
\label{eq:Fi}
\bF_i\,\equiv\,\{\,\PP_i,\, \FF_i,\, \FFt_i,\, \QQ_i\,\}\,, \qquad i=1,2,3 \,.
\end{equation}
Each function is associated with a bounded triangle region, which is systematically defined as follows: 
\beqs
\label{eq:function-forms-bub}
\begin{align}
\label{eq:P-123}
\text{Layer-0:} & \quad~ \triangle_{L_1L_2D_i} ~\longleftrightarrow~
\PP_i = \int (T_1T_2)^\vep\! \( \dlog\frac{L_1}{D_i} \wedge \dlog\frac{L_2}{D_i} \) 
\equiv \int\!\td\mu \, \bOme_{\PP_i}\,,
\\[1mm]
\label{eq:F-123}
\text{Layer-1:} & \quad~ \triangle_{T_1L_2D_i} ~\longleftrightarrow~
\FF_i = \int (T_1T_2)^\vep\! \( \dlog\frac{T_1}{D_i} \wedge \dlog\frac{L_2}{D_i} \) 
\equiv \int\!\td\mu \, \bOme_{\FF_i}\,,
\\[1mm]
\label{eq:Ft-123}
& \quad~ \triangle_{L_1T_2D_i} ~\longleftrightarrow~
\FFt_i = \int (T_1T_2)^\vep\! \( \dlog\frac{L_1}{D_i} \wedge \dlog\frac{T_2}{D_i} \) 
\equiv \int\!\td\mu \, \bOme_{\FFt_i}\,,
\\[1mm]
\text{Layer-2:} & \quad~ \triangle_{T_1T_2D_i} ~\longleftrightarrow~
\QQ_i = \int (T_1T_2)^\vep\! \( \dlog\frac{T_1}{D_i} \wedge \dlog\frac{T_2}{D_i} \) 
\equiv \int\!\td\mu \, \bOme_{\QQ_i}\,,
\end{align}
\eeqs
where we have re-defined the canonical forms as: $\Omega\equiv\bOme\,\td x_1\wedge\td x_2$\footnote{For simplicity, from now to the following analysis, we omit the superscript ``2'' of the 2-form $\Omega^2$.}, 
and further combined $\td x_1\wedge\td x_2$ with the prefactor $(T_1T_2)^\vep$ into a compact measure form: $\td\mu\equiv (T_1T_2)^\vep\hs\td x_1\!\wedge\!\td x_2$\,. We also omit writing the integration domain $\mathbb{R}^2_+$.\
To avoid confusion, we will hereafter refer to the coefficient $\bOme$ as the  canonical form.\
Then, we derive those canonical forms as: 
\beqs
\label{eq:forms-bub}
\begin{alignat}{3}
\label{eq:Ome-P-123}
\text{Layer-0:} \quad~ \bOme_{\PP_1} &= \frac{-2Y_2}{L_1L_2D_1}\,, \quad
&\bOme_{\PP_2} &= \frac{-2Y_1}{L_1L_2D_2}\,, \quad 
&\bOme_{\PP_3} &= \frac{-2(Y_1 \!+\! Y_2)}{L_1L_2D_3}\,,
\\[1mm] 
\label{eq:Ome-F-123}
\text{Layer-1:} \quad~ \bOme_{\FF_1} &= \frac{X_1 \!+\! Y_1 \!-\! Y_2}{T_1L_2D_1}\,, \quad 
&\bOme_{\FF_2} &= \frac{X_1 \!-\! Y_1 \!+\! Y_2}{T_1L_2D_2}\,, \quad 
&\bOme_{\FF_3} &= \frac{X_1 \!-\! Y_1 \!-\! Y_2}{T_1L_2D_3}\,, 
\\[1mm] 
\label{eq:Ome-Ft-123}
\bOme_{\FFt_1} &= \frac{X_2 \!+\! Y_1 \!-\! Y_2}{T_2L_1D_1}\,, \quad &\bOme_{\FFt_2} &= \frac{X_2 \!-\! Y_1 \!+\! Y_2}{T_2L_1D_2}\,,\quad 
&\bOme_{\FFt_3} &= \frac{X_2 \!-\! Y_1 \!-\! Y_2}{T_2L_1D_3}\,, 
\\[1mm] 
\label{eq:Ome-Q-123}
\text{Layer-2:} \quad~\bOme_{\QQ_1} &= \frac{X_1 \!+\! X_2 \!+\! 2Y_1}{T_1T_2D_1}\,, \quad 
&\bOme_{\QQ_2} &= \frac{X_1 \!+\! X_2 \!+\! 2Y_2}{T_1T_2D_2} \,,\quad 
&\bOme_{\QQ_3} &= \frac{X_1 \!+\! X_2}{T_1T_2D_3}\,.
\end{alignat}
\eeqs
Specifically, we define canonical forms with a positively oriented convention, ensuring that the aforementioned $\bOme$ are positive within the corresponding bounded region.\ 
In eqs.~\eqref{eq:function-forms-bub}--\eqref{eq:forms-bub}, we have categorized functions (forms) into three layers:
\begin{itemize}[leftmargin=*]
\item 
Layer-0: Functions in this layer do not involve any of the twist planes $T_1$ and $T_2$.

\item 
Layer-1: Functions in this layer are obtained by replacing the $L_1$- or $L_2$-plane in the corresponding layer-0 functions with the twist planes $T_1$ or $T_2$.
For example, $\FF_i,\FFt_i$ can be derived from the parent function as $\FF_i=\PP_i\big|{\substack{
\\[1.5mm] L_1\to T_1}}\,,~
\FFt_i=\PP_i\big|{\substack{
\\[1.5mm] L_2\to T_2}}$.

\item 
Layer-2: Functions in this layer are derived by further replacing the remaining $L_1$- or $L_2$-plane in the corresponding layer-1 functions with the twist planes $T_1$ or $T_2$ i.e., $\QQ_i=\FF_i\big|{\substack{
\\[1.5mm] L_2\to T_2}}=\FFt_i\big|{\substack{
\\[1.5mm] L_1\to T_1}}$.
\end{itemize}
Further, each set $\bF_i$ defined in \eqrefe{eq:Fi} contains four members, each corresponding to a canonical 2-form uniquely associated with a codimension-2 boundary, represented by the four black intersection points shown in fig.\,\ref{fig:2}.\ 
Thus, for each set of $\bF_i$, any form with that set of singularity can be decomposed into this basis by matching residues at these distinctive intersection points, which correspond to the dlog singularities of the form.

However, as noted earlier, the hyperplane arrangement analysis shows that the twelve triangles are linearly dependent and cannot fully span the differential system. Triangulating the bounded chambers reveals two linear relations among them:
\vspace*{-2mm}
\beqs
\label{eq:S1-S2}
\begin{align}
S_1=\,&\triangle_{T_1T_2D_1}-\triangle_{L_1T_2D_1}-\triangle_{L_2T_1D_1}-\triangle_{T_1T_2D_3}
\nn\\
=\,&\triangle_{L_1L_2D_3}-\triangle_{L_1T_2D_3}-\triangle_{L_2T_1D_3}-\triangle_{L_1L_2D_1}\,,
\\[0mm]
S_2=\,&\triangle_{T_1T_2D_2}-\triangle_{L_1T_2D_2}-\triangle_{L_2T_1D_2}-\triangle_{T_1T_2D_3}
\nn\\
=\,&\triangle_{L_1L_2D_3}-\triangle_{L_1T_2D_3}-\triangle_{L_2T_1D_3}-\triangle_{L_1L_2D_2}\,,
\end{align}
\eeqs
where $S_1$ and $S_2$ correspond to the blue-shaded and blue hexagonal regions in fig.\,\ref{fig:2}, respectively. Each region can be constructed by stitching together several triangles defined earlier, with two distinct ways of assembling them.\
Therefore, with a consistent choice of orientation, the constraints \eqref{eq:S1-S2} induce two linear relations among the canonical forms:
\vspace*{-1mm}
\beqs
\label{eq:Om-constraint}
\begin{align}
\bOme_{\QQ_1}-\bOme_{\FFt_1}-\bOme_{\FF_1}-\bOme_{\QQ_3} &\,=\,\bOme_{\PP_3}-\bOme_{\FFt_3}-\bOme_{\FF_3}-\bOme_{\PP_1} \,,
\\
\bOme_{\QQ_2}-\bOme_{\FFt_2}-\bOme_{\FF_2}-\bOme_{\QQ_3} &\,=\,\bOme_{\PP_3}-\bOme_{\FFt_3}-\bOme_{\FF_3}-\bOme_{\PP_2} \,,
\end{align}
\eeqs
where these conditions reduce the number of members of chosen basis from twelve to ten.

Given our focus on the original form $\bOme_{\PP}$ in \eqrefe{eq:psi-bub-frw-2}, we find that the forms $\{\bOme_{\PP_i}\}$ in \eqrefe{eq:P-123} can be combined through the following linear superposition:
\begin{equation}
\label{eq:P=P1+P2-P3}
\bOme_{\PP_\bub} = \bOme_{\PP_1} + \bOme_{\PP_2} - \bOme_{\PP_3}  \,,
\end{equation}
which allows us to select $\PP$\,\footnote{For convenience, we use the notation $\PP$ to represent $\PP_\bub$. Similarly, in section\,\ref{sec:5}, $\PP$ refers to $\PP_\sun$.}
as an independent basis element, rather than relying on the three parent functions $\{\PP_i\}$.\
Therefore, the ten independent members of our chosen basis belong to the following integral family:
\begin{equation}
\label{eq:Bub-I}
\bI_\bub \,=\, \big(\,\PP ,\, \FF_1 ,\, \FF_2 ,\, \FF_3 ,\, \FFt_1,\, \FFt_2,\, \FFt_3,\, \QQ_1,\,\QQ_2,\,\QQ_3\,\big)^T .
\end{equation}
Indeed, the canonical forms associated with the bounded chambers naturally provide an ``$\vep$-form'' \cite{Henn:2013pwa,Henn:2014qga} basis for an integral family $\bI$, where each integral has uniform transcendental weight (UT). Hence, $\bI\,$ satisfies a linear system of differential equations in a simple canonical form, governed by a first-order equation:
\begin{equation}
\label{eq:dI-UT}
\td\hs\bI\hs(\textbf{z},\vep) \,=\, \vep \hs A(\textbf{z}) \hs \bI\hs(\textbf{z},\vep) \,, 
\end{equation}
where $\textbf{z}$ is the set of all independent kinematic variables, for the two-site one-loop bubble case, $\textbf{z}=\{X_1,X_2,Y_1,Y_2\}$.\ And the matrix $A$ has the dlog form:
\begin{equation}
\label{eq:tA}
A(\textbf{z}) \,=\, \sum_i c_i \, \dlog[w_i(\textbf{z})]  \,\equiv\, \sum_i c_i \, l_i\,, 
\end{equation}
where the matrix $A$ obeys the following integrability conditions:
\begin{equation}
\label{eq:inte-con}
\td A=0 \,, \qquad A\wedge A=0\,.
\end{equation}
Further, in the above expression \eqref{eq:tA}, $\{c_i\}$ are the constant matrices, $\{w_i\}$ represent the ``(symbol) letters'' \cite{Goncharov:2010jf,Duhr:2011zq} which are the rational or algebraic functions of kinematic variables, and the set of all letters is called the ``alphabet''.\ For convenience, we have defined the dlog form of $w_i$ as $l_i$ in \eqrefe{eq:tA}.\ In the subsequent discussions, we refer to $l_i$ as the letter and the complete set of all independent $l_i$ as the alphabet.\
In the following sections, we will analyze the structure of $A$ arising in both bubble and tadpole cases.

\subsubsection{Derivation of Canonical DEs}
\label{sec:3.1.2}

In this section, we will calculate the canonical differential equations \eqref{eq:dI-UT} by using the approach of integration-by-parts (IBP)\,
\footnote{
Alternatively, a fully algebraic method using Gelfand-Kapranov-Zelevinsky (GKZ) system has been developed in ref.\,\cite{Fevola:2024nzj}.\ 
Upon restricting the GKZ system to the physical parameter space and applying a suitable gauge transformation, one can obtain the same
$\vep$-factorized canonical differential equations.
}.
We start by considering the first set of \eqrefe{eq:Fi} i.e., $\bF_1=\{\PP_1,\FF_1,\FFt_1,\QQ_1\}$,
where the total derivative for $\bF_1$ is given by
\begin{equation}
\td\bF_1\,=\,
\pd_{X_1}^{}\bF_1\td{X_1} + 
\pd_{X_2}^{}\bF_1\td{X_2} + \pd_{Y_1}^{}\bF_1\td{Y_1} +
\pd_{Y_2}^{}\bF_1\td{Y_2} \,.
\end{equation}
For the parent function $\PP_1$ in $\bF_1$, we observe that the differentiation of $\PP_1$ with respect to the external energy $X_1$ is equal to the derivative with respect to $x_1$. Thus, we can compute
\begin{align}
\label{eq:dX1-P1}
&\pd_{X_1}\PP_1 = \int\td\mu \, \pd_{x_1}\bOme_{\PP_1}
=\vep\!\int\!\td\mu\!\(-\frac{\bOme_{\PP_1}}{T_1}\)
\nn\\[1mm]
&=\vep\!\int\!\td\mu\,\Bigg\{\Bigg(
\rm{Res}\!\[\frac{-\bOme_{\PP_1}}{T_1}\]\!{\substack{\\[5.2mm]
L_1=0 \\ L_2=0}}~,~
\rm{Res}\!\[\frac{-\bOme_{\PP_1}}{T_1}\]\!{\substack{\\[5.2mm]
T_1=0 \\ L_2=0}}\,\Bigg)\!
\(\bOme_{\PP_1},\bOme_{\FF_1}\)^{T}
\Bigg\}
\nn\\[1mm]
&=\vep\!\int\!\td\mu\!\[\frac{1}{X_1\!+\!Y_1\!+\!Y_2}\,\bOme_{\PP_1}
+\frac{2Y_2}{(X_1\!+\!Y_1\!-\!Y_2)(X_1\!+\!Y_1\!+\!Y_2)}\,\bOme_{\FF_1}\]
\nn\\[1mm]
&=\vep\!\[\frac{1}{X_1\!+\!Y_1\!+\!Y_2}\,\PP_1
+\(\frac{1}{X_1\!+\!Y_1\!-\!Y_2}-\frac{1}{X_1\!+\!Y_1\!+\!Y_2}\)\!\FF_1\].
\end{align}
Similarly, the differentiation of $\PP_1$ with respect to $X_2$ proceeds analogously to that of $X_1$ and we simply need to replace $X_1\to X_2$ and $\FF_1\to\FFt_1$ in \eqrefe{eq:dX1-P1} and obtain
\begin{equation}
\pd_{X_2}\PP_1\,=\,\vep\!\[\frac{1}{X_2\!+\!Y_1\!+\!Y_2}\,\PP_1+\(\frac{1}{X_2\!+\!Y_1\!-\!Y_2}-\frac{1}{X_2\!+\!Y_1\!+\!Y_2}\)\!\FFt_1\].
\end{equation}
The calculation of differentiation of $\PP_1$ with respect to the internal energies $Y_1$ and $Y_2$ is non-trivial.\ 
However, these derivatives can be intuitively computed, yielding
\beqs
\begin{align}
\pd_{Y_1}\PP_1&= \int\!\td\mu\!\(\pd_{x_1}\bOme_{\PP_1}+\pd_{x_2}\bOme_{\PP_1}\)
\nn\\
&=\vep\[\(\frac{1}{X_1\!+\!Y_1+Y_2}+\frac{1}{X_2\!+\!Y_1\!+\!Y_2}\)\!\PP_1 + \(\frac{1}{X_1\!+\!Y_1\!-\!Y_2}-\frac{1}{X_1\!+\!Y_1\!+\!Y_2}\)\!\FF_1
\right.\nn\\
&\qquad\left.+\(\frac{1}{X_2\!+\!Y_1\!-\!Y_2}-\frac{1}{X_2\!+\!Y_1\!+\!Y_2}\)\!\FFt_1\],
\\[2mm]
\pd_{Y_2}\PP_1&=\int\!\td\mu\!\[\pd_{x_1}\!\(\frac{-x_1\!-\!X_1\!-\!Y_1}{Y_2}\,\bOme_{\PP_1}\)\!+\pd_{x_2}\!\(\frac{-x_2\!-\!X_2\!-\!Y_2}{Y_1}\,\bOme_{\PP_1}\)\!\]
\nn\\
&=\vep\[\(\frac{1}{X_1\!+\!Y_1\!+\!Y_2}+\frac{1}{X_2\!+\!Y_1\!+\!Y_2}\)\!\PP_1- \(\frac{1}{X_1\!+\!Y_1\!-\!Y_2}+\frac{1}{X_1\!+\!Y_1\!+\!Y_2}\)\!\FF_1
\right. \nn\\
&\qquad \left.-\(\frac{1}{X_2\!+\!Y_1\!-\!Y_2}+\frac{1}{X_2\!+\!Y_1\!+\!Y_2}\)\!\FFt_1\].
\end{align}
\eeqs
In terms of total derivative \eqref{eq:dP1} and expressing in dlog forms, we finally obtain the canonical differential equation for $\PP_1$ as follows:
\begin{align}
\label{eq:dP1}
\td\PP_1 =&\,\vep\Big[(\PP_1-\FF_1)\,\dlog(X_1\!+\!Y_1\!+\!Y_2)+(\PP_1-\FFt_1)\,\dlog(X_2\!+\!Y_1\!+\!Y_2)
\nn\\
&\quad+\, \FF_1\,\dlog(X_1\!+\!Y_1\!-\!Y_2)+ \FFt_1\,\dlog(X_2\!+\!Y_1\!-\!Y_2)\Big]\,.
\end{align}

Next, we examine the differentiation of the descendant functions $\FF_1, \FFt_1$ and $\QQ_1$.\ 
For $\FF_1$, the differential results with respect to external energies are computed as:
\beqs
\begin{align}
&\pd_{X_1}\FF_1 = \int\!\td\mu\!
\[\pd_{x_1}\!\(\frac{-T_1}{X_1\!+\!Y_1\!-\!Y_2}\,\bOme_{\FF_1}\)
+\pd_{x_2}\!\(\frac{-L_2}{X_1\!+\!Y_1\!-\!Y_2}\,\bOme_{\FF_1}\)\!\] 
\nn\\
&=\vep\!\int\!\!\td\mu\,
\Bigg\{\!\Bigg(\rm{Res}\!\[\!\frac{\bOme_{\FF_1}}{X_1\!+\!Y_1\!-\!Y_2}\!\]\!{\substack{\\[5.2mm]T_1=0 \\ L_2=0}} ~,~
\rm{Res}\!\[\!\frac{\bOme_{\FF_1}}{X_1\!+\!Y_1\!-\!Y_2}\!\]\!{\substack{\\[5.2mm]
T_1=0 \\ T_2=0}} \,\Bigg)\!
\(\bOme_{\FF_1},\bOme_{\QQ_1}\)^T
\nn\\
&\hspace*{1.25cm}
+\Bigg(\rm{Res}\!\[\!\frac{L_2\hs\bOme_{\FF_1}}{T_2(X_1\!+\!Y_1\!-\!Y_2)}\!\]\!{\substack{\\[5.2mm]T_1=0 \\ L_2=0}} ~,~
\rm{Res}\!\[\!\frac{L_2\hs\bOme_{\FF_1}}{T_2(X_1\!+\!Y_1\!-\!Y_2)}\!\]\!{\substack{\\[5.2mm]
T_1=0 \\ T_2=0}} \,\Bigg)\!
\(\bOme_{\FF_1},\bOme_{\QQ_1}\)^T \Bigg\}
\nn\\
&=\vep\!\[\frac{1}{X_1\!+\!Y_1\!-\!Y_2}\,\FF_1 + \frac{1}{X_1\!+\!X_2\!+\!2Y_1}\,\QQ_1\],
\\[2mm]
&\pd_{X_2}\FF_1 =\int\!\td\mu\,\pd_{x_2}\bOme_{\FF_1}
\nn\\
&=\vep\!\[\frac{1}{X_2\!+\!Y_1\!+\!Y_2}\,\FF_1 + \(\frac{1}{X_1\!+\!X_2\!+\!2Y_1}-\frac{1}{X_2\!+\!Y_1\!+\!Y_2}\)\!\QQ_1\],
\end{align}
\eeqs
and the results for differentiation with respect to the internal energies are
\beqs
\begin{align}
&\pd_{Y_1}\FF_1=\int\!\td\mu\!\[\pd_{x_1}\!\(\frac{-T_1}{X_1\!+\!Y_1\!-\!Y_2}\,\bOme_{\FF_1}\) \!+ \pd_{x_2}\!\(\frac{- T_1\!-\! 2L_2\!+\!D_1}{X_1\!+\!Y_1\!-\!Y_2}\,\bOme_{\FF_1}\)\!\]
\nn\\
&=\vep\!\[\(\frac{1}{X_1\!+\!Y_1\!-\!Y_2}+\frac{1}{X_2\!+\!Y_1\!+\!Y_2}\)\FF_1 + \(\frac{2}{X_1\!+\!X_2\!+\!2Y_1}-\frac{1}{X_2\!+\!Y_1\!+\!Y_2}\)\!\QQ_1\],
\\[2mm]
&\pd_{Y_2}\FF_1=\int\!\td\mu\!\[\pd_{x_1}\!\(\frac{T_1}{X_1\!+\!Y_1\!-\!Y_2}\,\bOme_{\FF_1}\) \!+ \pd_{x_2}\!\(\frac{-T_1\!+\!D_1}{X_1\!+\!Y_1\!-\!Y_2}\,\bOme_{\FF_1}\)\!\]
\nn\\
&=\vep\!\[\(-\frac{1}{X_1\!+\!Y_1\!-\!Y_2}+\frac{1}{X_2\!+\!Y_1\!+\!Y_2}\)\FF_1 - \frac{1}{X_2\!+\!Y_1\!+\!Y_2}\QQ_1\].
\end{align}
\eeqs
The calculation for $\FFt_1$ is similar to that of $\FF_1$. We will not provide the details.\ 
Hence, in terms of the total derivative, we can obtain the differential equations for $\FF_1$ and $\FFt_1$:
\beqs
\label{eq:dF1-dFt1}
\begin{align}
\td \FF_1&=\vep\big[\FF_1\hs\dlog(X_1\!+\!Y_1\!-\!Y_2)+ (\FF_1\!-\!\QQ_1)\hs\dlog(X_2\!+\!Y_1\!+\!Y_2)
\nn\\
&\qquad+\QQ_1\dlog(X_1\!+\!X_2\!+\!2Y_1) \big]\,,
\\[1mm]
\td \FFt_1&=\vep\big[\FFt_1 \hs \dlog(X_2\!+\!Y_1\!-\!Y_2)+ (\FFt_1\!-\!\QQ_1)\hs \dlog(X_1\!+\!Y_1\!+\!Y_2)
\nn\\
&\qquad+\QQ_1 \hs \dlog(X_1\!+\!X_2\!+\!2Y_1) \big]\,.
\end{align}
\eeqs
Finally, for the function $\QQ_1$, we can derive
\beqs
\begin{align}
\pd_{X_1}\QQ_1&=\int\!\td\mu\[\pd_{x_1}\!\!\(\frac{-T_1}{X_1\!+\!X_2\!+\!2Y_1}\,\bOme_{\QQ_1}\)\!+ \pd_{x_2}\!\!\( \frac{-T_2}{X_1\!+\!X_2\!+\!2Y_1}\,\bOme_{\QQ_1}\)\! \]
\nn\\
&=2\vep\!\int\!\td\mu \! \(\rm{Res}\!\[\!\frac{\bOme_{\QQ_1}}{X_1\!+\!X_2\!+\!2Y_1}\!\]\!{\substack{\\[5.2mm]
T_1=0 \\ T_2=0}}\,\bOme_{\QQ_1}\)
=2\vep\!\(\frac{1}{X_1\!+\!X_2\!+\!2Y_1}\)\!\QQ_1\,,
\\[1mm]
\pd_{X_2}\QQ_1&=\pd_{X_1}\QQ_1=2\vep\!\(\frac{1}{X_1\!+\!X_2\!+\!2Y_1}\)\!\QQ_1 \,,
\\[1mm]
\pd_{Y_1}\QQ_1&=\int\!\td\mu\[\pd_{x_1}\!\!\(\frac{-2\hs T_1}{X_1\!+\!X_2\!+\!2Y_1}\,\bOme_{\QQ_1}\)\!
+\pd_{x_2}\!\!\( \frac{-2\hs T_2}{X_1\!+\!X_2\!+\!2Y_1}\,\bOme_{\QQ_1}\)\! \]
\nn\\
&= \frac{4\hs\vep}{X_1 \!+\! X_2 \!+\! 2 Y_1}\QQ_1 \,,
\\[1mm]
\pd_{Y_2}\QQ_1 &=0 \,.
\end{align}
\eeqs
Therefore, the differential equation for $\QQ_1$ is given by
\begin{equation}
\label{eq:dQ1}
\td\QQ_1\,=\,2\hs\vep\hs\QQ_1\hs\dlog(X_1\!+\!X_2\!+\!2Y_1)\,.
\end{equation}

The canonical differential equations for $\bF_2$ can be directly derived from those of $\bF_1$ by substituting $Y_1\to Y_2$ in the dlog forms and relabeling the function subscripts $1 \to 2$.\
The resulting equations are summarized as follows:
\beqs
\label{eq:dPFFtQ-2}
\begin{align}
\label{eq:dP2}
\td \PP_2 &\,=\, \vep\hs\big[(\PP_2-\FF_2)\hs\dlog(X_1\!+\!Y_1\!+\!Y_2)+(\PP_2-\FFt_2)\hs\dlog(X_2\!+\!Y_1\!+\!Y_2)
\nn\\
&\hspace*{.9cm}+\!\FF_2\hs\dlog(X_1\!-\!Y_1\!+\!Y_2)+ \FFt_2\hs\dlog(X_2\!-\!Y_1\!+\!Y_2)\big]\,,
\\[1.5mm]
\td \FF_2 &\,=\, \vep\hs\big[\FF_2\hs \dlog(X_1\!-\!Y_1\!+\!Y_2)+ (\FF_2-\QQ_2)\hs\dlog(X_2\!+\!Y_1\!+\!Y_2)
\nn\\
&\hspace*{.9cm}+\!\QQ_2\hs\dlog(X_1\!+\!X_2\!+\!2Y_2) \big]\,,
\\[1.5mm]
\td \FFt_2 &\,=\, \vep\hs\big[\FFt_2\hs \dlog(X_2\!-\!Y_1\!+\!Y_2)+ (\FFt_2-\QQ_2)\hs \dlog(X_1\!+\!Y_1\!+\!Y_2)
\nn\\
&\hspace*{.9cm}+\!\QQ_2\hs \dlog(X_2\!+\!X_2\!+\!2Y_2) \big]\,,
\\[1.5mm]
\td \QQ_2 &\,=\, 2\hs\vep\hs\QQ_2 \hs \dlog(X_1\!+\!X_2\!+\!2Y_2)\,.
\end{align}
\eeqs
In appendix\,\ref{app:B1}, we provide the detailed derivation of the canonical differential equations for $\bF_3$.\ 
Here, we only summarize the results as follows:
\beqs
\label{eq:dPFFtQ-3}
\begin{align} 
\label{eq:dP3}
\td\PP_3 &\,=\, \vep\hs\big[(\PP_3-\FF_3)\hs\dlog(X_1\!+\!Y_1\!+\!Y_2)+(\PP_3-\FFt_3)\hs\dlog(X_2\!+\!Y_1\!+\!Y_2)
\nn\\
&\hspace*{.9cm}+\! \FF_3\hs\dlog(X_1\!-\!Y_1\!-\!Y_2)+ \FFt_3\hs\dlog(X_2\!-\!Y_1\!-\!Y_2)\big]\,,
\\[1.5mm]
\td \FF_3 &\,=\, \vep\hs[\FF_3\hs\dlog(X_1\!-\!Y_1\!-\!Y_2)\!+\!(\FF_3\!-\!\QQ_3)\hs\dlog(X_2\!+\!Y_1\!+\!Y_2)\!+\!\QQ_3\hs\dlog(X_1\!+\!X_2)] \,,
\\[1.5mm]
\td \FFt_3& \,=\,\vep\hs[\FFt_3\hs\dlog(X_2\!-\!Y_1\!-\!Y_2)\!+\!(\FFt_3\!-\!\QQ_3)\hs\dlog(X_1\!+\!Y_1\!+\!Y_2)\!+\!\QQ_3\hs\dlog(X_1\!+\!X_2)]\,,
\\[1.5mm]
\td \QQ_3 &\,=\,2\hs\vep\hs\QQ_3\hs\dlog(X_1\!+\!X_2) \,.
\end{align}
\eeqs
In addition, based on \eqrefe{eq:P=P1+P2-P3}, we combine \eqrefe{eq:dP1}, \eqrefe{eq:dP2} with \eqrefe{eq:dP3} to derive the final differential equation for $\PP$\,:
\begin{align} 
\label{eq:dP-merge}
\td\PP &=\vep\hs \Big[\Big(\PP\!-\! \sum_{i=1}^{3}\FF_i\Big)\hs\dlog(X_1\!+\!Y_1\!+\!Y_2)\!+\!\Big(\PP \!-\!\sum_{j=1}^{3}\FFt_j\Big)\hs\dlog(X_2\!+\!Y_1\!+\!Y_2)
\nn\\
&\qquad +\FF_1\hs\dlog(X_1\!+\!Y_1\!-\!Y_2) + \FFt_1\hs\dlog(X_2\!+\!Y_1\!-\!Y_2)
+\FF_2\hs\dlog(X_1\!-\!Y_1\!+\!Y_2) 
\nn\\[1mm]
&\qquad +\FFt_2\hs\dlog(X_2\!-\!Y_1\!+\!Y_2) +\FF_3\hs\dlog (X_1\!-\!Y_1\!-\!Y_2)\!+\!\FFt_3\hs\dlog (X_2\!-\!Y_1\!-\!Y_2)\Big]\,,
\end{align}
where we have rescaled\, $\FF_3\to-\FF_3\,,\,\FFt_3\to-\FFt_3$ and $\QQ_3\to-\QQ_3$.\ 
These rescalings are permissible due to the homogeneous nature of the system of canonical differential equations, which is invariant under sign changes of the components.

Then, by systematically compiling and analyzing the canonical differential equations in eqs.\,\eqref{eq:dP1}, \eqref{eq:dF1-dFt1}, \eqref{eq:dQ1}, \eqref{eq:dPFFtQ-2}, \eqref{eq:dPFFtQ-3} and \eqref{eq:dP-merge}, we have organized the results into a coherent structure. The final outcome can be summarized as follows:
\beqs
\label{eq:Bub-DEs-P-F-Ft-Q}
\begin{align} 
\label{eq:Bub-DEs-P}
\td\PP 
&=\vep\hs\big[\hs\PP(l_1+l_2)+\FF_1 (l_3-l_1) + \FFt_1 (l_4-l_2)+\FF_2 (l_5-l_1) +\FFt_2 (l_6 -l_2)
\nn\\
&\qquad+\FF_3 (l_7-l_1)+\FFt_3 (l_8-l_2)\hs\big]\,,
\\[1mm]
\label{eq:Bub-DEs-F-Ft-1}
\td \FF_1 &= \vep\hs\big[\FF_1 (l_2+l_3)+\QQ_1 (l_9-l_2)\hs\big] \,, \qquad~
\td \FFt_1 =\vep\hs\big[\FFt_1 (l_1+l_4)+\QQ_1 (l_9-l_1)\hs\big]\,,
\\
\label{eq:Bub-DEs-F-Ft-2}
\td \FF_2 &= \vep\hs\big[\FF_2 (l_2+l_5) +\QQ_2 (l_{10}-l_2)\hs\big]\,, \qquad
\td \FFt_2 =\vep\hs\big[\FFt_2 (l_1+l_6)+\QQ_2 (l_{10}-l_1)\hs\big] \,,
\\
\label{eq:Bub-DEs-F-Ft-3}
\td\FF_3 &= \vep\big[\FF_3 (l_2+l_7)+\QQ_3 (l_{11}-l_2)\big]\,, \qquad\,
\td\FFt_3 =\vep\big[\FFt_3 (l_1+l_8)+\QQ_3 (l_{11}-l_1)\big]\,,
\\
\label{eq:Bub-DEs-Q}
\td\QQ_1 &= 2\hs \vep\hs \QQ_1 l_9 \,, \qquad~~
\td\QQ_2 = 2\hs\vep\hs \QQ_2 l_{10} \,, \qquad~~
\td\QQ_3 = 2\hs\vep\hs \QQ_3 l_{11} \,.
\end{align}
\eeqs
Here, in the expressions of \eqrefe{eq:Bub-DEs-P-F-Ft-Q}, the letters are defined as: $l_i=\dlog(w_i)$ with $i=1,\ldots,11$, and the explicit forms of the letters are given by
\begin{alignat}{3}
\label{eq:bub-letters}
l_1 &=\dlog(X_1\!+\!Y_1\!+\!Y_2)\,, \qquad~ &
l_2 &=\dlog(X_2\!+\!Y_1\!+\!Y_2)\,, \qquad~ &
l_3 &=\dlog(X_1\!+\!Y_1\!-\!Y_2)\,,
\nn\\[1mm]
l_4 &=\dlog(X_2\!+\!Y_1\!-\!Y_2)\,, \qquad~ &
l_5 &=\dlog(X_1\!-\!Y_1\!+\!Y_2)\,, \qquad~ &
l_6 &=\dlog(X_2\!-\!Y_1\!+\!Y_2)\,,
\nn\\[1mm]
l_7 &=\dlog(X_1\!-\!Y_1\!-\!Y_2)\,, \qquad~ &
l_8 &=\dlog(X_2\!-\!Y_1\!-\!Y_2)\,, \qquad~ &
l_9 &=\dlog(X_1\!+\!X_2\!+\!2Y_1)\,,
\nn\\[1mm]
l_{10} &=\dlog(X_1\!+\!X_2\!+\!2Y_2)\,, \qquad~ &
l_{11} &=\dlog(X_1\!+\!X_2) \,. &
\end{alignat}
The solution to \eqrefe{eq:Bub-DEs-P-F-Ft-Q} closely parallels the tree-level procedure established in ref.\,\cite{Arkani-Hamed:2023bsv} and proceeds as follows:
(i).\,Solve the homogeneous equations for $\QQ_i$;
(ii).\,Substitute the resulting $\QQ_i$ into the equations for $\FF_i$ and $\FFt_i$ to determine their explicit forms;
(iii).\,Define the homogeneous component $\td\PP_{\text{hom}} \equiv \td\PP - \sum_{i=1}^3 (\td\FF_i + \td\FFt_i - \td\QQ_i)$ and solve for $\PP_{\text{hom}}$;
(iv).\,Finally, substitute the solutions for $\FF_i$, $\FFt_i$, and $\QQ_i$ into $\PP_{\text{hom}}$ to reconstruct the full solution for $\PP$.

We now turn to the structure of matrix $A$ appearing in \eqrefe{eq:dI-UT}. Based on the differential system given in \eqrefe{eq:Bub-DEs-P-F-Ft-Q}, the explicit form of $A$ for the current bubble configuration is
\begin{equation}
\label{eq:At-bub}
\hspace*{-3mm}
A_\bub =\!\(
\begin{array}{cccccccccc}
l_1\!+\!l_2 &~l_3\!-\!l_1 &~l_5\!-\!l_1 &~l_7\!-\!l_1 
&~l_4\!-\!l_2 &~l_6\!-\!l_2 &~l_8\!-\!l_2 &~0&~0&~0
\\[.5mm]
0&~l_2\!+\!l_3 &~0&~0&~0&~0&~0&~l_9\!-\!l_2 &~0&~0
\\[.5mm]
0&~0&~l_2\!+\!l_5 &~0&~0&~0&~0&~0&~l_{10}\!-\!l_2 &~0
\\[.5mm]
0&~0&~0&~l_2\!+\!l_7 &~0&~0&~0&~0&~0&~l_{11}\!-\!l_2 
\\[.5mm]
0&~0&~0&~0&~l_1\!+\!l_4 &~0&~0&~l_9\!-\!l_1 &~0&~0
\\[.5mm]
0&~0&~0&~0&~0&~l_1\!+\!l_6 &~0&~0&~l_{10}\!-\!l_1 &~0
\\[.5mm]
0&~0&~0&~0&~0&~0&~l_1\!+\!l_8 &~0&~0& ~l_{11}\!-\!l_1 
\\[.5mm]
0&~0&~0&~0&~0&~0&~0 &~2l_9 &~0&~0
\\[.5mm]
0&~0&~0&~0&~0&~0&~0&~0 &~2l_{10} &~0
\\[.5mm]
0&~0&~0&~0&~0&~0&~0&~0&~0 &~2l_{11} 
\end{array}
\)\!,
\end{equation}
where a careful verification confirms that $A_\bub$ satisfies the integrability conditions outlined in \eqrefe{eq:inte-con}, ensuring consistency within the underlying theoretical framework.

An intriguing aspect of our analysis lies in applying a finite sequence of elementary transformations to the matrix \eqref{eq:At-bub}, which leads to the following matrix:
\begin{equation}
A^{\pp}_\bub=\(\begin{array}{cccccccccc}
2l_{11}&~0&~0&~0&~0&~0&~0&~0&~0&~0
\\[.5mm]
0&~2l_9 &~0&~0&~0&~0&~0&~0&~0&~0
\\[.5mm]
0&~0&~2l_{10} &~0&~0&~0&~0&~0&~0&~0
\\ [.5mm]
l_1\!-\!l_8 &~0&~0&~l_8\!+\!l_1 &~0&~0&~0&~0&~0&~0
\\ [.5mm]
l_7\!-\!l_2 &~0&~0&~0&~l_2\!+\!l_7 &~0&~0&~0&~0&~0
\\[.5mm]
0&~0&~0&~0&~0&~l_1\!+\!l_2 &~0&~0&~0&~0
\\[.5mm]
0&~l_4\!-\!l_1 &~0&~0&~0&~0&~l_1\!+\!l_4 &~0&~0&~0
\\[.5mm]
0&~0&~l_6\!-\!l_1 &~0&~0&~0&~0&~l_1\!+\!l_6 &~0&~0
\\[.5mm]
0&~l_2\!-\!l_3 &~0&~0&~0&~0&~0&~0&~l_2\!+\!l_3 &~0
\\[.5mm]
0&~0&~l_2\!-\!l_5 &~0&~0&~0&~0&~0&~0&~l_2\!+\!l_5
\end{array}\)\!.
\end{equation}
This matches perfectly with the matrix derived using the (relative twisted) dual cohomology presented in ref.\,\cite{De:2023xue}.\
Therefore, for the first time, we have reproduced the two-site one-loop result using the method of twisted cohomology.\
This agreement provides additional confidence in our methods and highlights the robustness of our findings in relation to established theoretical frameworks.

Furthermore, we can demonstrate that in the limit $\vep\to 0$, the equations derived in FRW cosmology will reduce to the equations in dS spacetime.\
By imposing the following rescaling in \eqrefe{eq:Bub-DEs-P-F-Ft-Q}
\begin{equation}
\vspace*{-2mm}
\FF_i ~\to~ \FF_i/\vep \,, \qquad~~
\FFt_i ~\to~ \FFt_i/\vep \,, \qquad~~
\QQ_i ~\to~ \QQ_i/\vep^2 \,,
\end{equation}
and taking the limit $\vep\to 0$, we can obtain
\begin{align}
\label{eq:Bub-DEs-P-F-Ft-Q-dS}
\td\PP &=\FF_1 \hs(l_3\!-\!l_1) + \FFt_1 \hs(l_4\!-\!l_2)
+\FF_2 \hs(l_5\!-\!l_1) +\FFt_2 \hs(l_6 \!-\!l_2)
+\FF_3 \hs(l_7\!-\!l_1)+\FFt_3 \hs(l_8\!-\!l_2)\,,
\nn\\[1mm]
\td \FF_1 &= \QQ_1 \hs(l_9\!-\!l_2) \,, \quad
~\td \FFt_1 = \QQ_1 \hs(l_9\!-\!l_1)\,, \quad
\td \FF_2 = \QQ_2 \hs(l_{10}\!-\!l_2) \,, \quad
\td \FFt_2 = \QQ_2 \hs(l_{10}\!-\!l_1) \,,
\nn\\[1mm]
\td \FF_3 &= \QQ_3 \hs(l_{11}\!-\!l_2) \,, \quad
\td \FFt_3 = \QQ_3 \hs(l_{11}\!-\!l_1) \,, \quad
\td \QQ_1 = \td \QQ_2=\td \QQ_3=0\,,
\end{align}
where it is not difficult to check the matrix $A_\bub^{\rm{dS}}$ obeys the integrability conditions \eqref{eq:inte-con}.\
Subsequently, based on the differential equations \eqref{eq:Bub-DEs-P-F-Ft-Q-dS}, we can derive the symbol for two-site one-loop bubble correlator:
\begin{align}
\label{eq:Symbol-Bub}
\mS_{(2,1)}^\bub =&~ (w_9/w_2)\!\otimes\!(w_3/w_1)+(w_9/w_1)\!\otimes\!(w_4/w_2)
\nn\\[1mm]
&~+(w_{10}/w_2)\!\otimes\!(w_5/w_1)+(w_{10}/w_1)\!\otimes\!(w_6/w_2)
\nn\\[1mm]
&~+(w_{11}/w_2)\!\otimes\!(w_1/w_7)+(w_{11}/w_1)\!\otimes\!(w_2/w_8) \,,
\end{align}
where we revert the previously rescaled $\FF_3,\FFt_3$ and $\QQ_3$ back to their original forms.\
Further, by using the package of \textit{PolyLogTools} \cite{Duhr:2019tlz}, we can derive the transcendental function with transcendental degree-2 after integrating the symbol \eqref{eq:Symbol-Bub},
\begin{align}
f_{(2,1)}^\bub =&~-\pi^2/6+\rm{Li}_2(w_3/w_9)+\rm{Li}_2(w_4/w_9)+\rm{Li}_1(w_3/w_9) \rm{Li}_1(w_4/w_9) 
\nn\\
&+\rm{Li}_2(w_5/w_{10})+\rm{Li}_2(w_6/w_{10})+\rm{Li}_1(w_5/w_{10})\rm{Li}_1(w_6/w_{10}) 
\nn\\
&-\rm{Li}_2(w_7/w_{11})-\rm{Li}_2(w_8/w_{11})-\rm{Li}_1(w_7/w_{11}) \rm{Li}_1(w_8/w_{11})  \,.
\end{align}
Both $\mS_{(2,1)}^\bub$ and $f^\bub_{(2,1)}$ are consistent with the results presented in ref.\,\cite{Hillman:2019wgh}.\
This alignment reinforces the correctness of our method and demonstrates that our approach accurately captures the underlying physical relationships, contributing to a reliable interpretation of the two-site one-loop graph in de Sitter spacetime.

\subsection{Two-Site One-Loop Tadpole}
\label{sec:3.2}

\begin{figure}[t]
\centering
\begin{tikzpicture}[x=0.75pt,y=0.75pt,yscale=-1,xscale=1,scale=.8]
\draw  [fill={rgb, 255:red, 0; green, 0; blue, 0 }  ,fill opacity=1 ][line width=1.5]  (86.17,174.94) .. controls (86.17,172.73) and (87.96,170.94) .. (90.17,170.94) .. controls (92.38,170.94) and (94.17,172.73) .. (94.17,174.94) .. controls (94.17,177.15) and (92.38,178.94) .. (90.17,178.94) .. controls (87.96,178.94) and (86.17,177.15) .. (86.17,174.94) -- cycle ;
\draw [line width=1.5]    (160.85,174.94) -- (90.17,174.94) ;
\draw  [line width=1.5]  (162.85,174.25) .. controls (162.85,155.99) and (177.66,141.18) .. (195.92,141.18) .. controls (214.19,141.18) and (229,155.99) .. (229,174.25) .. controls (229,192.52) and (214.19,207.33) .. (195.92,207.33) .. controls (177.66,207.33) and (162.85,192.52) .. (162.85,174.25) -- cycle ;
\draw  [fill={rgb, 255:red, 0; green, 0; blue, 0 }  ,fill opacity=1 ][line width=1.5]  (159.25,174.25) .. controls (159.25,172.27) and (160.86,170.66) .. (162.85,170.66) .. controls (164.84,170.66) and (166.45,172.27) .. (166.45,174.25) .. controls (166.45,176.24) and (164.84,177.85) .. (162.85,177.85) .. controls (160.86,177.85) and (159.25,176.24) .. (159.25,174.25) -- cycle ;
\draw [color={rgb, 255:red, 0; green, 0; blue, 0 }  ,draw opacity=1 ][line width=1.2]    (60,106.19) -- (90.17,174.94) ;
\draw [color={rgb, 255:red, 0; green, 0; blue, 0 }  ,draw opacity=1 ][line width=1.2]    (97.75,105.69) -- (90.17,174.94) ;
\draw [color={rgb, 255:red, 0; green, 0; blue, 0 }  ,draw opacity=1 ][line width=1.2]    (178.5,105.19) -- (160.85,175.59) ;
\draw [color={rgb, 255:red, 0; green, 0; blue, 0 }  ,draw opacity=1 ][line width=1.2]    (143.25,105.69) -- (162.85,174.25) ;
\draw [color=gray  ,draw opacity=1 ][line width=1.5]    (14.5,105.69) -- (250.5,105.69) ;
\draw (143,184.5) node [anchor=north west][inner sep=0.75pt]  [font=\small] {$X_{2}$};
\draw (80.17,184.5) node [anchor=north west][inner sep=0.75pt]
[font=\small]  {$X_{1}$};
\draw (120.17,155.34) node [anchor=north west][inner sep=0.75pt]  [font=\small]{$Y_{1}$};
\draw (233.67,165.84) node [anchor=north west][inner sep=0.75pt]  [font=\small]{$Y_{2}$};
\draw (70,117) node [anchor=north west][inner sep=0.75pt]{$\cdots$};
\draw (150,117) node [anchor=north west][inner sep=0.75pt]{$\cdots$};
\end{tikzpicture}
\caption{Two-site one-loop tadpole diagram.}
\label{fig:3}
\end{figure}

At two-site one-loop, another topology is the tadpole contribution which is shown in fig.\,\ref{fig:3}.\
Following the Feynman rules given in section\,\ref{sec:2}, we can compute the wavefunction coefficient for the two-site one-loop tadpole graph in flat spacetime as follows:
\begin{align}
\label{eq:psi-tad-flat}
\tilde{\psi}_{(2,1)}^\tad &= \ii^2\int_{-\infty}^{0}\td\eta_1\int_{-\infty}^{0} \td\eta_2\,
K_1(X_1;\eta_1)\hs K_2(X_2;\eta_2)\hs G_1(Y_1;\eta_1,\eta_2)\hs G_2(Y_2;\eta_2,\eta_2)
\nn\\
&=\frac{X_1\!+\!Y_1\!+\! 2(X_2\!+\!Y_2)}
{(X_1\!+\!X_2)(X_1\!+\!Y_1)(X_2\!+\!Y_1)(X_1\!+\!X_2\!+\!2Y_2)(X_2\!+\!Y_1\!+\!2Y_2)}\,.
\end{align}
Similarly, the FRW wavefunction coefficient for tadpole can be obtained by imposing the energy shifts: $(X_1,\,X_2)\to(X_1+x_1,\,X_2+x_2)$ on its flat-space form.\
After incorporating the twister and normalization factors, we have 
\begin{align}
\label{eq:psi-1loop-tad}
\psi_{(2,1)}^\tad &= 4Y_1Y_2 \int_{\mathbb{R}^2_+}\td x_1\!\wedge\!\td x_2 \hs (x_1x_2)^\vep \hs\tilde{\psi}_{(2,1)}^\tad\Big|{\substack{
\nn\\[1.5mm]
X_1\to X_1+x_1 \\
X_2\to X_2+x_2}}
\nn\\
&= \int_{\mathbb{R}^2_+} \td x_1\!\wedge\!\td x_2\hs (T_1T_2)^\vep \!\[\frac{4Y_1Y_2}{L_1\bar{L}_2D_3}\!\(\frac{1}{L_2}+\frac{1}{D_2}\)\!\]\!
\equiv\!\int\!\td\mu\, \bOme_{\PP_\tad}\,,
\end{align}
where the hyperplanes are defined as follows:
\begin{align}
\label{eq:TLD-2}
T_1& = x_1 \,, \qquad T_2 = x_2 \,, 
\nn\\[.5mm]
L_1 &= x_1+X_1+Y_1 \,, \qquad
L_2 = x_2+X_2+Y_1 \,, \qquad
\bar{L}_2 = x_2+X_2+Y_1+2Y_2 \,,
\nn\\[.5mm]
D_2 &= x_1+x_2+X_1+X_2 +2Y_2\,, \qquad
D_3 = x_1+x_2+X_1+X_2  \,.
\end{align}

\subsubsection{Hyperplane Arrangements and Derivation of Canonical DEs}
\label{sec:3.2.1}

\begin{figure}[t]
\centering
\includegraphics[scale=.9]{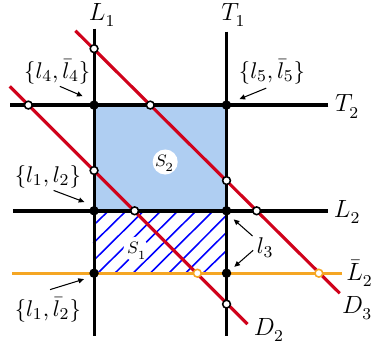}
\vspace*{-2mm}
\caption{Hyperplane arrangement for two-site one-loop tadpole wavefunction coefficient. All black and white intersection points represent codimension-2 boundaries. Among them, the six points $L_1\cap L_2$, $L_1\cap\bar{L}_2$, $L_1\cap T_2$, $L_2\cap T_1$, $\bar{L}_2\cap T_1$ and $T_1\cap T_2$ are used to determine the letters $\{l_i\}$ in the chosen basis.}
\label{fig:4}
\end{figure}

Similar to the analysis of bubble diagram, in \eqrefe{eq:TLD-2}, those hyperplanes can be arranged in $(x_1,x_2)$-plane as illustrated in fig.\,\ref{fig:4}.\
In this factorization, one term is exactly tree function and the other is a shifted tree with $X_2\to X_2+2Y_2$.\ 
This observation enables us to immediately read out the differential equations obeyed by the tadpole system, which is nothing but a linear combination of the two tree-level systems.\ 
Therefore, the total integral family $\bI_\tad$ contains eight independent members which can be divided into two sets:
\begin{equation}
\label{eq:I-tad-phy}
\bI_\tad = \big(\bF_2\cup\bF_3\big)^T ,
\end{equation}
with $\bF_2=\{\PP_2, \FF_2, \FFt_2,\QQ_2\}$ and $\bF_3=\{\PP_3, \FF_3, \FFt_3,\QQ_3\}$.\
The associated canonical forms for $\bF_2$ and $\bF_3$ are defined in the same way as \eqrefe{eq:forms-bub}:
\beqs
\label{eq:forms-tad-8}
\begin{alignat}{3}
\text{Layer-0:} \quad~ \bOme_{\PP_2} &= \frac{-2Y_1}{L_1\bar{L}_2D_2}\,, \hspace*{2.7cm}
&\bOme_{\PP_3} &= \frac{-2Y_1}{L_1L_2D_3}\,, 
\\[1mm] 
\text{Layer-1:} \quad~ \bOme_{\FF_2} &= \frac{X_1\!-\!Y_1}{T_1\bar{L}_2D_2} \,, \qquad~~
&\bOme_{\FF_3} &= \frac{X_1\!-\!Y_1}{T_1L_2D_3}\,, 
\\[1mm] 
\bOme_{\FFt_2} &= \frac{X_2\!-\!Y_1\!+\!2Y_2}{T_2L_1D_2}\,, \quad &\bOme_{\FFt_3} &= \frac{X_2\!-\!Y_1}{T_2L_1D_3}\,,
\\[1mm] 
\text{Layer-2:} \quad~\bOme_{\QQ_2} &= \frac{X_1\!+\!X_2\!+\!2Y_2}{T_1T_2D_2}\,, \qquad~~
&\bOme_{\QQ_3} &= \frac{X_1\!+\!X_2}{T_1T_2D_3}\,.
\end{alignat}
\eeqs
Notably, unlike the bubble case, the $D_1$-plane in the hyperplane arrangement (cf.\,fig.\,\ref{fig:3}) transforms into the hyperplane $\bar{L}_2$ which is parallel to hyperplanes $T_2$ and $L_2$ in the tadpole configuration (cf.\,fig.\,\ref{fig:4}).\ 
This transformation eliminates the presence of $\,\bF_1$ in the bubble case.\ 
Meanwhile, the definitions of canonical forms for $\bF_2$ and $\bF_3$ in \eqrefe{eq:forms-tad-8} remain similar to those in bubble case \eqref{eq:forms-bub}, albeit with some slight modifications to the definition of hyperplanes [cf.\,\eqrefe{eq:TLD-2}].

Then, using IBP method presented in section\,\ref{sec:3.2.1}, we can derive all the canonical differential equations associated with \eqrefe{eq:forms-tad-8}.\ The results are summarized below:
\beqs
\label{eq:Tad-DEs-P-F-Ft-Q}
\begin{align} 
\label{eq:tad-DE-P2P3}
\td\PP_2 &= \vep\hs\big[\hs\PP_2\hs(l_1+\bar{l}_2)+\FF_2 \hs (l_3-l_1) + \FFt_2 \hs(\bar{l}_4-\bar{l}_2)]\,,
\nn\\[1mm]
\td\PP_3 &= \vep\hs\big[\PP_3\hs(l_1+l_2)+\FF_3\hs (l_3-l_1) +\FFt_3 \hs(l_4 -l_2)]\,,
\\[1mm]
\label{eq:tad-DE-F2Ft2}
\td \FF_2 &= \vep\hs\big[\FF_2 \hs(\bar{l}_2+l_3)+\QQ_2\hs (\bar{l}_5-\bar{l}_2)\hs\big] \,, \qquad~
\td \FFt_2 =\vep\hs\big[\FFt_2 \hs(l_1+\bar{l}_4)+\QQ_2\hs (\bar{l}_5-l_1)\hs\big]\,,
\\[1mm]
\label{eq:tad-DE-F3Ft3}
\td \FF_3 &= \vep\hs\big[\FF_3\hs (l_2+l_3)+\QQ_3\hs (l_5-l_2)\hs\big] \,, \qquad~
\td \FFt_3 =\vep\hs\big[\FFt_3 \hs(l_1+l_4)+\QQ_3\hs (l_5-l_1)\hs\big]\,,
\\[1mm]
\label{eq:tad-DE-Q2Q3}
\td \QQ_2 &=2 \vep\hs \QQ_1 \hs\bar{l}_5 \,, \qquad
\td \QQ_3 =2 \vep\hs \QQ_2 \hs l_{5} \,, 
\end{align}
\eeqs
where the letters $l_i=\dlog (w_i)$ and $\bar{l}_i=\dlog (\bar{w}_i)$, which are given by
\begin{align}
\label{eq:tad-letters}
l_1 &=\dlog(X_1\!+\!Y_1)\,, \qquad~
l_2 =\dlog(X_2\!+\!Y_1)\,, \qquad~~
\bar{l}_2 =\dlog(X_2\!+\!Y_1\!+\!2Y_2)\,,
\nn\\[1mm]
l_3 &=\dlog(X_1\!-\!Y_1)\,, \qquad~
l_4 =\dlog(X_2\!-\!Y_1) \,, \qquad~~
\bar{l}_4 =\dlog(X_2\!-\!Y_1\!+\!2Y_2)\,,
\nn\\[1mm]
l_5 &=\dlog(X_1\!+\!X_2) \,,\qquad
\bar{l}_5 =\dlog(X_1\!+\!X_2\!+\!2Y_2) \,.
\end{align}
Note that a key difference arises compared to the bubble case. While the differential forms $\bOme_{\PP_\tad},\bOme_{\PP_2}$ and $\bOme_{\PP_3}$ satisfy the linear relation:
\begin{equation}
\label{eq:Omega-tad}
\bOme_{\PP_\tad}=\bOme_{\PP_2}-\bOme_{\PP_3} \,,
\end{equation}
it is not possible, within our chosen basis, to combine the differential equations of $\PP_2$ and $\PP_3$ into a single equation for $\PP_\tad$.\ 
This is because $\bOme_{\PP_2}$ and $\bOme_{\PP_3}$ are associated with different $L$\hs-pairs, leading to distinct letters, so that the corresponding differential system involves two separate parent functions.\
A detailed discussion will be provided in section\,\ref{sec:3.2.2}.\ 
Further, the wavefunction coefficient for a tadpole graph, analyzed through the positive geometry \cite{Arkani-Hamed:2017tmz}, corresponds to a cosmological polytope with the structure of a square pyramid.\ 
This polytope can be effectively triangulated into two smaller, distinct polytopes \cite{Benincasa:2018ssx}, each of which corresponds to a tree-level diagram.\ 
Among the two tree diagrams, the external energy $X_i$ in one tree diagram is shifted according to the internal line of the tadpole it connects to.\
Specifically, $\PP_2$ and $\PP_3$ correspond to the shifted and unshifted tree diagram.\
We will provide more details on higher-loop tadpole cases at the end of section\,\ref{sec:5}, which similarly involve multiple parent functions.\
Hence, when selecting the integral family $\bI_\tad$, we treat $\PP_2$ and $\PP_3$ as two independent elements.

Similarly, by taking the dS limit $\vep\!\to\!0$, canonical differential equations \eqref{eq:Tad-DEs-P-F-Ft-Q} will reduce to the following forms:
\beqs
\begin{alignat}{3}
\label{eq:tad-DE-P2P3-dS}
\td\PP_2 &=\FF_2\hs (l_3\!-\!l_1) 
+ \FFt_2\hs (\bar{l}_4\!-\!\bar{l}_2)\,, \qquad~
&&\td\PP_3 =\FF_3\hs (l_3\!-\!l_1) 
+\FFt_3\hs (l_4 \!-\!l_2) \,,
\\[1mm]
\label{eq:tad-DE-F2Ft2-dS}
\td \FF_2&= \QQ_2\hs (\bar{l}_5\!-\!\bar{l}_2) \,, \qquad~
&&\td \FFt_2= \QQ_2\hs (\bar{l}_{5}\!-\!l_1) \,,
\\[1mm]
\label{eq:tad-DE-F3Ft3-dS}
\td \FFt_2 &= 
\QQ_3\hs (l_{5}\!-\!l_2)\,, \qquad~
&&\td \FFt_3\hs = \QQ_3\hs (l_5\!-\!l_1) \,,
\\[1mm]
\label{eq:tad-DE-Q2Q3-dS}
\td \QQ_2 &= 0\,,
&&\td \QQ_3 = 0 \,, 
\end{alignat}
\eeqs
where we can write down the symbol results:
\begin{align}
\label{eq:symbol-tad}
\mS_{(2,1)}^\tad =
&~(\bar{w}_5/\bar{w}_2)\otimes(w_3/w_1)
+(\bar{w}_5/w_1)\otimes(\bar{w}_4/\bar{w}_2)
\nn\\
&~+(w_5/w_2)\otimes(w_3/w_1)
+(w_5/w_1)\otimes(w_4/w_2) \,,
\end{align}
and the transcendental function after integrating the symbol \eqref{eq:symbol-tad}
\begin{align}
f_{(2,1)}^\tad=&\,-\pi^2/6+ +\rm{Li}_2(w_3/\bar{w}_5)+\rm{Li}_2(\bar{w}_4/\bar{w}_5)+\rm{Li}_1(\bar{w}_3/\bar{w}_5) \rm{Li}_1(\bar{w}_4/\bar{w}_5)
\nn\\
&\,+\rm{Li}_2(w_3/w_5)+\rm{Li}_2(w_4/w_5)+\rm{Li}_1(w_3/w_5) \rm{Li}_1(w_4/w_5)  \,,
\end{align}
with $l_i=\dlog(w_i)$ and $\bar{l}_i=\dlog(\bar{w_i})$.

\subsubsection{New Features from Tadpole System and Triangle Basis}
\label{sec:3.2.2}

A closer look at the tadpole differential system (\ref{eq:Tad-DEs-P-F-Ft-Q}) reveals distinct features against the bubble system as discussed in section\,\ref{sec:3.1}.\ 
Importantly, unlike the bubble case, the two parent functions $\PP_2$ and $\PP_3$ in tadpole system cannot be merged into a single function $\PP$ without compromising the uniform transcendental property.\ 
This distinction arises because the calculation of the differential equations for $\PP_2$ and $\PP_3$ involves matching the residues at two different codimension-2 boundaries: the intersections $L_1 \cap L_2$ and $L_1 \cap \bar{L}_2$.\ 
Each of these intersections generates distinct dlog singularities i.e., $l_2$ and $\bar{l}_2$, preventing the two parent functions from being merged into single one.\ 
In contrast, for the bubble case, all parent functions share a single codimension-2 boundary, $L_1 \cap L_2$.

Another intriguing feature of the tadpole wavefunction coefficient is its selective involvement with the system, in contrast to the bubble case.\ 
Specifically, only eight basis functions contribute to the canonical differential equations for the tadpole system, despite the hyperplane arrangement suggesting a 10-dimensional vector space.\
This selective engagement with a subset of the full cohomology space appears to be a recurring pattern as the diagram's complexity increases.\ 
For instance, at tree level, the three-site diagram probes only a 16-dimensional subspace within the full 25-dimensional space given by cohomology \cite{Arkani-Hamed:2023kig}.\ 
This observation highlights that the physical observables possess unique properties absent in generic mathematical constructs.

In the two-site case, we can explicitly identify the components not probed by the tadpole function.\ The hyperplane arrangement yields twelve canonical forms, corresponding to all possible triangles bounded by the seven planes in \eqrefe{eq:TLD-2}, with $12 = 3 \times 2 \times 2$. Alongside the eight 2-forms specified in \eqrefe{eq:forms-tad-8}, the remaining four 2-forms, defined using the previously established positive orientation, are given by
\beqs
\label{eq:Pu23-Fu23}
\begin{align}
\label{eq:Pu23}
\text{Layer-0:} \quad~ \bOme_{\PP^{\uu}_2}&=\frac{-2(Y_1\!-\!Y_2)}{L_1 L_2 D_2}\,, \qquad~~~~ \bOme_{\PP^{\uu}_3}=\frac{-2(Y_1\!+\!Y_2)}{L_1 \bar{L}_2 D_3}\,,
\\[1mm]
\label{eq:Fu23}
\text{Layer-1:} \quad~\bOme_{\FF^{\uu}_2}&=\frac{X_1\!-\!Y_1\!+\!2Y_2}{T_1 L_2 D_2}, \qquad~~~ \bOme_{\FF^{\uu}_3}=\frac{X_1\!-\!Y_1\!-\!2Y_2}{T_1 \bar{L}_2 D_3} \,,
\end{align}
\eeqs
where the corresponding functions are defined as: $\PP^{\uu}_{2,3}\!=\int\!\td\mu\,\bOme_{\PP^{\uu}_{2,3}}\,,\, \FF^{\uu}_{2,3}\!=\int\!\td\mu\,\bOme_{\FF^{\uu}_{2,3}}$.\
As in the bubble case, the twelve 2-forms \eqref{eq:forms-bub} are actually overcomplete and subjected to two linear constraints induced by the corresponding triangulation relations \eqref{eq:Om-constraint}, which reduce the number of the independent basis down to ten.\ 
While in tadpole case, since the two disjoint tree systems have already provided eight independent basis functions, we will need two relations involving the above four extra 2-forms.\ Those two linear constraints are
\beqs
\label{eq:S1-S2-tad}
\begin{align}
S_1^{}=\,&\triangle_{L_1\bar{L}_2D_2}-\triangle_{L_1L_2D_2}+\triangle_{T_1L_2D_2}-\triangle_{T_1\bar{L}_2D_2}
\nn\\
=\,&\triangle_{L_1\bar{L}_2D_3}-\triangle_{T_1\bar{L}_2D_3}+\triangle_{T_1L_2D_3}-\triangle_{L_1L_2D_3}\,,
\\[1mm]
S_2^{}=\,&\triangle_{L_1L_2D_2}+\triangle_{T_1T_2D_2}-\triangle_{T_1L_2D_2}-\triangle_{T_2L_1D_2}
\nn\\
=\,&\triangle_{L_1L_2D_3}-\triangle_{T_1L_2D_3}-\triangle_{T_2L_1D_3}+\triangle_{T_1T_2D_3}\,,
\end{align}
\eeqs
where $S_1$ and $S_2$ correspond to the blue-shaded and blue rectangular regions in fig.\,\ref{fig:4}, respectively.\
Imposing the triangulation will give the following nontrivial relations:
\beqs
\label{eq:tad-constraints}
\begin{align}
\label{eq:tad-constraint-a}
\bOme_{\PP_2}-\bOme_{\PP^{\uu}_2}+\bOme_{\FF^{\uu}_2}-\bOme_{\FF_2}
&=\,\bOme_{\PP^{\uu}_3} -\bOme_{\FF^{\uu}_3} + \bOme_{\FF_3} -\bOme_{\PP_3} \,,
\\
\label{eq:tad-constraint-b}
\bOme_{\PP^{\uu}_2}+\bOme_{\QQ_2}-\bOme_{\FF^{\uu}_2}-\bOme_{\FFt_2}
&=\,\bOme_{\PP_3}-\bOme_{\FF_3}-\bOme_{\FFt_3}+\bOme_{\QQ_3} \,.
\end{align}
\eeqs
The relations in \eqrefe{eq:tad-constraint-a} imply that both $\Omega_{\PP_2^\uu}-\Omega_{\FF_2^\uu}$ and $\Omega_{\PP_3^\uu}-\Omega_{\FF_3^\uu}$ lie in the span of the eight forms in \eqrefe{eq:forms-tad-8}. 
Thus, with these eight forms retained, a basis for the full ten-dimensional triangle space at generic, nondegenerate kinematics is obtained by selecting one form from each of the pairs $(\Omega_{\PP_2^\uu},\Omega_{\FF_2^\uu})\,,\, (\Omega_{\PP_3^\uu},\Omega_{\FF_3^\uu})$.
Selecting both forms from the same pair would add only one independent direction and would therefore be insufficient to complete the basis.

These admissible choices lead to different coupling structures. For example, if $\Omega_{\PP_2^\uu}$ is selected, the corresponding function satisfies
\begin{equation}
\td\PP^{\uu}_2 \,\sim\, \PP^{\uu}_2+\FF^{\uu}_2+\FFt_2 \,,
\end{equation}
where $\FF^{\uu}_2$ can be further re-expressed in terms of basis functions according to the constraints \eqref{eq:tad-constraints}.\
Eliminating $\FF_2^\uu$ using \eqrefe{eq:tad-constraint-b} introduces $\PP_3,\FF_3,\FFt_3,\QQ_2$, and $\QQ_3$ into the equation for $\PP_2^\uu$, increasing the mixing across hierarchical layers and between the two tree systems. A similar effect occurs when $\Omega_{\PP_3^\uu}$ is selected.
To keep the coupling structure sparse, we therefore choose $\Omega_{\FF_2^\uu}$ and $\Omega_{\FF_3^\uu}$ as the two additional basis forms. Each corresponding function couples only to itself and its associated $\QQ$-function. Their differential equations, derived in appendix\,\ref{app:B2}, are:
\beqs
\begin{align}
\td\FF^{\uu}_2 =&~ \vep 
\big[\,\FF^{\uu}_2\hs \dlog(X_1\!-\!Y_1\!+\!2Y_2)+(\FF^{\uu}_2\!-\!\QQ_2)\hs\dlog(X_2\!+\!Y_1)
\nn\\
&\quad+\hsm\QQ_2\hs\dlog(X_1\!+\!X_2\!+\!2Y_2)\,\big]\,,
\\[1mm]
\td\FF^{\uu}_3 =&~ \vep\big[\,\FF^\uu_3\hs\dlog(X_1\!-\!Y_1\!-\!2Y_2)+(\FF^{\uu}_3\!-\!\QQ_3)\hs\dlog(X_2\!+\!Y_1\!+\!2Y_2)
\nn\\
&\quad+\hsm\QQ_3\hs\dlog(X_1\!+\!X_2)\,\big]\,.
\end{align}
\eeqs
Together, the basis $\{\PP_2 ,\, \FF_2 ,\, \FFt_2,\, \FF^{\uu}_2, \,\QQ_2,\, \PP_3 ,\, \FF_3 ,\, \FFt_3,\, \FF^{\uu}_3, \,\QQ_3\}$ defines the \textit{whole system} differential equations.\ 
The advantage of this choice of basis lies in its simplicity: each DE is of the UT type, and the mixing between functions is significantly minimized. Notably, the two additional basis functions are naturally associated with two disjoint tree systems, ensuring they remain decoupled from one another.

Insights from the study of complete hyperplane arrangements suggest the use of a \textit{tailored basis} for both the tadpole wavefunction and the whole system. The DE structure of $\{\FF^\uu_2,\FF^\uu_3\}$ being simpler than that of $\{\PP^\uu_2,\PP^\uu_3\}$ is expected from IBP. By trading the differentiation of external energy for that of the twist variables, the IBP operation of $\pd_{x_{1,2}}$ generates only $T_{1,2}$-poles.\ 
Given that $\FF^\uu_{2,3}$ already contain a single $T$-pole, IBP ensures that they primarily couple to themselves and $Q$-functions through residue matching. 
In contrast, since $\PP^\uu_{2,3}$ do not contain any $T$-poles, the $\FF^\uu$- and $\FFt$-functions can appear in their DEs.\
These observations highlight a guiding principle: whenever there is flexibility in choosing a basis, prioritizing canonical forms with more $T$-poles can lead to a more efficient and simplified formulation.

Furthermore, our approach of IBP on triangle canonical forms is a special case of the systematic procedure in ref.\,\cite{Arkani-Hamed:2023kig} using boundary-less linear combination of projective simplex forms.\ 
Namely, in integration space $\mathbb{R}^p$, we can define projective simplex $p$\hs-forms involving all possible $p$\hs-tuples of planes as follows:
\begin{equation}
\label{eq:projective-p-form}
[\hs\La_1 \cdots \La_p\hs] \,\equiv\, \dlog \La_1 \wedge \cdots \wedge \dlog \La_p \,,
\end{equation}
where the hyperplane $\La_I\equiv\{T_i,L_j,D_k\}$ and the exterior derivatives are with respect to the internal variables $x_\al$.\ 
In addition, to determine the combinations of \eqrefe{eq:projective-p-form}, we introduce the boundary operator
{\setlength{\abovedisplayskip}{5pt}
\setlength{\belowdisplayskip}{5pt}
\begin{equation}
\pd\hs[\hs\La_1 \cdots \La_p\hs] \,\equiv~ \sum^p_{r=1}(-1)^{r+1}[\hs\La_1 \cdots \widehat{\La}_r \cdots \La_p\hs] \,,
\end{equation}}
where the entry below $~\widehat{}~$ has been omitted.\ 
And the boundary-less means the form is closed under $\pd$.
Furthermore, these projective $p$\hs-forms are the canonical forms of the simplex formed from the intersection of the $p$ hyperplanes with the plane at infinity $\La_\infty$.\ 
It is straightforward to verify that the boundary operator satisfies $\pd^2=0$, and the linear combinations of projective $p$\hs-forms closed under $\pd$ correspond to bounded regions in $\mathbb{R}^p$, forming the basis of the twisted integral.\ 
Specifically, any $p$\hs-simplex (not projective simplex) in $\mathbb{R}^p$ defined by $p+1$ hyperplanes, is exact and hence closed:
\begin{equation}
\label{eq:p-simplex-form}
\Omega_{p\text{\hs-simplex}} \,=\, \pd[\La_1\cdots \La_{p+1}] \,,
\end{equation}
with simplest examples being triangle canonical forms: $\,\Omega_{\triangle}\!=\!\dlog(\La_i/\La_k)\wedge\dlog(\La_j/\La_k)$.
The iterative emerging of $T$-poles under IBP corresponds to a formula that holds inside of a twisted integral, which is proved in ref.\,\cite{Arkani-Hamed:2023kig}:
{\setlength{\abovedisplayskip}{8pt}
\setlength{\belowdisplayskip}{1pt}
\begin{equation}
\label{eq:derivative-proj-p-forms}
\td[\La_1 \cdots \La_p]=\vep \sum_a \pd\hs[\hs T_a \hs \La_1 \cdots \La_p\hs ]\,\dlog \bla \hat{T}_a \hs \hat{\La}_1 \cdots \hat{\La}_p\bra \,,
\end{equation}}
where we have defined
\begin{equation}
\La_a \equiv \hat{\La}_a \!\cdot\! P \,,\quad
P^I \!\equiv\!\(1, x_1, \ldots, x_p\),\quad
\bla\hat{\La}_1 \cdots \hat{\La}_{p+1}\bra \equiv \det(\hat{\La}_1, \ldots, \hat{\La}_{p+1})\,.
\end{equation}
The general strategy involves expressing the wavefunction as a closed linear combination of \eqrefe{eq:projective-p-form}, applying \eqrefe{eq:derivative-proj-p-forms}, and iteratively adding new functions to the basis as new letters appear.\ 
However, diverging from this comprehensive approach, we find it more practical in many cases to use bounded simplex regions as the basis.\ 
This choice minimizes mixing between different basis elements during differentiation.\ 
Explicitly, we employ a bottom-up strategy: 
we exhaust a priori the independent closed linear combinations of \eqrefe{eq:projective-p-form} with more twist $T$-planes and ensure that each form aligns with the simplex representation in \eqrefe{eq:p-simplex-form}.

The tailored basis for tadpole system suggests a \textit{preferred triangle basis} suitable for 2-dimensional two-site systems with general loop number and topology.\ We denote $\{\P,\,\F,\,\Q\}$ as collections of top, middle, and bottom layer functions respectively, where each represents a distinct type of function within the same category.\ Below, we state the procedure to exhaust the preferred triangle basis:
\begin{itemize}[leftmargin=*]

\item First of all, fully determine $\Q$\hs-basis with two $T$-planes, expressed as $\pd\hs[T_1 T_2 D_k]$, where $D_k$ iterates over all diagonally-placed lines.\ 
Since IBP generates only additional $T_{1,2}$-planes, these forms remain self-contained, interacting exclusively within the bottom layer, which ensures they serve as independent bottom-layer functions.

\item
Next, exhaust mid-layer $\F$\hs-basis with one $T$-plane and one $L$\hs-plane, expressed as $\pd\hs[T_iL_jD_k]$. This process involves substituting one $T$-plane in $\Q$\hs-basis with a parallel $L$\hs-plane.\ 
Under IBP, any new $T$-plane generated will be parallel to $L_j$, limiting interactions to forms with the same pole structure.\ These include the middle-layer functions themselves and the bottom-layer functions already exhausted.

\item 
Finally, identify the top-layer functions in $\P$\hs-basis by their exclusive $L$-poles after we exhaust all the $T$-pole basis, represented as $\pd\hs[L_iL_jD_k]$.\ 
The number of distinct $L$\hs-pairs in remaining basis determines the number of top-layer functions.\ After applying IBP, these functions generally interact within their layer and with corresponding $L$\hs-branches in the middle-layer $\F$\hs-functions.\ Consequently, they serve as parent functions governing the entire differential system.
\end{itemize}
The distinction in the preferred basis highlights why the tadpole system necessarily develops two parent functions, whereas the bubble system requires only one.\
For tadpole system, the presence of two types of $D$\hs-poles leads to the $\Q$\hs-basis being fully defined as $\{\QQ_2, \QQ_3\}$, consisting of two elements.\ The $\F$\hs-basis further splits into three distinct $L$\hs-branches, each associated with a common $L$\hs-pole:
$\{\FFt_2, \FFt_3\}$, each contains a $L_1$\hs-pole;
$\{\FF^\uu_2,\FF_3\}$, each contains a $L_2$\hs-pole;
$\{\FF_2, \FF^\uu_3\}$, each contains a $\bar{L}_2$\hs-pole.
\noindent
In addition, the existence of two unique $L$\hs-pairs, $(L_1,L_2)$ and $(L_1,\bar{L}_2)$, necessitates two parent functions to span the entire system.\ 
Specifically, $\PP_3$ interacts with the branches containing $L_1$\hs- and $L_2$\hs-poles, while $\PP_2$ interacts with the branches containing $L_1$\hs- and $\bar{L}_2$\hs-poles.\  
In contrast, the bubble system has only one type of $L$\hs-pair, $(L_1,L_2)$, meaning a single parent function suffices to span the system.\ Thus, in this case, we can select $\PP_1$, $\PP_2$, or $\PP_3$ as the parent function, as other two functions are linearly dependent by \eqrefe{eq:Om-constraint}.

One of the remarkable features of the preferred triangle basis is its ability to minimize mixing, leading to a maximally block-diagonalized form of $A$.\ 
For tadpole system, the basis is ordered as follows:
\begin{equation}
\label{eq:Tad-I}
\widehat{\bI}_{\tad} = \big(\bF_2\cup\bF_3\cup \{\FF^\uu_2, \FF^\uu_3\}\big)^T 
= \big(\PP_2 ,\, \FF_2 ,\, \FFt_2,\, \FF^\uu_2, \,\QQ_2,\, \PP_3 ,\, \FF_3 ,\, \FFt_3,\, \FF^\uu_3, \,\QQ_3\big)^T .
\end{equation}
The integration family \eqref{eq:Tad-I} obeys the differential equation:
$\td\widehat{\bI}_\tad = \vep\hs A_\tad \,\widehat{\bI}_\tad$,
where the corresponding matrix $A_\tad$ takes the form:
\begin{equation}
\label{eq:At-tad}
\hspace*{-3mm}
A_\tad =\!\(
\begin{array}{cccccccccc}
l_1\!+\!\bar{l}_2 &~l_3\!-\!l_1 &~\bar{l}_4\!-\!\bar{l}_2 &~0&~0&~0&~0&~0&~0&~0
\\[.5mm]
0&~\bar{l}_2\!+\!l_3 &~0&~0&~\bar{l}_5\!-\!\bar{l}_2 &~0&~0&~0&~0&~0
\\[.5mm]
0&~0&~l_1\!+\!\bar{l}_4 &~0&~\bar{l}_5\!-\!l_1 &~0&~0&~0&~0&~0
\\[.5mm]
0&~0&~0&~l_2\!+\!l_6 &~\bar{l}_5\!-\!l_2&~0&~0&~0&~0&~0
\\[.5mm]
0&~0&~0&~0&~2\bar{l}_5 &~0&~0&~0&~0&~0
\\[.5mm]
0&~0&~0&~0&~0&~l_1\!+\!l_2 &~l_3\!-\!l_1 &~l_4\!-\!l_2 &~0&~0
\\[.5mm]
0&~0&~0&~0&~0&~0&~l_2\!+\!l_3 &~0&~0&~l_5\!-\!l_2 
\\[.5mm]
0&~0&~0&~0&~0&~0&~0&~l_1\!+\!l_4 &~0&~l_5\!-\!l_1 
\\[.5mm]
0&~0&~0&~0&~0&~0&~0&~0&~l_2\!+\!l_7 &~l_5\!-\!\bar{l}_2 
\\[.5mm]
0&~0&~0&~0&~0&~0&~0&~0&~0&~2l_5 
\end{array}
\)\!.
\end{equation}
Further, in \eqrefe{eq:At-tad}, the two new letters $l_6$ and $l_7$ are given by
\begin{equation}
l_6=\dlog(X_1 \!-\! Y_1\!+\! 2Y_2)\,, \qquad~~~
l_7=\dlog(X_1 \!-\!Y_1\!-\!2Y_2) \,.
\end{equation}
The two tree systems are evidently integrable, and the two additional basis functions, along with their corresponding bottom-level functions $\{\FF^\uu_2,\FF^\uu_3,\QQ_2,\QQ_3\}$, are independently integrable.\ 
Therefore, the entire system satisfies the integrability condition.\ 
The block-diagonal structure offers a significant advantage:\ it reduces the integrability condition $A\wedge A=0$ to several simpler block-wise verifications.

To achieve a maximally block-diagonalized form of the matrix $A$, it is essential to remain in the triangular basis.\ 
This approach ensures that parent functions interact with the minimum number of $L$\hs-branch functions in the middle layer.\ Such a strategy enables us to efficiently determine whether the entire system is integrable.

As for the tadpole system, using the previously mentioned bottom-up method, we identified two sets of bases, $\{\PP_2,\FF_2,\FFt_2,\FF^\uu_2,\QQ_2\}$ and $\{\PP_3,\FF_3,\FFt_3,\FF^\uu_3,\QQ_3\}$.\ Due to the inability of their associated parent functions to merge, the matrix $A^\tad_{10\times10}$ is naturally decomposed into a $5 \times 5 \oplus 5 \times 5$ block-diagonal form [cf.\,\eqrefe{eq:At-tad}].\
While for the bubble system in section\,\ref{sec:3.1}, if we traverse from the bottom to the top layer and identify $\pd\hs[L_1L_2D_1]$ i.e., $\PP_1$ as the single parent function, the matrix $A^\bub_{10 \times 10}$ [cf.\,\eqrefe{eq:At-bub}] will be block-diagonalized into three submatrices:  
$4 \times 4 \oplus 3 \times 3 \oplus 3 \times 3$ with each block corresponding to $\{\PP_1,\FF_1,\FFt_1,\QQ_1\}, \{\FF_2,\FFt_2,\QQ_2\}$ and $\{\FF_3,\FFt_3,\QQ_3\}$, respectively.\ 
However, the chosen parent function $\PP_1$ in this case is unphysical, making it essential to identify an appropriate linear combination to determine the final, physically meaningful parent function  $\PP$.

In summary, our comprehensive investigation into loop-level hyperplane arrangements and their associated differential systems uncovers several unique features that are absent at tree level. These findings can be summarized as follows:
\begin{itemize}[leftmargin=*]
\item 
Ambiguity in Parent Function Definition:
At loop level, defining the parent function based solely on the hyperplane arrangement is ambiguous, and the differential system may involve multiple parent functions.\ 
Unlike tree-level systems, where a unique, boundary-less object directly corresponds to the cosmological wavefunction of interest and does not involve twisted planes, this uniqueness does not hold for loop-level cases such as the bubble or tadpole systems.\ 
For example, the tadpole system necessarily requires two parent functions to span the system.\ 
This ambiguity arises from the parallel $L$\hs- and $D$\hs-lines in hyperplane arrangements generated by loop diagrams, which contrasts with tree systems where all lines intersect.\ 
And this distinction is particularly evident in the tubing approach to wavefunction coefficients; see tree examples in ref.\,\cite{Arkani-Hamed:2023kig} and loop-level cases in \eqrefe{eq:psi-bub-ab} and \eqrefe{eq:psi-tad-ab}.

\item
Reduced Basis Size for CDEs:
At loop level, the number of basis functions relevant to the canonical differential equations is typically smaller than that at tree level with the same number of internal edges.\ 
For tree-level systems, the number of basis functions scales as $\,4^e\,$ where $e$ is the number of the internal edges.\ 
In contrast, our analysis shows that the one-loop bubble has 10 basis functions, the tadpole has 8, and the two-site two-loop sunrise has 22, two-site arbitrary loop in bubble type has $3 \cdot 2^e-2$ [cf.\,\eqrefe{eq:function-count}].\
The slower scaling of the relevant vector space dimension at loop level is expected, as the numerous parallel lines appearing in loop hyperplane arrangements make the hyperplane arrangements highly nontrivial, and indicate that the wavefunction of physical interest has special features that distinguish it from generic mathematical objects.
\end{itemize}

\begingroup
\raggedbottom
\setlength{\abovedisplayskip}{6pt plus 2pt minus 2pt}
\setlength{\belowdisplayskip}{6pt plus 2pt minus 2pt}
\setlength{\abovedisplayshortskip}{4pt plus 2pt minus 1pt}
\setlength{\belowdisplayshortskip}{6pt plus 2pt minus 2pt}
\interdisplaylinepenalty=10000
\widowpenalty=10000
\clubpenalty=10000
\newcommand{\flowdiagram}[2]{%
  \vcenter{\hbox{\includegraphics[scale=.52]{fig/#1/#2.pdf}}}%
}
\section{Kinematic Flow for Two-Site One-Loop Graph}
\label{sec:4}

In this section, we apply the tubing graph framework to study the differentiation system for two-site one-loop correlators.\
For the first time, we extend the tree-level tubing framework \cite{Arkani-Hamed:2023bsv,Arkani-Hamed:2023kig} to the loop-level.\ 
In addition, we introduce the graphical rules governing loop-level tubings and present a systematic methodology for deriving the associated canonical DEs which are aligned with the results derived in section\,\ref{sec:3}.

\subsection{Two-Site One-Loop Bubble}
\label{sec:4.1}

In ref.\,\cite{Arkani-Hamed:2017fdk}, the wavefunction coefficient for a given graph is obtained by summing over all possible ways of iteratively dividing the graph into connected tubes\,\footnote{Different conventions exist regarding whether the outermost tube, which contains all vertices, should be included. Such as  refs.\,\cite{carr2006coxeter,devadoss2020colorful,balduf2024tubings} choose to omit it, whereas in our work, we consistently include it.}, each tube graph associated with the inverse of the sum of its total energies.\ 
For two-site one-loop bubble case, there are two ways to divide the associated graph into two subgraphs:
\beqs
\label{eq:psi-bub-ab}
\begin{align}
\tilde{\psi}_{(2,1)}^{\bub,(a)} &=
\letterfig{Tubes}{bub-tube-a}{0.6}{0cm}
\hspace*{1.5cm}~=\frac{1}{X_1\!+\!Y_1\!+\!Y_2}
\!\times\!\frac{1}{X_2\!+\!Y_1\!+\!Y_2}
\!\times\!{\red\frac{1}{X_1\!+\!X_2\!+\!2Y_1}}
\!\times\!{\lgray\frac{1}{X_1\!+\!X_2}}\,,
\\[1mm]
\tilde{\psi}_{(2,1)}^{\bub,(b)} &=
\letterfig{Tubes}{bub-tube-b}{0.6}{0cm} \hspace*{1.5cm}~=\frac{1}{X_1\!+\!Y_1\!+\!Y_2}
\!\times\!\frac{1}{X_2\!+\!Y_1\!+\!Y_2}
\!\times\!{\lblu\frac{1}{X_1\!+\!X_2\!+\!2Y_2}}
\!\times\!{\lgray\frac{1}{X_1\!+\!X_2}}\,,
\end{align}
\eeqs
where each subgraph contains four tube graphs and the placement of external and internal energies $\{X_i,Y_j\}$, referring to fig.\,\ref{fig:1}. Here, for the sake of convenience, we have omitted the external legs and late-time boundary.\
Further, by summing over the two contributions in \eqrefe{eq:psi-bub-ab}, we can obtain the full wavefunction coefficient in flat spacetime, $\tilde{\psi}_{(2,1)}^\bub$, which aligns with \eqrefe{eq:psi-bub-flat}.\ 
Moreover, the tubes in \eqrefe{eq:psi-bub-ab} can be organized according to a two-dimensional associahedron with four vertices and four edges \cite{carr2011pseudograph}.

For the functions in the integral family \eqref{eq:Bub-I}, their diagrammatic representations are built using the marked tubing framework \cite{Arkani-Hamed:2023bsv,Arkani-Hamed:2023kig}, where the tubings in \eqref{eq:psi-bub-ab} are suitably modified to reflect different types of singularities.\ To facilitate the study of the associated differential system, it is convenient to consider the decomposition introduced in section\,\ref{sec:3.1}.\
And in the tubing framework, a marked graph is defined as a graph where each of the internal lines is labeled with a cross sign ``\,\letterfig{Tubes}{cross}{0.7}{0.3cm}'' to indicate internal energy, such as $Y_1$ and $Y_2$, while each of the nodes ``\,\letterfig{Tubes}{node}{0.7}{0.3cm}'' corresponds to external energy, such as $X_1$ and $X_2$.\
The relations among the functions in \eqrefe{eq:Bub-I} can be graphically expressed as:
%
\begin{equation}
\label{eq:Bub-functions}
\flowdiagram{Function-Bub}{Bub-I}
\end{equation}
Here, the parent function $\PP$ is obtained by merging $\PP_1$, $\PP_2$ and $\PP_3$ [cf.\,\eqrefe{eq:P=P1+P2-P3}], since they share a common codimension-2 boundary $L_1 \cap L_2$, and is represented by a marked graph without tubing.\
Each descendant function $\FF_i$ and $\FFt_i$ is associated with a marked graph containing a dashed-line tube ``\,\letterfig{Tubes}{dashed-tube}{0.55}{0.45cm}''.\ This tube encloses one external node, reflecting a twisted-plane substitution from $\PP_i$ [cf.\,eqs.\,\eqref{eq:Ome-P-123}--\eqref{eq:Ome-Ft-123}] and one cross sign, which encodes a ``folded singularity'' of the form $X_i \to Y_j \pm Y_{j'}$ appearing in the letter [cf.\,eqs.\,\eqref{eq:Bub-DEs-F-Ft-1}--\eqref{eq:Bub-DEs-F-Ft-3} and \eqref{eq:bub-letters}].\
Each function $\QQ_i$ corresponds to a graph with a merged tube,
which is formed by combining the dashed-line tube already present in
$\FF_i(\FFt_i)$ with an additional tube that encloses the second node.\ This second tube reflects another twisted plane substitution from $\FF_i(\FFt_i)$ [cf.\,\eqref{eq:Ome-Q-123}]. 
In other words, when the two tubes corresponding to $\FF_i$ and $\FFt_i$ share a common cross, their union produces a single, merged tube associated with $\QQ_i$.
Therefore, based on these points, \eqrefe{eq:Bub-functions} organizes all descendant functions into three groups, each corresponding to one of the subsystems $\bF_1,\bF_2$ and $\bF_3$ discussed in section\,\ref{sec:3.1.1}.\ The parent functions within each subsystem are combined accordingly.

The complete set of functions and their marked graphs are summarized in table\,\ref{tab:1}.\ 
Each graph is further dressed with dotted-line tubes ``\,\letterfig{Tubes}{dotted-tube}{0.55}{0.45cm}'', which do not represent new twist-plane substitutions but are introduced to complete the tubing configuration. This notion of complete tubing provides the foundation for constructing the ``family trees'' of the functions which will be discussed later.\ 
Moreover, in table\,\ref{tab:1}, each complete tube graph contains dashed-line tube, dotted-line tubes, or both.\ 
All of these are referred to as the ``inactive tubes''.\ By sequentially activating these tubes, which involves turning them into solid lines and shading them grey i.e., ``\,\letterfig{Tubes}{grey-tube}{0.55}{0.45cm}'', one generates new ``branches'' that collectively form a family tree.\ Each activated tube corresponds to a specific letter, as listed in table\,\ref{tab:2}, thereby enabling a diagrammatical prediction of the canonical differential equations directly from the structure of family trees.
\begin{table}[!tbp]
\centering
\begin{tabular}{c|clllll}
\hline\hline\xrowht{12mm}
Layer-0 
&\multicolumn{6}{c}{$\PP\,$\letterfig{Function-Bub}{layer0-P}{0.6}{1.2cm}}      
\\\hline\xrowht{13mm}
Layer-1 
&\multicolumn{1}{l}{$\FF_1\,$\letterfig{Function-Bub}{layer1-F1}{0.6}{1.2cm}} 
&\multicolumn{1}{l|}{$\FFt_1\,$\letterfig{Function-Bub}{layer1-Ft1}{0.6}{1.2cm}}
& $\FF_2\,$\letterfig{Function-Bub}{layer1-F2}{0.6}{1.2cm}
&\multicolumn{1}{l|}{$\FFt_2\,$\letterfig{Function-Bub}{layer1-Ft2}{0.6}{1.2cm}}
& $\FF_3\,$\letterfig{Function-Bub}{layer1-F3}{0.6}{1.2cm}       
& $\FFt_3\,$\letterfig{Function-Bub}{layer1-Ft3}{0.6}{1.2cm}      
\\\hline\xrowht{13mm}
Layer-2 
&\multicolumn{2}{c|}{$\QQ_1\,$\letterfig{Function-Bub}{layer2-Q1}{0.6}{1.2cm}}
&\multicolumn{2}{c|}{$\QQ_2\,$\letterfig{Function-Bub}{layer2-Q2}{0.6}{1.2cm}}  
&\multicolumn{2}{c}{$\QQ_3\,$\letterfig{Function-Bub}{layer2-Q3}{0.6}{1.2cm}} 
\\\hline\hline
\end{tabular}
\caption{Complete tubing configurations associated with the functions in integral family \eqref{eq:Bub-I} for the two-site one-loop bubble wavefunction coefficient. The functions in layer-1 and layer-2 are partitioned into three groups according to the subsystems $\bF_1,\bF_2$ and $\bF_3$.}
\label{tab:1}
\end{table}

Since tubes in the marked graphs can encode singularities, we now relate them to the letters appearing in the canonical differential equations \eqref{eq:Bub-DEs-P-F-Ft-Q}--\eqref{eq:bub-letters}.\ This correspondence has been established at tree level, where each letter is represented by a connected tube that encloses either nodes alone or nodes together with cross signs marked in the internal lines.\ 
This flow not only drives the deformation of tubes but also captures the folded singularities, providing a geometric picture of how the marked graphs encode analytic structure.\ The interplay between these features will be made explicit in the following discussion.
\begin{itemize}[leftmargin=*]
\item 
If a tube encircles only a single vertex (node), it corresponds to a dlog form of the sum of the energy associated with that vertex and all of the energies associated with the internal lines intersected by the tube.\
For example, consider a ``$\bm{\perp}$'' type graph with four nodes and three internal lines, where the tube only encloses the central vertex and intersects the three internal legs.\ 
Then, the letter associated with the tube is given by
%
\begin{equation}
\letterfig{Tubes}{tube-ex1}{0.68}{0cm}
\hspace*{2.9cm}
=\,\dlog(X_i+Y_j+Y_{j+1}+Y_k)\,.
\end{equation}

\item 
If a tube encircles both a vertex and its adjacent internal lines marked with cross signs, the corresponding dlog form requires flipping the sign of the energies associated with those marked internal lines.\
The following three examples illustrate different tubes and their corresponding letters:
\beqs
\begin{align}
\letterfig{Tubes}{tube-ex2-1}{0.68}{0cm}
\hspace*{2.9cm}
&=\,\dlog(X_i-Y_j+Y_{j+1}+Y_k)\,,
\\[1mm]
\letterfig{Tubes}{tube-ex2-2}{0.68}{0cm}
\hspace*{2.9cm}
&=\,\dlog(X_i-Y_j-Y_{j+1}+Y_k)\,,
\\[1mm]
\letterfig{Tubes}{tube-ex2-3}{0.68}{0cm}
\hspace*{2.9cm}
&=\,\dlog(X_i-Y_j-Y_{j+1}-Y_k)\,.
\end{align}
\eeqs

\item 
If a tube contains more than one vertex, then the energies associated with the marked internal lines between any two enclosed vertices will vanish in its associated dlog expression.\
In the example below, the corresponding letter is given as follows:
\begin{equation}
\letterfig{Tubes}{tube-ex3}{0.68}{0cm}
\hspace*{4.1cm}
=\,\dlog(X_{i-1}+X_{i}+Y_{j-1}-Y_{j+1}+Y_k)\,.
\end{equation}
\end{itemize}
Therefore, building on the correspondences and rules as discussed above, we apply the same strategy to the two-site one-loop bubble graph.\ Specifically, all the letters appearing in the differential equations \eqref{eq:Bub-DEs-P-F-Ft-Q}--\eqref{eq:bub-letters} can similarly be represented by marked tubings, which diagrammatically encode the analytic structure of each logarithmic singularity.\ These correspondences are systematically summarized in table\,\ref{tab:2}.
\begin{table}[!tbp]
\centering
\begin{tabular}{c|c|c|c|c|c}
\hline\hline
Letter & Tube & $\dlog$ Form & Letter & Tube & $\dlog$ Form
\\\hline\xrowht{12.5mm}
$l_1$ 
&\letterfig{Letter-Bub}{l1}{0.6}{1.2cm}
&$\dlog(X_1\!+\!Y_1\!+\!Y_2)$ 
&\multicolumn{1}{c|}{$l_2$}
&\multicolumn{1}{c|}{\letterfig{Letter-Bub}{l2}{0.6}{1.2cm}} 
&$\dlog(X_2\!+\!Y_1\!+\!Y_2)$ 
\\\hline\xrowht{12.5mm}
$l_3$
&\letterfig{Letter-Bub}{l3}{0.6}{1.2cm} 
&$\dlog(X_1\!+\!Y_1\!-\!Y_2)$
&\multicolumn{1}{c|}{$l_4$}
&\multicolumn{1}{c|}{\letterfig{Letter-Bub}{l4}{0.6}{1.2cm}}   
&$\dlog(X_2\!+\!Y_1\!-\!Y_2)$
\\\hline\xrowht{12.5mm}
$l_5$
&\letterfig{Letter-Bub}{l5}{0.6}{1.2cm}  
&$\dlog(X_1\!-\!Y_1\!+\!Y_2)$
&\multicolumn{1}{c|}{$l_6$}
&\multicolumn{1}{c|}{\letterfig{Letter-Bub}{l6}{0.6}{1.2cm}} 
&$\dlog(X_2\!-\!Y_1\!+\!Y_2)$
\\\hline\xrowht{12.5mm}
$l_7$
&\letterfig{Letter-Bub}{l7}{0.6}{1.2cm} 
&$\dlog(X_1\!-\!Y_1\!-\!Y_2)$
& \multicolumn{1}{c|}{$l_8$}
&\multicolumn{1}{c|}{\letterfig{Letter-Bub}{l8}{0.6}{1.2cm}}  
&$\dlog(X_2\!-\!Y_1\!-\!Y_2)$
\\\hline\xrowht{12.5mm}
$l_9$
&\letterfig{Letter-Bub}{l9}{0.6}{1.2cm}  
&$\dlog(X_1\!+\!X_2\!+\!2Y_1)$
&\multicolumn{1}{c|}{$l_{10}$}
&\multicolumn{1}{c|}{\letterfig{Letter-Bub}{l10}{0.6}{1.2cm}}   
&$\dlog(X_1\!+\!X_2\!+\!2Y_2)$  
\\\hline\xrowht{12.5mm}
$l_{11}$
&\letterfig{Letter-Bub}{l11}{0.6}{1.2cm}  
& $\dlog(X_1\!+\!X_2)$ &  
\multicolumn{3}{l}{} 
\\[5mm]\hline\hline
\end{tabular}
\caption{The alphabet for two-site one-loop bubble wavefunction coefficient.}
\label{tab:2}
\end{table}

We now turn to the complete tubing graph associated with the parent function $\PP$ to illustrate how the marked graph framework generates the corresponding family tree and allows for a diagrammatic prediction of the canonical differential equation.\
The construction of family tree for $\PP$ proceeds step by step. In the first step, we activate the inactive tubes in its complete tubing configuration.\ This activation gives rise to two distinct branches, each corresponding to a different letter
\begin{equation}
\label{eq:FT-P-1}
\flowdiagram{FT-Bub}{P-1}
\end{equation}
where for each branch, we have also assigned the functions associated with the tube graph.\
Next, each activated tube in \eqrefe{eq:FT-P-1} that does not contain a cross sign can ``grow'' to enclose all of its adjacent crosses in all possible ways [cf.\,\eqrefe{eq:Bub-functions} or table\,\ref{tab:1}].\ 
Each growth pathway generates a new branch.\
Collectively, we obtain the complete family tree for $\PP$:
\begin{equation}
\label{eq:FT-P-Full}
\flowdiagram{FT-Bub}{P-Full}
\end{equation}
where we have assigned the descendant functions $\{\FF_i\}$ and $\{\FFt_i\}$ associated with the tubes in the third layer of \eqrefe{eq:FT-P-Full}.

To obtain the differential equation from the family tree \eqref{eq:FT-P-Full}, we multiply each tube graph (letter) by the difference between its corresponding function and the sum of the functions associated with all its descendant tube graphs, scaled by a factor $\vep$\,:
\begin{align}
&\td \PP = \vep\[
\(\PP\!-\!\sum_{i=1}^{3}\FF_i\)
\letterfig{Letter-Bub}{l1}{0.6}{0cm}\hspace*{1.3cm}
+\(\PP\!-\!\sum_{j=1}^{3}\FFt_j\)\letterfig{Letter-Bub}{l2}{0.6}{0cm}\hspace*{1.2cm}
+\FF_1~\letterfig{Letter-Bub}{l3}{0.6}{0cm}\hspace*{1.2cm}
~+\FF_2~\letterfig{Letter-Bub}{l5}{0.6}{0cm}\hspace*{1.2cm}
\right.\nn\\[1mm]
&\left.\hspace*{1.5cm}
+\,\FF_3~\letterfig{Letter-Bub}{l7}{0.6}{0cm}\hspace*{1.2cm}
+\,\FFt_1~\letterfig{Letter-Bub}{l4}{0.6}{0cm}\hspace*{1.2cm}
+\FFt_2~\letterfig{Letter-Bub}{l6}{0.6}{0cm}\hspace*{1.2cm}
+\FFt_3~\letterfig{Letter-Bub}{l8}{0.6}{0cm}\hspace*{1.2cm}\],
\end{align}
where it is consistent with the differential equation given in \eqrefe{eq:Bub-DEs-P}.\
As we proceed with the activation and growth processes, we can find each newly activated tube generates additional branches, expanding the family tree and manifesting the structure underlying the canonical differential equations through the graphical representation.

Next, we investigate the two descendant functions, $\FF_1$ and $\FFt_1$. Focusing on the graph associated with $\FF_1$, the two tubes can be activated separately, yielding
\begin{equation}
\label{eq:FT-F1-1}
\flowdiagram{FT-Bub}{F1-1}
\end{equation}
where in the first branch, we omit the inactive tube on the right side, since this branch is fully evolved.\
In the second branch, the activated tube is allowed to continue growing, enclosing its adjacent cross signs in three possible ways:
\begin{equation}
\label{eq:FT-F1-2}
\flowdiagram{FT-Bub}{F1-2}
\end{equation}
Here, in the second and third branches, the activated tube encloses the cross sign above it or both the cross signs above and below, respectively. During this process, the activated tube ``merges'' with its adjacent dashed-line tube, resulting in a larger activated tube.\
However, as shown in \eqrefe{eq:FT-F1-2}, we have applied red arrows and red crosses to mark the first and third branches, emphasizing that the activated tube cannot evolve further along these two branches.\ 
This restriction arises for two reasons:
First, the most immediate observation is that the tube graph and function $\QQ_1$ in the second branch belong to the same family $\bF_1$ as $\FF_1$ [see the first group $\{ \FF_1, \FFt_1, \QQ_1 \}$ in \eqrefe{eq:Bub-functions} or table\,\ref{tab:1}]. The graphs in the first and third branches clearly do not belong to $\bF_1$.\
Second, the newly activated tube in the first branch of \eqrefe{eq:FT-F1-2} is allowed to merge with its neighboring tube, forming a larger one.\ Then, the resulting tubes in both the first and third branches evolve into equivalent configurations,
\begin{equation}
\label{eq:FT-F1-3}
\flowdiagram{FT-Bub}{F1-3}
\end{equation}
where the tubes in the first and second layer correspond to the same letter i.e., $\dlog(X_1\!+\!X_2)$. Moreover, the tube in the second layer is also associated with the function $\QQ_3$, which should correspond to the ultimate tube in the family trees generated by $\FF_3$ and $\FFt_3$ (will be discussed later).\ Obviously, $\QQ_3$ is not the member of $\bF_1$.

For the function $\FFt_1$, we can construct its associated family tree using a method similar to that of $\FF_1$, so we will ignore the detailed process here.\
Consequently, the complete family trees for $\FF_1$ and $\FFt_1$ are constructed as follows:
%
\begin{equation}
\label{eq:FT-F1Ft1}
\flowdiagram{FT-Bub}{F1Ft1-full}
\end{equation}
Hence, the canonical differential equations associated with the functions $\FF_1$ and $\FFt_1$ can be directly read from the family trees \eqref{eq:FT-F1Ft1}:
\beqs
\label{eq:DE-FT-F1Ft1}
\begin{align}
\td \FF_1 &= \vep\[
\big(\FF_1-\QQ_1\big)~\letterfig{Letter-Bub}{l2}{0.6}{0cm}
\hspace*{1.2cm}
+\FF_1~\letterfig{Letter-Bub}{l3}{0.6}{0cm}\hspace*{1.2cm}
+\QQ_1~\letterfig{Letter-Bub}{l9}{0.6}{0cm}\hspace*{1.4cm}\],
\\[0mm]
\td \FFt_1 &= \vep\[
\big(\FFt_1-\QQ_1\big)~\letterfig{Letter-Bub}{l1}{0.6}{0cm}\hspace*{1.2cm}
+\FFt_1~\letterfig{Letter-Bub}{l4}{0.6}{0cm}\hspace*{1.2cm}
+\QQ_1~\letterfig{Letter-Bub}{l9}{0.6}{0cm}\hspace*{1.4cm}\],
\end{align}
\eeqs
which align with the differential equation given in \eqrefe{eq:Bub-DEs-F-Ft-1}.

Due to the presence of symmetry, the complete family trees for $\FF_2$ and $\FFt_2$ are
%
\begin{equation}
\label{eq:FT-F2Ft2}
\flowdiagram{FT-Bub}{F2Ft2-full}
\end{equation}
and the differential equations for $\FF_2$ and $\FFt_2$ can be directly read from \eqrefe{eq:FT-F2Ft2}:
\beqs
\label{eq:DE-FT-F2Ft2}
\begin{align}
&\hspace*{-1cm}\td \FF_2 = \vep\[
\big(\FF_2-\QQ_2\big)~\letterfig{Letter-Bub}{l2}{0.6}{0cm}\hspace*{1.2cm}
+\FF_2~\letterfig{Letter-Bub}{l5}{0.6}{0cm}\hspace*{1.2cm}
+\QQ_2~\letterfig{Letter-Bub}{l10}{0.6}{0cm}\hspace*{1.4cm}\],
\\[0mm]
&\hspace*{-1cm}\td \FFt_2 = \vep\[
\big(\FFt_2-\QQ_2\big)~\letterfig{Letter-Bub}{l1}{0.6}{0cm}\hspace*{1.2cm}
+\FFt_2~\letterfig{Letter-Bub}{l6}{0.6}{0cm}\hspace*{1.2cm}
+\QQ_2~\letterfig{Letter-Bub}{l10}{0.6}{0cm}\hspace*{1.4cm}\],
\end{align}
\eeqs
where they are coincident with \eqrefe{eq:Bub-DEs-F-Ft-2}.

We then study the functions $\FF_3$ and $\FFt_3$.\ 
For the graph associated with $\FF_3$, the two inactive tubes can be activated separately, forming two branches:
\begin{equation}
\label{eq:FT-F3-1}
\flowdiagram{FT-Bub}{F3-1}
\end{equation}
where in the second branch, the activated tube can continue to grow and enclose the cross sign above, below, or both, thereby forming three possible branches:
\begin{equation}
\label{eq:FT-F3-2}
\flowdiagram{FT-Bub}{F3-2}
\end{equation}
where the three branches are topologically equivalent, uniquely represented by the second-branch graph.

A similar method applies to $\FFt_3$.\
The full family trees for $\FF_3$ and $\FFt_3$ are
\begin{equation}
\label{eq:FT-F3Ft3}
\flowdiagram{FT-Bub}{F3Ft3-full}
\end{equation}
Thus, the canonical differential equations associated with the functions $\FF_3$ and $\FFt_3$ can be obtained from the family tree \eqref{eq:FT-F3Ft3} as:
\beqs
\label{eq:FT-DE-F3Ft3}
\begin{align}
&\hspace*{-1cm}\td \FF_3 = \vep\[
\big(\FF_3-\QQ_3\big)~\letterfig{Letter-Bub}{l2}{0.6}{0cm}\hspace*{1.2cm}
+\FF_3~\letterfig{Letter-Bub}{l7}{0.6}{0cm}\hspace*{1.2cm}
+\QQ_3~\letterfig{Letter-Bub}{l11}{0.6}{0cm}\hspace*{1.4cm}\],
\\[0mm]
&\hspace*{-1cm} \td \FFt_3 = \vep\[
\big(\FFt_3-\QQ_3\big)~\letterfig{Letter-Bub}{l1}{0.6}{0cm}\hspace*{1.2cm}
+\FFt_3~\letterfig{Letter-Bub}{l8}{0.6}{0cm}\hspace*{1.2cm}
+\QQ_3~\letterfig{Letter-Bub}{l11}{0.6}{0cm}\hspace*{1.4cm}\],
\end{align}
\eeqs
where they are coincident with \eqrefe{eq:Bub-DEs-F-Ft-3}.

Finally, we analyze the last three descendant functions $\QQ_1, \QQ_2$ and $\QQ_3$.\ 
In this scenario, the graphs associated with $\{\QQ_i\}$ are enclosed by a single tube that becomes activated:
\begin{equation}
\label{eq:FT-Q}
\flowdiagram{FT-Bub}{Q-full}
\end{equation}
where the factor 2 in the second line of \eqrefe{eq:FT-Q} emerges from the enclosure of two vertices within the activated tube.\
Hence, by referencing the family trees in \eqrefe{eq:FT-Q}, we can derive the canonical DEs for $\{\QQ_i\}$ as follows:
\begin{equation} 
\label{eq:FT-DE-Q}
\td \QQ_1 = 2\hs\vep\hs \QQ_1~\letterfig{Letter-Bub}{l9}{0.6}{0cm}\hspace*{1.5cm},\hspace*{.8cm}
\td \QQ_2 =2 \hs \vep\hs \QQ_2 ~\letterfig{Letter-Bub}{l10}{0.6}{0cm}\hspace*{1.5cm},\hspace*{.8cm}
\td \QQ_3 =2 \hs \vep\hs \QQ_3 ~\letterfig{Letter-Bub}{l11}{0.6}{0cm}\hspace*{1.5cm},
\end{equation}
where they are consistent with \eqrefe{eq:Bub-DEs-Q}.

\subsection{Two-Site One-Loop Tadpole}
\label{sec:4.2}

Similar to the bubble case, for the two-site one-loop tadpole graph, there are also two ways to divide the graph into two subgraphs composed by connected tubes:
\beqs
\label{eq:psi-tad-ab}
\begin{alignat}{3}
\tilde{\psi}_{(2,1)}^{\tad,(a)} &=\,
\letterfig{Tubes}{tad-tube-a}{0.6}{0cm}
&\hspace*{1.8cm}&
\quad~~=\frac{1}{X_1\!+\!Y_1} \!\times\!
\frac{1}{X_2\!+\!Y_1\!+\!2 Y_2} \!\times\!
{\red\frac{1}{X_1\!+\!X_2\!+\!2 Y_2}}\!\times\!
{\lgray\frac{1}{X_1\!+\!X_2}}\,,
\\[1mm]
\tilde{\psi}_{(2,1)}^{\tad,(b)} &=\,
\letterfig{Tubes}{tad-tube-b}{0.6}{0cm}&&
\quad~~=\frac{1}{X_1\!+\!Y_1} \!\times\!
\frac{1}{X_2\!+\!Y_1\!+\!2 Y_2} \!\times\!
{\lblu\frac{1}{X_2\!+\!Y_1}}\!\times\!
{\lgray\frac{1}{X_1\!+\!X_2}}\,,
\end{alignat}
\eeqs
where each subgraph contains four tubes.
Summing over the two contributions in \eqrefe{eq:psi-tad-ab}, we can obtain the full wavefunction coefficient in flat spacetime,
$\tilde{\psi}_{(2,1)}^\tad$, which is consistent with \eqrefe{eq:psi-tad-flat}.\
And the tubes in \eqrefe{eq:psi-tad-ab} are related to a two-dimensional associahedron with three vertices, two edges and two rays \cite{carr2011pseudograph}.
\begin{table}[!tbp]
\centering
\begin{tabular}{c|cl|cl}
\hline\hline\xrowht{9.5mm}
Layer-0 
&\multicolumn{2}{c|}{$\PP_2$\,\letterfig{Function-Tad}{layer0-P2}{0.6}{2.3cm}}  
&\multicolumn{2}{c}{$\PP_3$\,\letterfig{Function-Tad}{layer0-P3}{0.6}{1.7cm}}   
\\\hline\xrowht{9.5mm}
Layer-1 
&\multicolumn{1}{l}{$\FF_2$\,\letterfig{Function-Tad}{layer1-F2}{0.6}{2.3cm}} 
&$\FFt_2$\,\letterfig{Function-Tad}{layer1-Ft2}{0.6}{2.3cm}
&\multicolumn{1}{l}{$\FF_3$\,\letterfig{Function-Tad}{layer1-F3}{0.6}{1.5cm}}
&\multicolumn{1}{l}{$\FFt_3$\,\letterfig{Function-Tad}{layer1-Ft3}{0.6}{1.5cm}} 
\\\hline\xrowht{9.5mm}
Layer-2 
&\multicolumn{2}{c|}{$\QQ_2$\,\letterfig{Function-Tad}{layer2-Q2}{0.6}{2.3cm}} 
&\multicolumn{2}{c}{$\QQ_3$\,\letterfig{Function-Tad}{layer2-Q3}{0.6}{1.7cm}}   
\\\hline\hline
\end{tabular}
\caption{Complete tubing configurations associated with the integral family \eqref{eq:I-tad-phy} for the two-site one-loop tadpole wavefunction coefficient.\ Across all layers, the functions are partitioned into two groups based on $\bF_2$ and $\bF_3$.}
\label{tab:3}
\end{table}

The tubing framework is well-suited for analyzing the physical functions and it does not naturally extend to unphysical components.\ Accordingly, we consider the integral family \eqref{eq:I-tad-phy} composed of physical functions.\
And the relations among marked tubings associated with the functions in \eqrefe{eq:I-tad-phy} are illustrated as:
\begin{equation}
\label{eq:Tad-functions}
\flowdiagram{Function-Tad}{Tad-I}
\end{equation}
where in the marked graphs associated with $\bF_3$, the tadpole loop has been omitted.\ This simplification is justified by the observation that, based on \eqrefe{eq:forms-tad-8}, the hyperplanes involved in the forms of $\bF_3$ depend only on the internal energy $Y_1$, and thus the graph behaves effectively as a tree.\ As a result, the tadpole loop in $\bF_3$ does not need to be marked with a cross sign, and the graph admits the following equivalent configuration:
\begin{equation}
\hspace*{5mm}
\raisebox{4mm}{{\small$Y_1$}\hspace*{-0.85cm}}
\letterfig{Function-Tad}{P3-1}{0.6}{2.3cm}
\widesim~~
\letterfig{Function-Tad}{P3-2}{0.6}{0.5cm}
\raisebox{4mm}{{\small$Y_1$}\hspace*{1cm}}
\end{equation}
In addition, unlike \eqrefe{eq:Bub-functions}, the two parent functions $\PP_2$ and $\PP_3$ in \eqrefe{eq:Tad-functions} cannot be merged into a single one, since they have different codimension-2 boundaries, $L_1 \cap L_2$ and $L_1 \cap \bar{L}_2$ (cf.\,section\,\ref{sec:3.2.2}).\ 
Further, the complete tubing graphs associated with the functions in \eqrefe{eq:I-tad-phy} are summarized in table\,\ref{tab:3}.

In table\,\ref{tab:4}, we have used marked tubings to represent all relevant letters appearing in the differential equation for one-loop tadpole graph [cf.\,eqs.\,\eqref{eq:Tad-DEs-P-F-Ft-Q}--\eqref{eq:tad-letters}].\ 
The letters $l_6$ and $l_7$ are not included in this table as they are irrelevant to the physical system.\
In addition, in table\,\ref{tab:4}, the letters $l_1, \ldots, l_5$ have two equivalent tubing representations, both of which will appear in the differential equations for $\bF_2$ and $\bF_3$ discussed later.
\begin{table}[!tbp]
\centering
\begin{tabular}{c|ccc}
\hline\hline
Letter
& \multicolumn{2}{c|}{Tube} 
&\multicolumn{1}{c}{$\dlog$ Form}
\\\hline\xrowht{10mm}
$l_1$
&\multicolumn{1}{c|}{\letterfig{Letter-Tad}{l1a}{0.6}{2.2cm}}
&\multicolumn{1}{c|}{\letterfig{Letter-Tad}{l1b}{0.6}{1.5cm}}
&$\dlog(X_1\!+\!Y_1)$
\\\hline\xrowht{13mm}
$l_2$
&\multicolumn{1}{c|}{\letterfig{Letter-Tad}{l2a}{0.6}{2.2cm}}
&\multicolumn{1}{c|}{\letterfig{Letter-Tad}{l2b}{0.6}{1.5cm}}
&$\dlog(X_2\!+\!Y_1)$
\\\hline\xrowht{10mm}
$l_3$
&\multicolumn{1}{c|}{\letterfig{Letter-Tad}{l3}{0.6}{2.2cm}}
&\multicolumn{1}{c|}{\letterfig{Letter-Tad}{l3b}{0.6}{1.5cm}}
&$\dlog(X_1\!-\!Y_1)$
\\\hline\xrowht{13mm}
$l_4$
&\multicolumn{1}{c|}{\letterfig{Letter-Tad}{l4a}{0.6}{2.2cm}}
&\multicolumn{1}{c|}{\letterfig{Letter-Tad}{l4b}{0.6}{1.5cm}}
&$\dlog(X_2\!-\!Y_1)$
\\\hline\xrowht{13mm}
$l_5$
&\multicolumn{1}{c|}{\,\letterfig{Letter-Tad}{l5a}{0.6}{2.4cm}}
&\multicolumn{1}{c|}{\,\letterfig{Letter-Tad}{l5b}{0.6}{1.6cm}}
&\multicolumn{1}{c}{$\dlog(X_1\!+\!X_2)$}
\\\hline\xrowht{10mm}
$\bar{l}_2$
&\multicolumn{2}{c|}{\letterfig{Letter-Tad}{bl2}{0.6}{2.2cm}}
&$\dlog(X_2\!+\!Y_1\!+\!2Y_2)$ 
\\\hline\xrowht{10mm}
$\bar{l}_4$
&\multicolumn{2}{c|}{\letterfig{Letter-Tad}{bl4}{0.6}{2.2cm}}
&$\dlog(X_2\!-\!Y_1\!+\!2Y_2)$
\\\hline\xrowht{10mm}
$\bar{l}_5$
&\multicolumn{2}{c|}{\letterfig{Letter-Tad}{bl5}{0.6}{2.2cm}}
&$\dlog(X_1\!+\!X_2\!+\!2Y_2)$
\\ \hline\hline
\end{tabular}
\caption{The alphabet for two-site one-loop tadpole wavefunction coefficient where the letters $l_6$ and $l_7$ are excluded since they do not contribute to the physical system.}
\label{tab:4}
\end{table}

\begin{samepage}
Now, let us analyze the kinematic flow structure of the tadpole case.  
First, for the two parent functions $\PP_2$ and $\PP_3$, each of them contains two dashed-line tubes.\ These tubes can be activated and then enclose their adjacent cross signs, forming two distinct branches:
\begin{equation}
\label{eq:FT-P2-P3-tad}
\flowdiagram{FT-Tad}{P23-Full}
\end{equation}
\end{samepage}
In the second branch of $\PP_2$, the activated tube can only enclose the left-side cross sign. Similar to the case in \eqrefe{eq:FT-F1-2}, if the tube encloses the right-side cross sign or both the left- and right-side cross signs simultaneously, the resulting graphs would not belong to any of the graphs associated with $\bF_2$ in \eqrefe{eq:Tad-functions} or table\,\ref{tab:3}.

Therefore, the canonical differential equations for $\PP_2$ and $\PP_3$ are obtained according to the family trees \eqref{eq:FT-P2-P3-tad}
\beqs
\begin{align}
&\hspace*{-2cm}
\td \PP_2 = \vep\[
\big(\PP_2-\FF_2\big)~\letterfig{Letter-Tad}{l1a}{0.6}{0cm}\hspace*{2.35cm}
+\FF_2~\letterfig{Letter-Tad}{l3a}{0.6}{0cm}\right.
\nn\\
&\hspace*{-.8cm}\left.
+\,\big(\PP_2-\FFt_2\big)~~\letterfig{Letter-Tad}{bl2}{0.6}{0cm}
\hspace*{2.2cm}
+\FFt_2~~\letterfig{Letter-Tad}{bl4}{0.6}{0cm}\hspace*{2.2cm}\],
\\[1.5mm]
&\hspace*{-2cm}\td \PP_3 = \vep\[
\big(\PP_3-\FF_3\big)~\letterfig{Letter-Tad}{l1b}{0.6}{0cm}\hspace*{1.5cm}
~+\,\FF_3~\letterfig{Letter-Tad}{l3b}{0.6}{0cm}\hspace*{1.5cm}
\right.\nn\\
&\hspace*{-.82cm}\left.
+\,\big(\PP_3-\FFt_3\big)~~\letterfig{Letter-Tad}{l2b}{0.6}{0cm}\hspace*{1.5cm}
+\,\FFt_3~\letterfig{Letter-Tad}{l4b}{0.6}{0cm}\hspace*{1.5cm}\],
\end{align}
\eeqs
where they are consistent with \eqrefe{eq:tad-DE-P2P3}.\
From this, we can find that the letters associated with $\PP_2$ and $\PP_3$ are not entirely the same as they contain \letterfig{Letter-Tad}{bl2}{0.35}{1.3cm} and \letterfig{Letter-Tad}{l2b}{0.35}{0.9cm} respectively.\ 
This difference prevents the functions $\PP_2$ and $\PP_3$ from being merged into a single function unlike the bubble case, where all parent functions share the same letters \letterfig{Letter-Bub}{l1}{0.35}{0.65cm} and 
\letterfig{Letter-Bub}{l2}{0.35}{0.65cm} [or see eqs.\,\eqref{eq:dP1}, \eqref{eq:dP2} and \eqref{eq:dP3}].

Next, for the descendant functions $\FF_2$ and $\FFt_2$, their corresponding family trees are given as follows:
%
\begin{equation}
\label{eq:FT-F2-Ft2-tad}
\flowdiagram{FT-Tad}{F2Ft2-Full}
\end{equation}
Similarly, in the second branch of $\FF_2$, the activated tube encloses only the cross sign on its left and merges with the dashed-line tube; any other configuration produces graphs that do not belong to $\bF_2$.\
Further, the canonical differential equations can be read from the family trees \eqref{eq:FT-F2-Ft2-tad}
\beqs
\begin{align}
\td \FF_2 &= \vep\[
\big(\FF_2-\QQ_2\big)~\letterfig{Letter-Tad}{bl2}{0.6}{0cm}\hspace*{2.2cm}
+\FF_2~\letterfig{Letter-Tad}{l3}{0.6}{0cm}\hspace*{2.35cm}
+\QQ_2~\letterfig{Letter-Tad}{bl5}{0.6}{0cm}\hspace*{2.4cm}\],
\\[0mm]
\td \FFt_2 &= \vep\[
\big(\FFt_2-\QQ_2\big)~\letterfig{Letter-Tad}{l1}{0.6}{0cm}\hspace*{2.35cm}
+\FFt_2~\letterfig{Letter-Tad}{bl4}{0.6}{0cm}\hspace*{2.2cm}
+\QQ_2~\letterfig{Letter-Tad}{bl5}{0.6}{0cm}\hspace*{2.4cm}\],
\end{align}
\eeqs
where they are in agreement with \eqrefe{eq:tad-DE-F2Ft2}.


For $\FF_3$ and $\FFt_3$, their corresponding family trees are given by
\begin{equation}
\label{eq:FT-F3-Ft3-tad}
\flowdiagram{FT-Tad}{F3Ft3-Full}
\end{equation}
and the canonical differential equations can be read from the \eqrefe{eq:FT-F3-Ft3-tad} as:
\beqs
\begin{align}
\td \FF_3 &= \vep\big[\,
\big(\FF_2-\QQ_3\big)~\letterfig{Letter-Tad}{l2b}{0.6}{0cm}\hspace*{1.5cm}
+\FF_3~\letterfig{Letter-Tad}{l3b}{0.6}{0cm}\hspace*{1.5cm}
+\QQ_3~\letterfig{Letter-Tad}{l5b}{0.6}{0cm}\hspace*{1.7cm}\big]\,,
\\[1mm]
\td \FFt_3 &= \vep\big[\,
\big(\FFt_3-\QQ_3\big)~\letterfig{Letter-Tad}{l1b}{0.6}{0cm}\hspace*{1.5cm}
+\FFt_3~\letterfig{Letter-Tad}{l4b}{0.6}{0cm}\hspace*{1.5cm}
+\QQ_3~\letterfig{Letter-Tad}{l5b}{0.6}{0cm}\hspace*{1.7cm}\big]\,,
\end{align}
\eeqs
where they are consistent with \eqrefe{eq:tad-DE-F3Ft3}.

For the last two descendant functions $\QQ_2$ and $\QQ_3$, each graph associated with them only contains a single tube which can be activated as:
\begin{equation}
\label{eq:FT-Q2-Q3-tad}
\flowdiagram{FT-Tad}{Q23-Full}
\end{equation}
and the canonical differential equations can be obtained from the family tree \eqref{eq:FT-Q2-Q3-tad} as:
\begin{equation} 
\td \QQ_2 =2\hs\vep\hs \QQ_2 ~\letterfig{Letter-Tad}{bl5}{0.6}{0cm}\hspace*{2.5cm},\hspace*{1.2cm}
\td \QQ_3 =2 \hs \vep\hs \QQ_3 ~\letterfig{Letter-Tad}{l5b}{0.6}{0cm}\hspace*{1.7cm},
\end{equation}
where they are compatible with \eqrefe{eq:tad-DE-Q2Q3}.

In summary, the consistency among all differential equations confirms the validity of our approach and demonstrates the effectiveness of the kinematic flow method in capturing the loop-level structure.\
It also suggests that the framework can be extended to more complex cases in future studies.

\endgroup

\section{Analysis for Two-Site Higher Loop Graphs}
\label{sec:5}

In sections\,\ref{sec:3}--\ref{sec:4}, we have presented a detailed analysis of one-loop cases.\ In this section, we briefly examine several two-site $L$-loop examples with $L \geqq 2$, illustrating how our approach to the two-site one-loop diagram naturally extends to higher-loop orders.\
Furthermore, through these examples, we uncover a systematic shift pattern that leads to a compact and general formula for two-site $L$-loop diagrams, which accurately reflects the underlying structure of the associated differential systems.

\begin{figure}[t]
\centering
\begin{subfigure}{0.31\textwidth}
\raisebox{5mm}{\hspace*{-3mm}
\includegraphics[scale=.85]{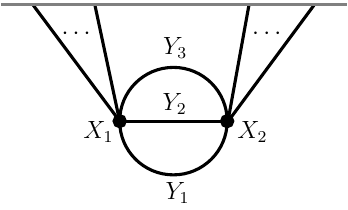}}
\caption{}
\label{fig:5a}
\end{subfigure}%
\hspace*{1.5cm}
\begin{subfigure}{0.31\textwidth}
\includegraphics[scale=.85]{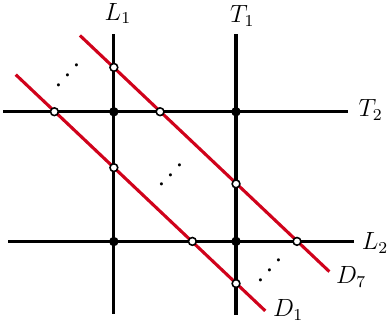}
\caption{}
\label{fig:5b}
\end{subfigure}
\vspace*{-2mm}
\caption{(a).\,Two-site two-loop ``sunset'' diagram.\
(b).\,The configuration of hyperplane arrangement for the two-site two-loop sunset wavefunction coefficient.} \label{fig:5}
\end{figure}

We first study the cases of bubble-like diagrams.\ Importantly, their wavefunction coefficients and associated differential systems serve as an unshifted central building block in our generic two-site $L$-loop formula.\ We begin by a two-site two-loop ``sunset'' diagram (cf.\,fig.\,\ref{fig:5a}), whose wavefunction coefficient is computed as follows:
{\setlength{\abovedisplayskip}{12pt}
\setlength{\belowdisplayskip}{5pt}
\begin{align}
\label{eq:Psi-2loop-sun}
\psi_{(2,2)}^\sun &= 8Y_1Y_2Y_3\!\int_{\mathbb{R}^2_+}\!\!\frac{(T_1T_2)^\vep\td x_1\!\wedge\!\td x_2}{L_1 L_2 D_7}\! \[\!\frac{1}{D_1}\!\(\frac{1}{D_4}\!+\!\frac{1}{D_5}\)\!+\!\frac{1}{D_2}\!\(\frac{1}{D_4}\!+\!\frac{1}{D_6}\)\!+\!\frac{1}{D_3}\!\(\frac{1}{D_5}\!+\!\frac{1}{D_6}\)\!\]
\nn\\
&\equiv \int\!\td\mu\,\bOme_{\PP_\sun}\,,
\end{align}}
where $\td\mu\equiv (T_1T_2)^\vep\hs\td x_1\!\wedge\!\td x_2$ and the divisors (associated with relative singularities) are defined as follows:
\begin{align}
L_1& = x_1+X_1+Y_1+Y_2+Y_3 \,, \hspace*{2cm}
L_2 = x_2+X_2+Y_1+Y_2+Y_3 \,, 
\nn\\
D_1 &= x_1+x_2+X_1+X_2 +2Y_1\,, \hspace*{1.7cm} D_2 =x_1+x_2+X_1+X_2+2Y_2 \,,
\nn\\
D_3 &= x_1+x_2+X_1+X_2 +2Y_3 \,, \hspace*{1.7cm} D_4 =x_1+x_2+X_1+X_2+2(Y_1+Y_2) \,,
\nn\\
D_5 &= x_1+x_2+X_1+X_2 +2(Y_1+Y_3) \,, \quad\,~ D_6 =X_1+X_2+x_1+x_2+2(Y_2+Y_3) \,,
\nn\\
D_7 &= x_1+x_2+X_1+X_2 \,.
\end{align}
And the Feynman diagram and the associated hyperplane arrangement for \eqrefe{eq:Psi-2loop-sun} are illustrated in fig.\,\ref{fig:5}.\
Inspecting fig.\,\ref{fig:5b}, there are 28 bounded triangles in total, each one corresponding to a unique function as can be categorized into one of the three layers.\ 
In layer-0, there are 7 parent functions $\PP_i$, each associated with a canonical form defined as $\Omega_{\PP_i}\!\!=\!\dlog(L_1/D_i)\wedge\dlog(L_2/D_i)\equiv\bOme_{\PP_i}\td\mu$ with $i=1,\ldots,7$.\
But similar to the one-loop bubble case, the 7 parent functions share one common $L$\hs-pair and can be combined into a single parent function via the following linear relation:
\begin{equation}
\label{eq:Omega-sun}
\bOme_{\PP_\sun} = \sum_{a=1}^{3}\bOme_{\PP_a} - \sum_{b=4}^{7}\bOme_{\PP_b} \,.
\end{equation}
Thus, for the sunset system, there are 22 independent functions organized into three layers, as presented in table\,\ref{Atab:1} of appendix\,\ref{app:C}, forming the integration family:
\begin{equation}
\label{eq:Sun-I}
\bI_\sun = ({\red\PP} ,\, {\blu \FF_1 ,\, \ldots ,\, \FF_7 ,\, \FFt_1,\, \ldots,\, \FFt_7},\,  {\org \QQ_1,\,\ldots,\QQ_7})^T \,.
\end{equation}
Here, in \eqrefe{eq:Sun-I}, layer-0 comprises a single parent function $\PP$, marked in red, layer-1 consists of 14 descendant functions $\{\FF_i,\FFt_i\}$, marked in blue, and layer-2 includes 7 descendant functions $\{\QQ_i\}$, marked in orange.\
Evidently, this basis choice subjected to \eqrefe{eq:Omega-sun} is a preferred triangle basis according to the algorithm discussed in section\,\ref{sec:3.2.2}.

The canonical differential equations can be straightforwardly constructed from the family trees generated via tube graphs. We omit the details.\ 
Here, we summarize the results as follows:
\beqs
\label{eq:DE-2loop}
\begin{align}
\td{\red\PP} &= \vep\hs\Big [{\red\PP}\hs(l_1+l_2)+\sum_{i=1}^{7}{\blu\FF_i}\hs(l_{2i+1}-l_1)+\sum_{j=1}^{7}{\blu\FFt_j}\hs(l_{2j+2}-l_2)\Big]\,,
\\[-1mm]
\td{\blu\FF_k} &= \vep\hs\big[{\blu\FF_k}\hs(l_{2k+1}+l_2) + {\org\QQ_k}\hs (l_{k+16}-l_2)\hs\big]\,,
\\[1.5mm]
\td{\blu\FFt_k} &= \vep\hs\big[{\blu\FFt_k}\hs(l_{2k+2}+l_1) + {\org\QQ_k}\hs (l_{k+16}-l_1)\hs\big]\,,
\\[1.5mm]
\td{\org\QQ_k} &= 2\hs\vep\hs{\org\QQ_k}\hs l_{k+16}\,,
\end{align}
\eeqs
where $k=1,\ldots,7$ and the letters $\{l_1,\ldots,l_{23}\}$ are summarized in table\,\ref{Atab:2} of appendix\,\ref{app:C}.\
Further, as referred to eqs.\,\eqref{eq:dI-UT}, \eqref{eq:Sun-I} and \eqref{eq:DE-2loop}, the matrix $A$ of two-site sunset case is a $22 \times 22$ matrix, presented below:
%
\begin{equation}
A_\sun\,=\(\!\!
\begin{array}{c}\hspace*{6cm}
\raisebox{0.em}{\hbox{\begin{minipage}{3.5cm}
\hspace*{-5.9cm}
\includegraphics[scale=.4]{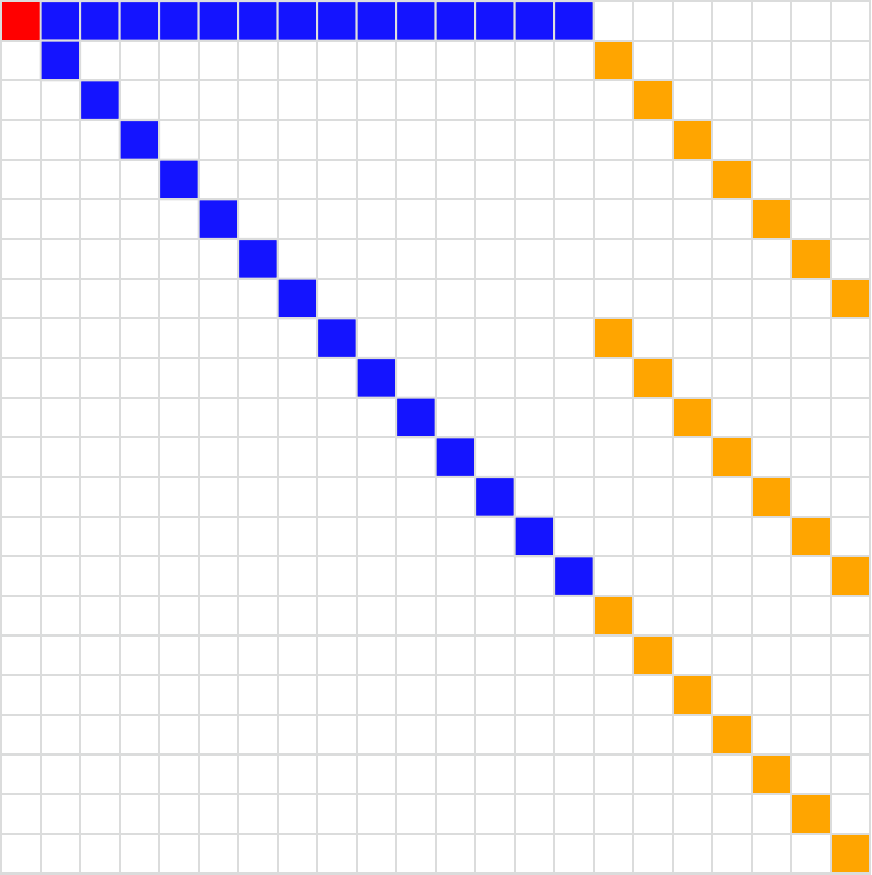}
\end{minipage}}}
\end{array}\hspace*{-3.3cm}
\)_{\!22\times22},
\end{equation}
where each colored cell corresponds to the letters associated with the functions in \eqrefe{eq:Sun-I}.\ 
In addition, it is not difficult to check that $A_\sun$ obeys the integrability conditions \eqref{eq:inte-con}.

Moreover, we discuss a general 2-site graph with $\NN\!\in\!\mathbb{Z}^+$ bulk-to-bulk propagators between two vertices, as shown in fig.\,\ref{fig:6}.\ 
In this case, the kinematic flow approach reveals that the number of independent functions is given by

\begin{equation}
\label{eq:function-count}
1+ 3\hs\sum_{i=1}^{\NN} \binom{\NN\,}{i} =\, 3\cdot2^{\hs\NN}-2\,,
\end{equation}
where on the left-hand side, factor 1 represents one single parent function, the index $i$ counts the number of cross signs contained within the tube for each type of function $\FF_k$ and the overall factor 3 reflects the fact that functions $\FF_k,\,\FFt_\ell$ and $\QQ_q$ appear in equal quantities.
Alternatively, by analyzing the hyperplane arrangement and applying the algorithm for a preferred triangle basis, the number of functions can be determined as follows:
\begin{equation}
\label{eq:D-count-1}
4\mZ - (\mZ-1) = 3\hs\mZ+1\,,
\end{equation}
where $\mZ$ denotes the number of diagonally-placed hyperplanes $\{D_i\}$.\ 
The total number of letters is exactly one greater than the number of functions.\
Combining \eqrefe{eq:function-count} with \eqrefe{eq:D-count-1}, we can obtain:
\begin{equation}
\label{eq:D-count}
\mZ = 2^{\hs\NN} - 1 \,,
\end{equation}
which reveals how many parent functions we can identify before they are merged.
\begin{figure}[t]
\centering
\tikzset{every picture/.style={line width=0.5pt}}
\begin{tikzpicture}[x=0.75pt,y=0.75pt,yscale=-.8,xscale=.8]
\draw  [line width=1.5]  (141.25,191.94) .. controls (141.25,167.89) and (160.74,148.39) .. (184.79,148.39) .. controls (208.84,148.39) and (228.33,167.89) .. (228.33,191.94) .. controls (228.33,215.98) and (208.84,235.48) .. (184.79,235.48) .. controls (160.74,235.48) and (141.25,215.98) .. (141.25,191.94) -- cycle ;
\draw  [fill={rgb, 255:red, 0; green, 0; blue, 0 }  ,fill opacity=1 ][line width=1.5]  (137.25,191.94) .. controls (137.25,189.73) and (139.04,187.94) .. (141.25,187.94) .. controls (143.46,187.94) and (145.25,189.73) .. (145.25,191.94) .. controls (145.25,194.15) and (143.46,195.94) .. (141.25,195.94) .. controls (139.04,195.94) and (137.25,194.15) .. (137.25,191.94) -- cycle ;
\draw  [fill={rgb, 255:red, 0; green, 0; blue, 0 }  ,fill opacity=1 ][line width=1.5]  (224.33,191.94) .. controls (224.33,189.73) and (226.12,187.94) .. (228.33,187.94) .. controls (230.54,187.94) and (232.33,189.73) .. (232.33,191.94) .. controls (232.33,194.15) and (230.54,195.94) .. (228.33,195.94) .. controls (226.12,195.94) and (224.33,194.15) .. (224.33,191.94) -- cycle ;
\draw  [line width=1.5]  (142.08,191.19) .. controls (142.08,172.96) and (161.39,158.19) .. (185.21,158.19) .. controls (209.03,158.19) and (228.33,172.96) .. (228.33,191.19) .. controls (228.33,209.41) and (209.03,224.19) .. (185.21,224.19) .. controls (161.39,224.19) and (142.08,209.41) .. (142.08,191.19) -- cycle ;
\draw  [line width=1.5]  (141.25,191.44) .. controls (141.25,179.98) and (160.56,170.69) .. (184.38,170.69) .. controls (208.19,170.69) and (227.5,179.98) .. (227.5,191.44) .. controls (227.5,202.9) and (208.19,212.19) .. (184.38,212.19) .. controls (160.56,212.19) and (141.25,202.9) .. (141.25,191.44) -- cycle ;
\draw (108,185) node [anchor=north west][inner sep=0.75pt]  [font=\small]  {$X_1$};
\draw (235,185) node [anchor=north west][inner sep=0.75pt]  [font=\small]  {$X_2$};
\draw (176,123) node [anchor=north west][inner sep=0.75pt]  [font=\small]  {$Y_{\hsm\NN}$};
\draw (177,241) node [anchor=north west][inner sep=0.75pt]  [font=\small]  {$Y_1$};
\draw (192,178) node [anchor=north west][inner sep=0.75pt] [font=\large,rotate=-90]{$\cdots$};
\end{tikzpicture}
%
\caption{The diagram for a general two-site bubble-like wavefunction coefficient involves $\NN$ internal lines.\ The late-time boundary and external legs are omitted.}
\label{fig:6}
\end{figure}

\vs

Finally, for the general 2-site ($\NN\!-\!1$)-loop diagram illustrated in fig.\,\ref{fig:6}, the form of its single parent function can be constructed by shifting the form corresponding to the parent function of a 2-site tree-level diagram.\ 
Specifically, we define the canonical form associated with the two-site tree-level parent function $\PP_\tree$ as:
\begin{equation}
\label{eq:Omega-tree}
\bOme_{\PP_\tree} =\, - \frac{L_1\!+\!L_2\!-\!D}{L_1\hs L_2\hs D}\,,
\end{equation}
where the hyperplanes are given by
\begin{equation}
L_1 =x_1\!+\!X_1\!+\!Y \,,\qquad
L_2 =x_2\!+\!X_2\!+\!Y \,,\qquad
D =x_1\!+\!x_2\!+\!X_1\!+\!X_2\,.
\end{equation}
Thus, the form associated with the parent function for a 2-site ($\NN\!\!-\!\!1$)-loop bubble-like correlator is expressed as follows:
\begin{equation}
\label{eq:2-site-LC}
\bOme_{\PP_{\rm{bll}}} \,= 
\sum_{\SS \in P(\YY)\backslash\{\YY\}} (-1)^{|\SS|+1}\, \bOme_{\PP_\tree} \bigg|{\substack{\\[1.5mm]
\hspace*{-6.5mm}Y\to\summ_{Y_i\in\YY} Y_i\\[1mm]
D \hs\to D + 2\hs\summ_{Y_j\in\SS} Y_j}} ~,
\end{equation}
where $\YY\!=\!\{Y_1, \ldots, Y_\NN\}$ is the set of internal energies and $P(\YY)$ denotes the power set of $\YY$.\
Additionally, the right-hand side of \eqrefe{eq:2-site-LC} involves a summation over $\mZ \!=\! 2^{\hs\NN} \!-\! 1$ possible reparametrization(s), each associated with a distinct parent function to be merged. These contributions are ultimately consolidated into a single parent function through the summation in \eqrefe{eq:2-site-LC}.

At the two-site two-loop level, three additional topologies emerge, each involving one or more tadpoles.\ 
As highlighted previously, these cases exemplify scenarios where the differential equations unavoidably involve multiple parent functions and fail to fully span the hyperplane system.\ 
In the rest of this section, we demonstrate that the preferred triangle basis offers a systematic and transparent approach to address these more intricate cases.\ 
It enables us to reveal tree-level factorization, quantify the number of functions in each layer, and systematically organize the structure of the $A$ matrix.\
Finally, by appropriately regrouping the to-be-merged parent functions, these examples yield a compact formula for generic two-site graphs.

\begin{figure}[t]
\centering
\begin{subfigure}{0.31\textwidth}
\hspace*{-3mm}
\raisebox{12mm}{
\includegraphics[scale=.85]{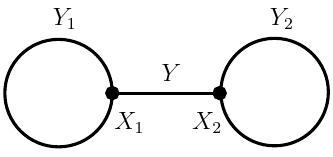}}
\caption{}
\label{fig:7a}
\end{subfigure}%
\hspace*{1.5cm}
\begin{subfigure}{0.31\textwidth}
\includegraphics[scale=.85]{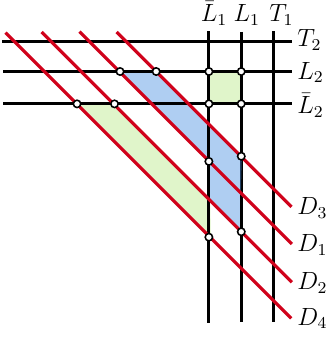}
\caption{}
\label{fig:7b}
\end{subfigure}
\vspace*{-2mm}
\caption{(a).\,Two-site two-tadpole ``dumbbell'' diagram where the late-time boundary and external legs are omitted.\ 
(b).\,The hyperplane arrangement of dumbbell system in the configuration $X\!\gg\!Y$\hs\protect\footnotemark\,where
all non-vanishing codimension-2 boundaries of $\bOme_{\PP_\dbl}$ are marked as the white intersection points, allowing us to extract the triangulation corresponding to the tree factorization.\ 
The triangulation yields a net contribution by summing over all triangular areas achieved through the matching of residues on codimension-2 boundaries.\ The net contribution corresponds to the colored regions, with blue areas assigned a weight of $+1$ and green areas a weight of $-1$. Summing the forms associated with colored regions, weighted accordingly, reproduces $\bOme_{\PP_\dbl}$.} \label{fig:7}
\end{figure}
\footnotetext{This configuration is chosen for convenience so that one does not need to deal with the relative position between $L$ and $D$ planes.}

\paragraph{Dumbbell} 
One of the two-loop systems demonstrating tree-level factorization analogous to the one-loop tadpole case is the two-tadpole ``dumbbell'' diagram, depicted in fig.\,\ref{fig:7a}. 
The wavefunction coefficient is computed as follows:
\begin{align}
\label{eq:Psi-2loop-dbl}
\psi_{(2,2)}^{\dbl} &= 8YY_1Y_2\!\int_{\mathbb{R}^2_+}\!\frac{(T_1T_2)^\vep\td x_1\!\wedge\!\td x_2}{\bar{L}_1 \bar{L}_2 D_3}\! \[\!\frac{1}{L_1L_2}+ \frac{1}{D_1}\!\(\frac{1}{L_2}\!+\!\frac{1}{D_4}\)+\frac{1}{D_2}\!\(\frac{1}{L_1}\!+\!\frac{1}{D_4}\)\!\]
\nn\\[0mm]
&\equiv \int\!\td\mu\, \bOme_{\PP_\dbl}\,,
\end{align}
where the hyperplanes are defined as follows:
\begin{alignat}{3}
\label{eq:hyperplanes-dbl}
L_1& = x_1+X_1+Y \,, \qquad
&&L_2 = x_2+X_2+Y \,, 
\nn\\
\bar{L}_1& = x_1+X_1+Y+2Y_1 \,,\qquad
&&\bar{L}_2 = x_2+X_2+Y+2Y_2 \,, 
\nn\\
D_1 &= x_1+x_2+X_1+X_2 +2Y_1\,, \qquad
&&D_2 =x_1+x_2+X_1+X_2+2Y_2 \,,
\nn\\
D_3 &= x_1+x_2+X_1+X_2 \,, 
&&D_4 =x_1+x_2+X_1+X_2+2(Y_1+Y_2) \,.
\end{alignat}
As the hyperplane arrangement becomes increasingly complicated at higher loops, we briefly describe how to obtain a similar tree-level factorization as the one-loop case \eqref{eq:Omega-tad}.\
Evaluating every codimension-2 residue of the integrand within hyperplane arrangement (indicated by white intersection points in fig.\,\ref{fig:7b}), we identify the shaded region whose canonical form is $\bOme_{\PP_\dbl}$.\
The shaded regions provide a visual representation of the physical wavefunction coefficient on the hyperplane configuration.\ 
Its triangulation further indicates that the wavefunction coefficient can be decomposed into a linear combination of four tree-level subsystems as:
\begin{equation}
\label{eq:Omega-dbl}
\bOme_{\PP_\dbl} = \bOme_{\PP_1} + \bOme_{\PP_2} - \bOme_{\PP_3} - \bOme_{\PP_4}  \,,
\end{equation}
where each $\bOme_{\PP_i}$ can be expressed in terms of \eqrefe{eq:hyperplanes-dbl} as follows:
\begin{equation}
\label{eq:dbl-bOme-Pi}
\bOme_{\PP_1} \!=\! \frac{-2Y}{\bar{L}_1L_2D_1}\,, \qquad~
\bOme_{\PP_2} \!=\! \frac{-2Y}{L_1\bar{L}_2D_2} \,, \qquad~
\bOme_{\PP_3} \!=\! \frac{-2Y}{L_1L_2D_3} \,,\qquad~
\bOme_{\PP_4} \!=\! \frac{-2Y}{\bar{L}_1\bar{L}_2D_4} \,.
\end{equation}
Further, we find that each form $\bOme_{\PP_i}$ of subsystem can also be interpreted as a shifted version of \eqrefe{eq:Omega-tree}:
\begin{alignat}{3}
\label{eq:dbl-shift}
\bOme_{\PP_1} &=\bOme_{\PP_\tree}\big|{\substack{
\\[1.5mm]
X_1\to X_1+2Y_1}} \,, \qquad~~~
&&\bOme_{\PP_2} = \bOme_{\PP_\tree}\big|{\substack{
\\[1.5mm]
X_2\to X_2+2Y_2}} \,,
\nn\\[0mm]
\bOme_{\PP_3} &= \bOme_{\PP_\tree}\,,
&&\bOme_{\PP_4}=\bOme_{\PP_\tree}\Big|{\substack{
\\[1.5mm]
X_1\to X_1+2Y_1 \\
X_2\to X_2+2Y_2}}\,.
\end{alignat}
Now, inspecting \eqrefe{eq:dbl-bOme-Pi} or \eqrefe{eq:dbl-shift}, it becomes evident that each $\PP_i$ is associated with a distinct $L$\hs-pair.\
This distinction validates these functions as top-layer components within the preferred triangular basis.\ 
In essence, the four parent functions cannot be merged into a single one because the corresponding letters, arising from their shifts, are inherently different.\
Specifically, the differential equation for each parent function takes the form:
\begin{equation}
\td\PP_i = \vep\hs\big[\PP_i\hs(l_{1,i}+l_{2,i})+\cdots]\,,
\end{equation}
where $\{l_{1,i}\}$ and $\{l_{2,i}\}$ (with $i\!=\!1,\ldots,4$) are the letters under the shift operations \eqref{eq:dbl-shift}.\ 
Within this framework, each shifted tree system appears as an integrable 
$4\times4$ tree block in the matrix $A_\dbl$.\
From the perspective of the preferred triangle basis, it is clear that the entire system, consisting of 24 bounded regions, is spanned by 4 $\P$-functions, 16 $\F$\hs-functions and 4 $\Q$-functions, while the physical two-loop dumbbell system utilizes only 8 of these 16 $\F$\hs-functions.

In short, the wavefunction coefficient and the associated differential system of the dumbbell graph decompose into a direct sum of multiple tree-like subsystems.\ The tree system serves as an unshifted central building block, while the tadpoles induce shifts and sign factors acting directly on the tree.

\begin{figure}[t]
\centering
\begin{subfigure}{0.31\textwidth}
\hspace*{-3mm}
\raisebox{12mm}{
\includegraphics[scale=.85]{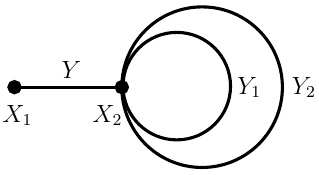}}
\caption{}
\label{fig:8a}
\end{subfigure}%
\hspace*{1.5cm}
\begin{subfigure}{0.31\textwidth}
\raisebox{-1mm}{\includegraphics[scale=.85]{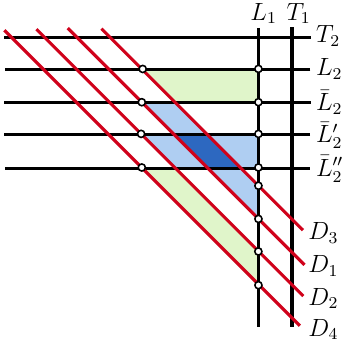}}
\vspace*{-6mm}
\caption{}
\label{fig:8b}
\end{subfigure}
\vspace*{-2mm}
\caption{(a).\,Two-site two-tadpole ``lollipop'' diagram.\ 
(b).\,The associated hyperplane arrangement in the configuration $X\!\!\gg\!\!Y$, where the net contribution corresponds to the colored regions, with dark blue area assigned a weight of $+2$, blue areas $+1$ and green areas $-1$.\
Summing the forms associated with the weighted regions reproduces $\bOme_{\PP_\lop}$.}
\label{fig:8}
\end{figure}

\paragraph{Lollipop} 
Another example with a similar tree-level factorization is a diagram representing a ``lollipop'' with two tadpoles, as illustrated in fig.\,\ref{fig:8a}.\
The corresponding wavefunction coefficient is given by
\begin{align}
\label{eq:Psi-2loop-lop}
\psi_{(2,2)}^{\lop} &= 8YY_1Y_2\!\int_{\mathbb{R}^2_+}\!
\frac{(T_1T_2)^\vep\td x_1\!\wedge\!\td x_2}{L_1 \bar{L}^{\pp\pp}_2 D_3}\! \[\!
\frac{1}{D_1}\!\(\frac{1}{D_4}\!+\!\frac{1}{\bar{L}_2}\)
+\frac{1}{D_2}\!\(\frac{1}{D_4}\!+\!\frac{1}{\bar{L}_2^\pp}\)
+\frac{1}{L_2}\!\(\frac{1}{\bar{L}_2}\!+\!\frac{1}{\bar{L}_2^\pp}\)
\!\]
\nn\\[-1mm]
&\equiv \int\!\td\mu\, \bOme_{\PP_\lop}\,,
\end{align}
where the hyperplanes are defined as follows:
\begin{alignat}{3}
\label{eq:hyperplanes-lop}
L_1 &= x_1+X_1+Y\,, \qquad
&&L_2 = x_2+X_2+Y \,,   \qquad~~ 
\bar{L}_2 = x_2+X_2+Y+2Y_1 \,, 
\nn\\
\bar{L}_2^\pp &= x_2+X_2+Y+2Y_2\,,
&&\bar{L}_2^{\pp\pp} = x_2+X_2+Y+2(Y_1+Y_2) \,,
\nn\\
D_1 &= x_1+x_2+X_1+X_2 +2Y_1\,, \quad
&&D_2 =x_1+x_2+X_1+X_2+2Y_2 \,,
\nn\\
D_3 &= x_1+x_2+X_1+X_2 \,, 
&&D_4 =x_1+x_2+X_1+X_2+2(Y_1+Y_2) \,.
\end{alignat}

A similar analysis in its hyperplane arrangement (cf.\,fig.\,\ref{fig:8b}) shows that the tree factorization for $\bOme_{\PP_\lop}$ is given by
\begin{equation}
\label{eq:Omega-lop}
\bOme_{\PP_\lop}= \bOme_{\PP_1} + \bOme_{\PP_2} - \bOme_{\PP_3} - \bOme_{\PP_4}  \,,
\end{equation}
where $\bOme_{\PP_i}$ can be expressed in terms of \eqrefe{eq:hyperplanes-lop} as follows:
\begin{equation}
\label{eq:lop-bOme-Pi}
\bOme_{\PP_1} \!=\! \frac{-2Y}{L_1\bar{L}_2D_1}\,, \qquad~
\bOme_{\PP_2} \!=\! \frac{-2Y}{L_1\bar{L}^\pp_2D_2} \,, \qquad~
\bOme_{\PP_3} \!=\! \frac{-2Y}{L_1L_2D_3} \,,\qquad~
\bOme_{\PP_4} \!=\! \frac{-2Y}{L_1\bar{L}^{\pp\pp}_2D_4} \,.
\end{equation}
And each form $\bOme_{\PP_i}$ can be regarded as a shifted version of \eqrefe{eq:Omega-tree}:
\begin{alignat}{3}
\label{eq:lop-shift}
\bOme_{\PP_1} &=\bOme_{\PP_\tree}\big|{\substack{
\\[1.5mm]
X_2\to X_2+2Y_1}} \,, \qquad
&&\bOme_{\PP_2} = \bOme_{\PP_\tree}\big|{\substack{
\\[1.5mm]
X_2\to X_2+2Y_2}} \,,
\nn\\[1mm]
\bOme_{\PP_3} &= \bOme_{\PP_\tree}\,,
&&\bOme_{\PP_4}=\bOme_{\PP_\tree}\big|{\substack{
\\[1.5mm]
X_2\to X_2+2(Y_1+Y_2)}}\,.
\end{alignat}
Inspecting \eqrefe{eq:lop-bOme-Pi} or \eqrefe{eq:lop-shift}, each $\PP_i$ has a distinct $L$\hs-pair, serving as a distinct top-layer component.\ 
Similar to the dumbbell case, the matrix $A_\lop$ has four $4\times4$ blocks.\ 
The whole system in the preferred basis contains 4 $\P$-functions, 20 $\F$\hs-functions and 4 $\Q$-functions, corresponding to a total of 28 bounded regions, while the physical two-loop lollipop system only involves 8 of these 20 $\F$\hs-functions.

Once again, the lollipop system reduces to a direct sum of central tree systems with appropriate shifts and sign factors induced by the attached two tadpoles.

\begin{figure}[t]
\centering
\begin{subfigure}{0.31\textwidth}
\hspace*{-3mm}
\raisebox{7mm}{
\includegraphics[scale=.8]{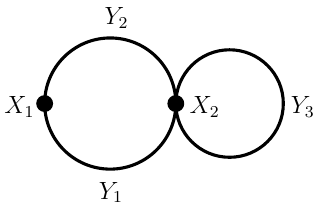}}
\caption{}
\label{fig:9a}
\end{subfigure}%
\hspace*{1.5cm}
\begin{subfigure}{0.31\textwidth}
\raisebox{-1mm}{\includegraphics[scale=.85]{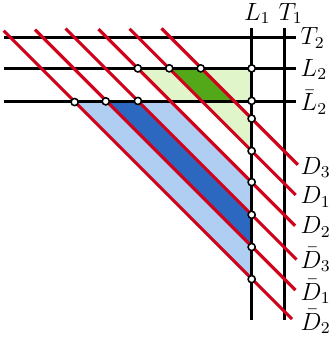}}
\vspace*{-7mm}
\caption{}
\label{fig:9b}
\end{subfigure}
\vspace*{-2mm}
\caption{(a).\,Two-site two-loop ``gourd'' diagram with one bubble and one tadpole.\ 
(b).\,The associated hyperplane arrangement where the net contribution corresponds to the colored regions, with dark blue area assigned a weight of $+2$, blue areas $+1$, dark green area $-2$ and green areas $-1$.\
Summing the forms associated with the weighted regions reproduces $\bOme_{\PP_\grd}$.}
\label{fig:9}
\end{figure}

\paragraph{Gourd}
The third example is the two-site two-loop ``gourd'' diagram shown in fig.\,\ref{fig:9a}, which exhibits a mixed structure incorporating features of both the bubble and tadpole systems.\ This allows for a nontrivial regrouping of parent functions. The corresponding wavefunction coefficient is given by
\begin{align}
\label{eq:Psi-2loop-gourd}
\psi_{(2,2)}^{\grd} &\!=\! -8Y_1Y_2 Y_3\!\int_{\mathbb{R}^2_+}\!\!\!\frac{(T_1T_2)^\vep\td x_1\!\wedge\!\td x_2}{L_1 \bar{L}_2 D_3}\! \[\!\frac{1}{L_2}\!\(\frac{1}{D_1}\!+\!\frac{1}{D_2}\)\!+\!\frac{1}{\bar{D}_1}\!\(\frac{1}{D_1}\!+\!\frac{1}{\bar{D}_3}\)\!+\!\frac{1}{\bar{D}_2}\!\(\frac{1}{D_2}\!+\!\frac{1}{\bar{D}_3}\)\!\]
\nn\\[-1mm]
&\!\equiv \int\!\td\mu\, \bOme_{\PP_\grd}\,,
\end{align}
where the hyperplanes are defined as follows:
%
\begin{alignat}{3}
\label{eq:hyperplanes-grd}
L_1 &= x_1+X_1+Y_1+Y_2\,, \qquad
&&
\nn\\ 
L_2 &= x_2+X_2+Y_1+Y_2 \,,  \qquad
&&\bar{L}_2 = x_2+X_2+Y_1+Y_2+2Y_3 \,, 
\nn\\
D_1 &= x_1+x_2+X_1+X_2 +2Y_1\,, 
&&\bar{D}_1 = x_1+x_2+X_1+X_2+2(Y_1+Y_3) \,,
\nn\\
D_2 &= x_1+x_2+X_1+X_2+2Y_2 \,, \qquad
&& \bar{D}_2 =x_1+x_2+X_1+X_2+2(Y_2+Y_3) \,,
\nn\\
D_3 &= x_1+x_2+X_1+X_2 \,,  \qquad
&&\bar{D}_3 =x_1+x_2+X_1+X_2+2Y_3 \,.
\end{alignat}

Next, inspecting fig.\,\ref{fig:9b} and evaluating each residue on codimension-2 boundaries associated with $\PP_\grd$, we find that tree factorization for $\bOme_{\PP_\grd}$ is given by
\begin{equation}
\label{eq:grd-ome-shift}
\bOme_{\PP_\grd} = \bOme_{\PP_1^\pp} + \bOme_{\PP_2^\pp} - \bOme_{\PP_3^\pp} -\big(\,\bOme_{\PP_1}+\bOme_{\PP_2}-\bOme_{\PP_3}\,\big) \,,
\end{equation}
where in terms of \eqrefe{eq:hyperplanes-grd}, $\bOme_{\PP_i},\bOme_{\PP_i^\pp}$ can be further expressed as follows:
\beqs
\begin{alignat}{3}
\label{eq:grd-bOme-Pi}
\bOme_{\PP_1} &= \frac{-2Y_2}{L_1{L}_2{D}_1}\,,\qquad~~
&\bOme_{\PP_2} = \frac{-2Y_1}{L_1{L}_2{D}_2}\,,\qquad~~
&\bOme_{\PP_3} = \frac{-2(Y_1+Y_2)}{L_1{L}_2{D}_3}\,,
\\[1mm]
\label{eq:grd-bOme-Ppi}
\bOme_{\PP^\pp_1} &= \frac{-2Y_2}{L_1\bar{L}_2\bar{D}_1}\,,\qquad~~
&\bOme_{\PP^\pp_2} = \frac{-2Y_1}{L_1\bar{L}_2\bar{D}_2}\,,\qquad~~
&\bOme_{\PP^\pp_3} = \frac{-2(Y_1+Y_2)}{L_1\bar{L}_2\bar{D}_3}\,.
\end{alignat}
\eeqs
Similarly, each $\bOme_{\PP_i}$ and $\bOme_{\PP_i^\pp}$ is represented as a shifted version of \eqrefe{eq:Omega-tree}:
\begin{alignat}{3}
\label{eq:grd-shift}
\bOme_{\PP_1} &=\bOme_{\PP_\tree}\Big|{\substack{
\\[.5mm]
\hspace*{-5mm}Y\to Y_2\\
X_1\to X_1+Y_1\\
X_2\to X_2+Y_1}} \,, \hspace*{2cm}
&&\bOme_{\PP_1^\pp} =\bOme_{\PP_\tree}\Big|{\substack{
\\[.5mm]
\hspace*{-12mm}Y\to Y_2\\
\hspace*{-7mm}X_1\to X_1+Y_1\\
X_2\to X_2+Y_1+2Y_3}} \,,
\nn\\[0mm]
\bOme_{\PP_2} &=\bOme_{\PP_\tree}\Big|{\substack{
\\[.5mm]
\hspace*{-5mm}Y\to Y_1\\
X_1\to X_1+Y_2\\
X_2\to X_2+Y_2}} \,, 
&&\bOme_{\PP_2^\pp} =\bOme_{\PP_\tree}\Big|{\substack{
\\[.5mm]
\hspace*{-12mm}Y\to Y_1\\
\hspace*{-7mm}X_1\to X_1+Y_2\\
X_2\to X_2+Y_2+2Y_3}} \,,
\nn\\
\bOme_{\PP_3} &=\bOme_{\PP_\tree}\big|{\substack{
\\[1.5mm]
Y\to Y_1+Y_2}} \,, 
&&\bOme_{\PP_3^\pp} =\bOme_{\PP_\tree}\Big|{\substack{
\\[1.5mm]
X_2\to X_2+2Y_3\\
\hspace*{-1mm}Y\to Y_1+ Y_2}} \,.
\end{alignat}
At this stage, all $\PP_i$ share the same $(L_1,L_2)$-pair, while $\PP^\pp_i$ share the same $(L_1,\bar{L}_2)$-pair. This observation suggests that the six parent functions can be merged into just two parent functions.\ 
And they can be grouped into a one-loop bubble and a shifted one-loop bubble:
\begin{equation}
\label{eq:grd-shift-2}
\bOme_{\PP_1} + \bOme_{\PP_2} - \bOme_{\PP_3}=\bOme_{\PP_\bub}  \,, \qquad \bOme_{\PP^\pp_1}+\bOme_{\PP^\pp_2}-\bOme_{\PP^\pp_3}=\bOme_{\PP_\bub}\big|{\substack{\\[1.5mm]
X_2\to X_2+2Y_3}}\,,
\end{equation}
where $\bOme_{\PP_\bub}$ is defined in \eqrefe{eq:P=P1+P2-P3}.\ 
Hence, the matrix $A_{\grd}$ takes a block-diagonal form:
\begin{equation}
A^\grd_{20\times20} \,=\, A^\bub_{10\times10} \,\oplus\, A^\bub_{10\times10}\big|{\substack{\\[1.5mm]
X_2\to X_2+2Y_3}} \,,
\end{equation}
where $A^\bub_{10\times10}$ is defined in \eqrefe{eq:At-bub}. This can be further diagonalized into two $4\times4$ plus four $3\times3$ blocks by choosing either one of $\PP_i$ and one of $\PP^\pp_i$ as the two top-layer functions in the preferred triangle basis.\ 
The whole system in the preferred basis has 2 $\P$-functions, 18 $\F$\hs-functions, and 6 $\Q$-functions corresponding to a total of 26 bounded regions, while the physical two-loop gourd system selectively involves 12 of these 18 $\F$-functions.

The gourd system provides a notable example where the tree-level factorization is redundant, as in the bubble case, but the merging of components is partially obstructed, similar to the tadpole case.\ As a result, a minimal representation of the associated differential system involves the direct sum of a bubble system and its shifted counterpart.\ Crucially, in this case, the tadpole induces a shift and sign on the entire bubble structure, rather than on the individual central tree components as in previous cases.

Taken together, these tadpole examples exhibit a recurring structural pattern: each system decomposes into a direct sum of the corresponding central building block, with shift and sign factors arising combinatorially from the tadpole attachments.\ In practice, this pattern becomes apparent by first applying possibly redundant tree-level factorization and then performing an appropriate recombination based on a preferred triangle basis. This procedure yields a minimal representation of the full differential system, compactly encoding all the shift structures within the central building blocks, as illustrated by comparing \eqrefe{eq:grd-shift-2} with \eqrefe{eq:grd-shift}.

\vs

\begin{figure}[t]
\centering
\tikzset{every picture/.style={line width=0.3pt}}
\begin{tikzpicture}[x=0.75pt,y=0.75pt,yscale=-.9,xscale=.9,scale=0.8]
\draw  [line width=1.5]  (291.25,209.27) .. controls (291.25,182.14) and (313.24,160.14) .. (340.38,160.14) .. controls (367.51,160.14) and (389.5,182.14) .. (389.5,209.27) .. controls (389.5,236.4) and (367.51,258.39) .. (340.38,258.39) .. controls (313.24,258.39) and (291.25,236.4) .. (291.25,209.27) -- cycle ;
\draw  [fill={rgb, 255:red, 0; green, 0; blue, 0 }  ,fill opacity=1 ][line width=1.5]  (287.25,209.27) .. controls (287.25,207.06) and (289.04,205.27) .. (291.25,205.27) .. controls (293.46,205.27) and (295.25,207.06) .. (295.25,209.27) .. controls (295.25,211.48) and (293.46,213.27) .. (291.25,213.27) .. controls (289.04,213.27) and (287.25,211.48) .. (287.25,209.27) -- cycle ;
\draw  [fill={rgb, 255:red, 0; green, 0; blue, 0 }  ,fill opacity=1 ][line width=1.5]  (385.5,209.27) .. controls (385.5,207.06) and (387.29,205.27) .. (389.5,205.27) .. controls (391.71,205.27) and (393.5,207.06) .. (393.5,209.27) .. controls (393.5,211.48) and (391.71,213.27) .. (389.5,213.27) .. controls (387.29,213.27) and (385.5,211.48) .. (385.5,209.27) -- cycle ;
\draw  [line width=1.5]  (291.25,209.27) .. controls (291.25,193.14) and (313.06,180.06) .. (339.96,180.06) .. controls (366.86,180.06) and (388.67,193.14) .. (388.67,209.27) .. controls (388.67,225.4) and (366.86,238.48) .. (339.96,238.48) .. controls (313.06,238.48) and (291.25,225.4) .. (291.25,209.27) -- cycle ;
\draw  [line width=1.5]  (389.5,209.27) .. controls (389.5,188.29) and (406.51,171.28) .. (427.49,171.28) .. controls (448.47,171.28) and (465.48,188.29) .. (465.48,209.27) .. controls (465.48,230.25) and (448.47,247.26) .. (427.49,247.26) .. controls (406.51,247.26) and (389.5,230.25) .. (389.5,209.27) -- cycle ;
\draw  [line width=1.5]  (389.5,209.27) .. controls (389.5,199.74) and (397.23,192.01) .. (406.76,192.01) .. controls (416.29,192.01) and (424.02,199.74) .. (424.02,209.27) .. controls (424.02,218.8) and (416.29,226.53) .. (406.76,226.53) .. controls (397.23,226.53) and (389.5,218.8) .. (389.5,209.27) -- cycle ;
\draw  [line width=1.5]  (257.56,209.27) .. controls (257.56,199.74) and (265.29,192.01) .. (274.82,192.01) .. controls (284.36,192.01) and (292.08,199.74) .. (292.08,209.27) .. controls (292.08,218.8) and (284.36,226.53) .. (274.82,226.53) .. controls (265.29,226.53) and (257.56,218.8) .. (257.56,209.27) -- cycle ;
\draw  [line width=1.5]  (215.27,209.27) .. controls (215.27,188.29) and (232.28,171.28) .. (253.26,171.28) .. controls (274.24,171.28) and (291.25,188.29) .. (291.25,209.27) .. controls (291.25,230.25) and (274.24,247.26) .. (253.26,247.26) .. controls (232.28,247.26) and (215.27,230.25) .. (215.27,209.27) -- cycle ;
\draw (333,136) node [anchor=north west][inner sep=0.75pt]  
[font=\small]  {$Y_{\hsm\NN}$};
\draw (333,265) node [anchor=north west][inner sep=0.75pt]  
[font=\small]  {$Y_1$};
\draw (346,197) node [anchor=north west][inner sep=0.75pt]  [font=\large,rotate=-90]  {$\cdots$};
\draw (430,205) node [anchor=north west][inner sep=0.75pt]  [font=\large,rotate=0]  {$\cdots$};
\draw (225,205) node [anchor=north west][inner sep=0.75pt]  [font=\large,rotate=0]  {$\cdots$};
\draw (427,182) node [anchor=north west][inner sep=0.75pt]  
[font=\small]  {$\tmY_1$};
\draw (462,167) node [anchor=north west][inner sep=0.75pt]  
[font=\small]  {$\tmY_\JJ$};
\draw (236,182) node [anchor=north west][inner sep=0.75pt]  
[font=\small]  {$\mY_1$};
\draw (195,167) node [anchor=north west][inner sep=0.75pt]  
[font=\small]  {$\mY_\II$};
\draw (297,203) node [anchor=north west][inner sep=0.75pt]  
[font=\small]  {$X_1$};
\draw (355,203) node [anchor=north west][inner sep=0.75pt]  
[font=\small]  {$X_2$};
\end{tikzpicture}
\caption{A general 2-site diagram with arbitrary loops.\ The central diagram contains $\NN$ internal lines and its two vertices are connected to $\II$ and $\JJ$ tadpole loops, respectively.}
\label{fig:10}
\end{figure}

Building on the above discussions, the parent function for a general 2-site diagram with arbitrary loops can also be analyzed.\ 
As illustrated in fig.\,\ref{fig:10}, such a 2-site diagram can be divided into two parts: (i).\,The central part of the diagram, which contains $\NN$ internal lines with $\NN\in\mathbb{Z}^+$, which is same as fig.\,\ref{fig:6}.\
(ii).\,The two vertices of the central diagram, each connects to $\II$ and $\JJ$ internal lines with $\II,\JJ\in\mathbb{N}^0$, forming two tadpole structures.\
Hence, for a two-site $L$\hs-loop correlator with $L=\NN+\II+\JJ-1$, its associated canonical form can be decomposed into an unshifted form $\bOme_{\PP_\ctr}$, corresponding to the central diagram and all possible shifted forms of $\bOme_{\PP_\ctr}$ generated by the tadpole loops combinatorially,
\begin{equation}
\label{eq:2-site-N-loop}
\bOme_{\PP} \,= 
\sum_{\SS_\L \in P(\YY_\L)}\sum_{\SS_\R \in P(\YY_\R)} (-1)^{|\SS_\L\cup\,\SS_\R|+1}\, \bOme_{\PP_\ctr}\bigg|{\substack{\\[1.5mm]
\! X_1\to X_1 + 2\sum_{\mY_i \in \SS_\L}\!\mY_i\\[-0.5mm]
X_2\to X_2 + 2\sum_{\tmY_j \in \SS_\R}\!\tmY_j}} ~,
\end{equation}
where $\YY_\L =\{\mY_1, \ldots, \mY_\II\}$ and $\YY_\R=\{\tmY_1, \ldots, \tmY_\JJ\}$ are the sets of internal energies of tadpole loops associated with vertices 1 and 2, while $P(\YY_L)$ and $P(\YY_R)$ denote their power sets.\
Thus, all the examples discussed in sections\,\ref{sec:3}--\ref{sec:5} can be systematically summarized as specific cases of \eqrefe{eq:2-site-N-loop} as follows:

\begin{itemize}[leftmargin=*]

\item 
$\NN\!=\!1$: The unshifted form is $\bOme_{\PP_\ctr}\!\!=\bOme_{\PP_\tree}$.\ 
For $(\II,\JJ)\!=\!(0,1)$, \eqrefe{eq:2-site-N-loop} yields the one-loop tadpole correlator \eqref{eq:Omega-tad}; 
For $(\II,\JJ)\!=\!(1,1)$, it corresponds to the two-loop dumbbell correlator \eqref{eq:Omega-dbl}; 
For $(\II,\JJ)\!=\!(0,2)$, it corresponds to the two-loop lollipop correlator \eqref{eq:Omega-lop}.

\item 
$\NN\!=\!2$: The unshifted form is $\bOme_{\PP_\ctr}\!\!=\bOme_{\PP_\bub}$, where $\bOme_{\PP_\bub}$ has already been discussed in section\,\ref{sec:3.1}, and a more general expression is provided in \eqrefe{eq:2-site-LC}.\
Then, for $(\II,\JJ)\!=\!(0,0)$, \eqrefe{eq:2-site-N-loop} reproduces the two-loop bubble correlator \eqref{eq:P=P1+P2-P3}; 
For $(\II,\JJ)\!=\!(0,1)$, it reproduces the two-loop gourd correlator \eqref{eq:grd-ome-shift}.

\item 
$\NN\!=\!3$: The unshifted form is $\bOme_{\PP_\ctr}\!\!=\bOme_{\PP_\sun}$.\ 
For $(\II,\JJ)\!=\!(0,0)$, \eqrefe{eq:2-site-N-loop} gives the two-loop sunset correlator \eqref{eq:Omega-sun}.

\end{itemize}
Additionally, we consider a more intricate example with $(\NN,\II,\JJ)=(3,1,2)$.\
In this case, the unshifted form is $\bOme_{\PP_\ctr}\!\!=\bOme_{\PP_\sun}$, and the power sets of $\YY_\L,\YY_\R$ are given by $P(\YY_\L)\!=\!\{\varnothing,\{\mY_1\}\}$ and $P(\YY_\R)=\{\varnothing,\{\tmY_1\},\{\tmY_2\},\{\tmY_1,\tmY_2\}\}$.\ 
Substituting these into \eqrefe{eq:2-site-N-loop}, we can obtain:
\begin{align} 
\bOme_{\PP} = &\, -\bOme_{\PP_\sun}
+\bOme_{\PP_\sun}\big|{\substack{\\[1.5mm]
X_1\to X_1+2\mY_1}}
+\bOme_{\PP_\sun}\big|{\substack{\\[1.5mm]
X_2\to X_2+2\tmY_1}}
+\bOme_{\PP_\sun}\big|{\substack{\\[1.5mm]
X_2\to X_2+2\tmY_2}}
\nn\\
&\,-\bOme_{\PP_\sun}\big|{\substack{\\[1.5mm]
X_2\to X_2+2(\tmY_1+\tmY_2)}}
-\bOme_{\PP_\sun}\Big|{\substack{\\[.5mm]
X_1\to X_1+2\mY_1 \\
X_2\to X_2+2\tmY_1}}
-\bOme_{\PP_\sun}\Big|{\substack{\\[.5mm]
X_1\to X_1+2\mY_1 \\
X_2\to X_2+2\tmY_2}}
\nn\\
&\,+\bOme_{\PP_\sun}\Big|{\substack{\\[.5mm]
\hspace*{-8mm}X_1\to X_1+2\mY_1 \\
X_2\to X_2+2(\tmY_1+\tmY_2)}}\,.
\end{align}

Therefore, according to the decomposition relation \eqref{eq:2-site-N-loop}, the differential equations for all two-site $L$\hs-loop diagrams can be derived.\ 
The key lies in determining $\bOme_{\ctr}$, which takes one of the two forms: a tree diagram or a bubble-like structure.\ For both cases, the differential system can be analyzed using the perspectives of twisted cohomology (hyperplane arrangements) and kinematic flow framework\,\footnote{%
In section\,\ref{sec:4.2}, when discussing the kinematic flow for one-loop tadpole correlator, we used red and hollow crosses marked on the tadpole internal lines to distinguish between the shifted and unshifted tree diagrams consistent with \eqrefe{eq:2-site-N-loop}.\ 
This approach is equally applicable to higher tadpole-loop scenarios.}. 
By systematically accounting for all possible shifts in external energies $X_1$ and $X_2$, we can obtain the complete system of differential equations.  
Moreover, the matrix $A$ of the differential system for an arbitrary loop diagram can be block-decomposed based on \eqrefe{eq:2-site-N-loop}:
\begin{equation}
\label{eq:2-site-N-loop-tA}
A \,= 
\bigoplus_{\SS_\L \in P(\YY_\L)}\bigoplus_{\SS_\R \in P(\YY_\R)} A_\ctr\bigg|{\substack{\\[1.5mm]
\! X_1\to X_1 + 2\sum_{\mY_i \in \SS_\L}\!\mY_i\\[-0mm]
X_2\to X_2 + 2\sum_{\tmY_j \in \SS_\R}\!\tmY_j}} ~,
\end{equation}
where matrix $A_{\ctr}$ corresponds to the differential system of unshifted form $\bOme_{\ctr}$.\ Each block matrix in \eqrefe{eq:2-site-N-loop-tA} satisfies the integrability conditions \eqref{eq:inte-con}, ensuring that the full matrix $A$ also satisfies these conditions.\
The overall sign in front of each block matrix is not particularly significant, as we can always rescale $A$ to have a positive sign through the homogeneous nature of the differential equations.\ This adjustment does not affect the final conclusions.\ 
The extension of this framework to multi-site multi-loop diagrams is left for our future work.

\section{Conclusion and Outlook}
\label{sec:6}

In this study, we systematically analyze the differential systems of loop-level wavefunction coefficients in conformally-coupled scalar field theory with non-conformal polynomial interactions in a power-law FRW universe.
By employing the mathematical tools of twisted cohomology, hyperplane arrangements and IBP techniques, we successfully derive the canonical differential equations for one-loop bubble and tadpole diagrams, revealing distinct structural features in each case.\
We extend the kinematic flow framework to the loop level for the first time.\ 
By representing singularities with marked tubing graphs, we are able to derive all the canonical differential equations from the family trees formed by the evolution of the marked tubes.\ Furthermore, we introduce a generalized formula for analyzing the canonical form associated with the two-site arbitrary-loop wavefunction coefficients, marking a significant advancement in applying the kinematic flow method beyond tree-level systems. 

To elaborate, our analysis provides several crucial insights into the structures of cosmological correlators, uncovering novel geometric and topological connections.\ 
First, we have found a fundamental distinction between tadpole and bubble systems, with the former exhibiting multiple parent functions.\ This difference arises from the presence of distinct $L$\hs-pairs associated with each parent function in tadpole system, while the multiple parent functions in bubble system can be merged into a single one due to the same $L$\hs-pairs shared with all parent functions.\ 
Second, we show that the wavefunction coefficients of the tadpole system interact with a constrained subset of the full vector space, despite the higher-dimensional structure implied by their associated hyperplane arrangement.\
We introduce a tailored triangle basis for two-site tadpole system.\ 
The procedure for constructing the preferred triangle basis includes exhaustively determining bottom, middle, and top-layer functions through systematic interactions of $T$- and $L$-planes.\ 
In addition, we show that the basis size at loop level is smaller than that at tree level, due to numerous non-trivial parallel lines in the loop-level hyperplane arrangements.\
Third, the extension of kinematic flow framework to the loop-level correlators plays a key role in predicting the canonical differential equations by utilizing marked tubing graphs.\
These marked tubings represent the singularities of the differential system in dlog form and enable the construction of family trees that capture the hierarchical structure of the entire system.\ Following the simple rules, we have successfully obtained the canonical DEs for one-loop bubble and tadpole correlators which are consistent with the DEs computed through IBP.

Further, building on the analysis of the one-loop case, we propose a general formula for analyzing two-site diagrams with arbitrary loops.\ This formula demonstrates that the canonical form associated with a two-site arbitrary-loop wavefunction coefficient can be decomposed into a combination of unshifted and shifted components, with the shifting procedure determined combinatorially by the existence of tadpole loops.\ 
In the absence of tadpole loops, i.e., bubble-like case, we can start with the pure two-site tree-level correlator and shift its sum of external energies $X_1\!+\!X_2$ according to the internal energies $Y_i$ from the bubble-like diagram, summing all possible shifts accordingly.\ 
In the case involving tadpole loops, when a tree diagram connects to the tadpole loops, its external energies $X_1,X_2$ should be shifted according to the internal energies $\mY_i,\tmY_j$ of the tadpoles.\ 
If a bubble-like diagram connects to tadpole loops, we first reapply the shifts of tree diagram to obtain the form of bubble-like diagram without any tadpoles.\ Then, we can reproduce the shifts for $X_1,X_2$ according to the internal energies of the tadpole loops connected to the vertices of the resulting bubble-like diagram.\ Finally, we sum over all possible shifts to obtain the final result.\
As a result, any two-site $L$-loop diagram can be broken down into its fundamental tree and bubble-like structures, along with their shifted counterparts.\
Moreover, this general formula enables the establishment of a block-wise decomposition rule for the matrix $A$ in the differential systems, ensuring consistency with integrability conditions across all components.\
Such insights and methodologies lay the foundation for exploring even more intricate multi-site systems in our future studies.

As we look to the future, our research will focus on extending these methodologies to higher-loop configurations with more sites and exploring massive correlators.\ 
Additionally, the study of massive scalar fields presents an even more intriguing avenue of exploration.\ In ref.\,\cite{Gasparotto:2024bku}, the differential equations for two-site tree-level wavefunction coefficients in dS spacetime were analyzed.\ Introducing an additional mass term into \eqrefe{eq:S-FRW}, the action becomes
\begin{equation}
S_{\rm{FRW}} \,=\, -\int\td^d x \int^{0}_{-\infty}\td\eta\hs \sqrt{-g}\, \Bigg[\,\frac{1}{2}\hs g^{\ab} \pd_{\al}^{}\phi\hs \pd_{\be}^{}\phi +\frac{1}{2}\xi R\hs \phi^2 + \frac{1}{2}m^2 \phi^2 +\sum_{n\geqq3} \frac{\,\lam_n\,}{n!}\hs \phi^n \Bigg]\,.
\end{equation}
This modification complicates the equation of motion for $\phi$, yielding solutions that involve the Hankel function of the second kind, $H^{(2)}_\nu(-E\eta)$ with $\nu \!\!=\!\! \frac{\sqrt{1\!-\!4m^2}}{2}$ in the conformally coupled case \cite{Benincasa:2019vqr}.\ 
Hence, it will result in a more intricate hyperplane configuration.\ For instance, in the two-site tree-level case, there are four twist variables $\{x_1, x_2, t_1, t_2\}$.\ 
Despite these complexities, the hyperplane arrangements and kinematic flow for both massive tree-level and loop-level scenarios remain a compelling area for further investigation.

\begin{acknowledgments}
We sincerely thank Paolo Benincasa, Giacomo Brunello, J.\,J.\,Carrasco, Zong-Zhe Du, Jianyu Gong, Xu Wang and Yu Wu for their insightful discussions and valuable suggestions.\ 
YH and CS would like to acknowledge the Northwestern University Amplitudes and Insight group, Department of Physics and Astronomy, and Weinberg College for support.
\end{acknowledgments}


\appendix

\section{A Brief Review of Twisted Cohomology}
\label{app:A}

In this appendix, we provide a brief overview of twisted cohomology, emphasizing the results relevant to the main text. We also include a short discussion on its connection to relative twisted dual cohomology at the end.

The integral representation of cosmological correlators described in the main text belongs to the class of Aomoto-Gel'fand hypergeometric integrals \cite{aomoto1977structure,gelfand1986general}:
\begin{equation}
I = \int_\Ga u(\textbf{x})\hs\varphi(\textbf{x}) \,, 
\label{eq:twistedintdef}
\end{equation}
where $u(\textbf{x})$ is a multivalued function and $\varphi(\textbf{x})$ is a holomorphic differential $n$-form, with $\textbf{x}=(x_1,\ldots,x_n)$.
In general, we can choose $u(\textbf{x})$ and $\varphi(\textbf{x})$ as follows:
\begin{equation}
u(\textbf{x})=\prod_i [T_i(\textbf{x})]^{\vep_i}\,,\qquad~~
\varphi(\textbf{x}) = \frac{N(\textbf{x})\,\td^n\textbf{x}}{~\prod_i [T_i(\textbf{x})]^{a_i}\prod_j[S_j(\textbf{x})]^{b_j}~} \,,
\end{equation}
where $\{T_i,S_j\}$ define a collection of hyperplanes in $n$-dimensional complex space $\mathbb{C}^n$ and $N(\textbf{x})$ is an arbitrary polynomial.\ 
The parameters $\vep_i\in\mathbb{C}\setminus\mathbb{Z}$ and $(a_i,b_j)\in\mathbb{Z}$.\ 
Moreover, the form $\varphi$ is holomorphic on the open manifold $X\!=\!\mathbb{C}^n\setminus\big(\bigcup_i\{T_i\!=0\}\big)\bigcup\hs\big(\bigcup_j \{S_j\!=0\}\big)\hs$.

A key point is that such integrals, and hence the associated differential forms, are not all linearly independent due to integration-by-parts identities.\ Consider an ($n\!-\!1$)-form $\zeta$, applying the Stokes' theorem yields:
\begin{equation}
0= \int_{\pd\Ga} u\hs\zeta = \int_\Ga \td(u\hs\zeta) = \int_\Ga (\td u \wedge \zeta + u\hs\td\zeta)\equiv \int_\Ga u \nabla_\omega\hs\zeta\,,
\end{equation}
where we assume that $u$ vanishes on the boundary $\Ga$, i.e. $u|_{\pd\Ga}\!=\! 0$, so that no boundary terms survive. Here, we have introduced covariant derivative (connection) $\nabla_\omega$:
\begin{equation}
\label{Aeq:nab-omega}
\nabla_\omega \equiv \td + \omega \, \wedge\,, \qquad~~~
\omega \equiv \dlog u \,,
\end{equation}
where it diﬀers from the ordinary derivative by the 1-form $\omega$.
And two $n$-forms are said to be equivalent if they differ by a covariant derivative of a $(n-1)$-form $\zeta$:
\begin{equation}
\bla\varphi\hs\big|:~~ \varphi \,\sim\, \varphi + \nabla_\omega \zeta \,,
\end{equation}
which defines an equivalence class $\bla\varphi\hs\big|$. The space of such equivalence classes forms the twisted cohomology group:
\begin{equation}
H^n(X,\nabla_\omega) \,=\, \frac{\{\varphi\in\Omega^n(X,\nabla_\omega)|\nabla_\omega\varphi=0\}}{\{\nabla_\omega\hs\theta\,|\,\theta\in\Omega^{n-1}(X,\nabla_\omega)\}}\,,
\end{equation}
where $\Omega^n(X,\nabla_\omega)$ denotes the space of holomorphic differential $n$-forms on $X$.

Since $H^n(X,\nabla_\omega)$ is a vector space, it admits a natural dual vector space, given by the relative twisted dual cohomology:
\begin{equation}
\check{H}^n(X, \nabla_\omega) = H^n(\check{X}, \check{X} \cap S, \check{\nabla}_{\check{\omega}}) \,,
\end{equation} 
where $\check{X}\!=\!\mathbb{C}^n\setminus\bigcup_i\{T_i\!=\!0\}$ and $S\!=\!\bigcup_j\{S_j\!=\!0\}$.\
The dual covariant derivative $\check{\nabla}_{\check{\omega}}$ is defined as follows:
\begin{equation}
\check{\nabla}_{\check{\omega}}\equiv\td+\check{\omega} \wedge \,,\qquad~~~
\check{\omega}\equiv\dlog\check{u}=\dlog(1/u) = -\omega \,.
\end{equation}
In addition, the dual space is given by the equivalence classes:
\begin{equation}
\big|\check{\varphi}\bra:~~ \check{\varphi} \,\sim\, \check{\varphi} +\check{\nabla}_{\check{\omega}}\hs \zeta \,.
\end{equation}
The bilinear pairing between a twisted cocycle and its dual is the intersection number $\bla \varphi \hs \big| \check{\varphi} \bra$, whose detailed properties will not be discussed here (cf.\,ref.\,\cite{Mastrolia:2018uzb,Frellesvig:2019kgj,Caron-Huot:2021xqj,Caron-Huot:2021iev}).

\section{The Derivation of Two-Site One-Loop DEs from IBP}
\label{app:B}

In this appendix, we provide additional details on the derivation of the differential equations for certain functions that were omitted in the main text.

\subsection{Two-Site One-Loop Bubble}
\label{app:B1}

In \eqrefe{eq:Fi} of section\,\ref{sec:3.1},  $\bF_3=\{\PP_3,\FF_3,\FFt_3,\QQ_3\}$ and the total derivative for $\bF_3$ is
\begin{equation}
\label{Aeq:dP3}
\td\bF_3\,=\,
\pd_{X_1}^{}\bF_3\td{X_1} + 
\pd_{X_2}^{}\bF_3\td{X_2} + \pd_{Y_1}^{}\bF_3\td{Y_1} +
\pd_{Y_2}^{}\bF_3\td{Y_2} \,.
\end{equation}
The derivative of the parent function $\PP_3$ with respect to the external energy $X_1$ can be calculated as follows:
\begin{align}
\label{Aeq:dX1-P3}
&\pd_{X_1}\PP_3 =\int\!\td\mu\, \pd_{x_1}\bOme_{\PP_3}
=\vep\!\int\!\td\mu\(-\frac{\bOme_{\PP_3}}{T_1}\)
\nn\\[1mm]
&=\vep\!\int\td\mu\,\Bigg\{\Bigg(
\rm{Res}\!\[\frac{-\bOme_{\PP_3}}{T_1}\!\]\!{\substack{\\[5.2mm]
L_1=0 \\ L_2=0}}~,~
\rm{Res}\!\[\frac{-\bOme_{\PP_3}}{T_1}\!\]\!{\substack{\\[5.2mm]
T_1=0 \\ L_2=0}}\,\Bigg)\!
\(\bOme_{\PP_3},\bOme_{\FF_3}\)^{T}
\Bigg\}
\nn\\[1mm]
&=-\vep\!\int\td\mu\!\[\frac{1}{X_1\!+\!Y_1\!+\!Y_2}\,\bOme_{\PP_3}
+\frac{2(Y_1\!+\!Y_2)}{(X_1\!-\!Y_1\!-\!Y_2)(X_1\!+\!Y_1\!+\!Y_2)}\,\bOme_{\FF_3}\]
\nn\\[1mm]
&=-\vep\!\[\frac{1}{X_1\!+\!Y_1\!+\!Y_2}\,\PP_3+\(\frac{1}{X_1\!-\!Y_1\!-\!Y_2}-\frac{1}{X_1\!+\!Y_1\!+\!Y_2}\)\!\FF_3\].
\end{align}
Similarly, the differentiation with respect to 
$X_2$ follows the same steps as for $X_1$. We only need to make the following substitutions: $X_1\!\to\!X_2$ and $\FF_3\!\to\!\FFt_3$ in \eqrefe{Aeq:dX1-P3}:
\begin{equation}
\pd_{X_2}\PP_3 =\vep\!\[\frac{1}{X_2\!+\!Y_1\!+\!Y_2}\,\PP_3+\(\frac{1}{X_2\!-\!Y_1\!-\!Y_2}-\frac{1}{X_2\!+\!Y_1\!+\!Y_2}\)\!\FFt_3\].
\end{equation}

The derivative of the parent function $\PP_3$ with respect to the internal energy $Y_1$ can be computed as follows:
%
\begin{align}
\label{Aeq:dY1-P3}
&\pd_{Y_1}\PP_3 =\int\!\td\mu\[\pd_{x_1}\!\(\frac{-x_1\!-\!X_1}{Y_1\!+\!Y_2}\,\bOme_{\PP_3}\)\!+\pd_{x_2}\!\(\frac{-x_2\!-\!X_2}{Y_1\!+\!Y_2}\,\bOme_{\PP_3}\)\!\]
\nn\\[1mm]
&=\vep\!\int\!\td\mu\,\Bigg\{\Bigg(
\rm{Res}\!\[\!\frac{(x_1\!+\!X_1)\bOme_{\PP_3}}{T_1(Y_1\!+\!Y_2)}\!\]\!{\substack{\\[5.2mm]
L_1=0 \\ L_2=0}}~,~
\rm{Res}\!\[\!\frac{(x_1\!+\!X_1)\hs\bOme_{\PP_3}}{T_1(Y_1\!+\!Y_2)}\!\]\!{\substack{\\[5.2mm]T_1=0 \\ L_2=0}} \,\Bigg)\!
\(\bOme_{\PP_3},\bOme_{\FF_3}\)^{T}
\nn\\
&\hspace*{1.35cm}+\Bigg(
\rm{Res}\!\[\!\frac{(x_1\!+\!X_1)\bOme_{\PP_3}}{T_2(Y_1\!+\!Y_2)}\!\]\!{\substack{\\[5.2mm]
L_1=0 \\ L_2=0}}~,~
\rm{Res}\!\[\!\frac{(x_1\!+\!X_1)\hs\bOme_{\PP_3}}{T_2(Y_1\!+\!Y_2)}\!\]\!{\substack{\\[5.2mm]T_2=0 \\ L_1=0}} \,\Bigg)\!
\(\bOme_{\PP_3},\bOme_{\FFt_3}\)^{T}\Bigg\}
\nn\\[1mm]
&=\vep\bigg[\!\(\frac{1}{X_1\!+\!Y_1\!+\!Y_2}+\frac{1}{X_2\!+\!Y_1\!+\!Y_2}\)\!\PP_3 - \(\frac{1}{X_1\!+\!Y_1\!+\!Y_2}+\frac{1}{X_1\!-\!Y_1\!-\!Y_2}\)\!\FF_3
\nn\\
&\qquad~-\(\frac{1}{X_2\!+\!Y_1\!+\!Y_2}+\frac{1}{X_2\!-\!Y_1\!-\!Y_2}\)\!\FFt_3\bigg]\,.
\end{align}
%
By replacing $Y_1\to Y_2$ in \eqrefe{Aeq:dY1-P3}, it is not difficult to derive the result for $Y_2$\,:
\begin{align}
\pd_{Y_2}\PP_3 & = \vep\[\!\(\frac{1}{X_1\!+\!Y_1\!+\!Y_2}+\frac{1}{X_2\!+\!Y_1\!+\!Y_2}\)\!\PP_3 - \(\frac{1}{X_1\!+\!Y_1\!+\!Y_2}+\frac{1}{X_1\!-\!Y_1\!-\!Y_2}\)\!\FF_3
\right. \nn\\
&\hspace*{1.cm}-\left.\(\frac{1}{X_2\!+\!Y_1\!+\!Y_2}+\frac{1}{X_2\!-\!Y_1\!-\!Y_2}\)\!\FFt_3\] =\pd_{Y_1}\PP_3  \,.
\end{align}
Thus, in terms of the total derivative \eqref{Aeq:dP3} and writing in dlog forms, we finally obtain the full differential equation for $\PP_3$\,:
\begin{align}
\td\PP_3 \,=\, &\,-\vep\Big[(\PP_3-\FF_3)\hs\dlog(X_1\!+\!Y_1\!+\!Y_2)+(\PP_3-\FFt_3)\hs\dlog(X_2\!+\!Y_1\!+\!Y_2)
\nn\\
& \quad+ \FF_3\hs\dlog(X_1\!-\!Y_1\!-\!Y_2)+ \FFt_3\hs\dlog(X_2\!-\!Y_1\!-\!Y_2)\Big]\,.
\end{align}

Next, we examine the differentiation of the descendant functions $\FF_3$ and $\FFt_3$. For $\FF_3$, the results with respect to the external energies are given by
\beqs
\begin{align}
&\pd_{X_1}\FF_3 = \int\!\td\mu\!
\[\pd_{x_1}\!\!\(\frac{-T_1}{X_1\!-\!Y_1\!-\!Y_2}\,\bOme_{\FF_3}\)
\!+\pd_{x_2}\!\!\(\frac{-L_2}{X_1\!-\!Y_1\!-\!Y_2}\,\bOme_{\FF_3}\)\!\] 
\nn\\
&=\vep\!\int\!\!\td\mu\,
\Bigg\{\!\Bigg(\rm{Res}\!\[\!\frac{\bOme_{\FF_3}}{X_1\!-\!Y_1\!-\!Y_2}\!\]\!{\substack{\\[5.2mm]T_1=0 \\ L_2=0}} ~,~
\rm{Res}\!\[\!\frac{\bOme_{\FF_3}}{X_1\!-\!Y_1\!-\!Y_2}\!\]\!{\substack{\\[5.2mm]
T_1=0 \\ T_2=0}} \,\Bigg)\!
\(\bOme_{\FF_3},\bOme_{\QQ_3}\)^T
\nn\\
&\hspace*{1.25cm}+\Bigg(\rm{Res}\!\[\!\frac{L_2\hs\bOme_{\FF_3}}{T_2(X_1\!-\!Y_1\!-\!Y_2)}\!\]\!{\substack{\\[5.2mm]T_1=0 \\ L_2=0}} ~,~
\rm{Res}\!\[\!\frac{L_2\hs\bOme_{\FF_3}}{T_2(X_1\!-\!Y_1\!-\!Y_2)}\!\]\!{\substack{\\[5.2mm]
T_1=0 \\ T_2=0}} \,\Bigg)\!
\(\bOme_{\FF_3},\bOme_{\QQ_3}\)^T \Bigg\}
\nn\\
&=\vep\[\frac{1}{X_1\!-\!Y_1\!-\!Y_2}\,\FF_3 + \frac{1}{X_1\!+\!X_2}\,\QQ_3\],
\\[1.5mm]
&\pd_{X_2}\FF_3 =\int\!\td\mu\,\pd_{x_2}\!\bOme_{\FF_3}
\nn\\
&=\vep\!\[\frac{1}{X_2\!+\!Y_1\!+\!Y_2}\,\FF_3 + \(\frac{1}{X_1+X_2}-\frac{1}{X_2+\!\!Y_1\!+\!Y_2}\)\!\QQ_3\],
\end{align}
\eeqs
and the differentiation with respect to the internal energies is computed as:
\beqs
\begin{align}
\pd_{Y_1}\FF_3 &=\int\!\td\mu\!\[\pd_{x_1}\!\!\(\frac{x_1}{X_1\!-\!Y_1\!-\!Y_2}\,\bOme_{\FF_3}\)\!
+\pd_{x_2}\!\!\(\frac{x_2\!+\!X_1\!+\!X_2}{X_1\!-\!Y_1\!-\!Y_2}\,\bOme_{\FF_3}\)\!\]
\nn\\[1mm]
&=\vep\[\!\(-\frac{1}{X_1\!-\!Y_1\!-\!Y_2}+\frac{1}{X_2\!+\!Y_1\!+\!Y_2}\)\!\FF_3 - \frac{1}{X_2\!+\!Y_1\!+\!Y_2}\QQ_3\],
\\[1mm]
\pd_{Y_2}\FF_3&=\pd_{Y_1}\FF_3\,.
\end{align}
\eeqs
We omit the derivation for $\FFt_3$.
In terms of the total derivative, the complete canonical differential equations for $\FF_3$ and $\FFt_3$ are obtained as:
\beqs
\begin{align}
\td\FF_3 &=\vep\big[\FF_3\hs\dlog(X_1\!-\!Y_1\!-\!Y_2)\!+\! (\FF_3\!-\!\QQ_3)\hs \dlog(X_2\!+Y_1\!+\!Y_2)+\QQ_3\hs\dlog(X_1\!+\!X_2)\big]\,,
\\[1mm]
\td\FFt_3 &=\vep\big[\FFt_3\hs\dlog(X_2\!-\!Y_1\!-\!Y_2)\!+\! (\FFt_3\!-\!\QQ_3)\hs \dlog(X_1\!+Y_1\!+\!Y_2)+\QQ_3\hs\dlog(X_1\!+\!X_2)\big]\,.
\end{align}
\eeqs
Finally, for the descendant function $\QQ_3$, we can derive
\beqs
\begin{align}
\pd_{X_1}\QQ_3 &=\!\!\int\!\td\mu\!\[\!\pd_{x_1}\!\!\(\frac{-x_1}{X_1+X_2}\bOme_{\QQ_3}\)
\!+\!\pd_{x_2}\!\(\frac{-x_2}{X_1+X_2}\bOme_{\QQ_3}\!\)\! \] 
\nn\\[1mm]
&=
2\vep\!\int\!\td\mu \, \rm{Res}\!\[\!\frac{\bOme_{\QQ_3}}{X_1\!-\!X_2}\!\]\!{\substack{\\[5.2mm]
T_1=0 \\ T_2=0}}\,\bOme_{\QQ_3} = 2\vep\,\frac{\QQ_3}{X_1\!+\!X_2}\,,
\\[1mm]
\pd_{X_2}\QQ_3 &= \pd_{X_1}\QQ_3 \,,
\\[1mm]
\pd_{Y_1}\QQ_3 &= \pd_{Y_2}\QQ_3=0 \,,
\end{align}
\eeqs
where the differential equation for $\QQ_3$ is obtained as:
\begin{equation}
\td \QQ_3=2\vep\hs\QQ_3 \hs \dlog(X_1\!+\!X_2)\,.
\end{equation}

\subsection{Two-Site One-Loop Tadpole}
\label{app:B2}

The definitions for $\FF^\uu_2$ and $\FF^\uu_3$ can be found in \eqrefe{eq:Fu23} of section\,\ref{sec:3.2}.\ The differentiation for the function $\FF^\uu_2$ with respect to the external and internal energies is computed as:
\beqs
\begin{align}
\pd_{X_1}\FF^\uu_2&=\int\td\mu\, \[\pd_{x_1}\(\frac{-x_1}{X_1-Y_1+2Y_2}\bOme_{\FF^\uu_2}\)+\pd_{x_2}\(\frac{-L_2}{X_1-Y_1+2Y_2}\bOme_{\FF^\uu_2}\)\]  
\nn\\
&=\vep\[\frac{1}{X_1-Y_1+2Y_2}\FF^\uu_2+\frac{1}{X_1+X_2+2Y_2}\QQ_2\],
\\[1mm]
\pd_{X_2}\FF^\uu_2&=\int\td\mu\, \pd_{x_2}\bOme_{\FF^\uu_2}=\vep\[\frac{1}{X_2+Y_1}\FF^\uu_2+\(\frac{1}{X_1+X_2+2Y_2}-\frac{1}{X_2+Y_1}\)\QQ_2\],
\\[1mm]
\pd_{Y_1}\FF^\uu_2&=\int\td\mu\, \[\pd_{x_1}\(\frac{x_1}{X_1-Y_1+2Y_2}\bOme_{\FF^\uu_2}\)+\pd_{x_2}\(\frac{x_2+X_1+X_2+2Y_2}{X_1-Y_1+2Y_2}\bOme_{\FF^\uu_2}\)\] 
\nn\\
&=\vep \[\(\frac{1}{X_2+Y_1}-\frac{1}{X_1-Y_1+2Y_2}\)\FF^\uu_2-\frac{1}{X_2+Y_1}\QQ_2\],
\\[1mm]
\pd_{Y_2}\FF^\uu_2&=\int\td\mu\, \[\pd_{x_1}\(\frac{-2x_1}{X_1-Y_1+2Y_2}\bOme_{\FF^\uu_2}\)+\pd_{x_2}\(\frac{-2L_2}{X_1-Y_1+2Y_2}\bOme_{\FF^\uu_2}\)\]  
\nn\\
&=\vep \[\frac{2}{X_1-Y_1+2Y_2}\FF^\uu_2+\frac{2}{X_1+X_2+2Y_2}\QQ_2\].
\end{align}
\eeqs
Collectively, the canonical differential equation for $\FF^\uu_2$ is given by
\begin{align}
\td\FF^\uu_2=\vep\hs 
\big[&\FF^\uu_2\hs \dlog(X_1\!-\!Y_1\!+\!2Y_2)+\QQ_2\hs \dlog(X_1\!+\!X_2\!+\!2Y_2)
\nn\\
&+(\FF^\uu_2-\QQ_2)\hs \dlog(X_2\!+\!Y_1)\big]\,.
\end{align}

The differentiation for the function $\FF^\uu_3$ with respect to the external and internal energies is computed as:
\beqs
\begin{align}
\pd_{X_1}\FF^\uu_3&=\int\td\mu \[\pd_{x_1}\(\frac{-x_1}{X_1-Y_1-2Y_2}\bOme_{\FF^\uu_3}\)+\pd_{x_2}\(\frac{-x_2-X_2-Y_1-2Y_2}{X_1-Y_1-2Y_2}\bOme_{\FF^\uu_3}\)\]  
\nn\\
&=\vep\[\frac{1}{X_1-Y_1-2Y_2}\bOme_{\FF^\uu_3}+\frac{1}{X_1+X_2}\bOme_{\QQ_3}\],
\\[1mm]
\pd_{X_2}\FF^\uu_3&=\int\td\mu\, \pd_{x_2}\bOme_{\FF^\uu_3}
=\int\td\mu\(-\frac{\bOme_{\FF^\uu_3}}{T_2}\) 
\nn\\
&=\vep\[\frac{1}{X_2+Y_1+2Y_2}\bOme_{\FF^\uu_3}+\(\frac{1}{X_1+X_2}-\frac{1}{X_2+Y_1+2Y_2}\)\bOme_{\QQ_3}\],
\\[1mm]
\pd_{Y_1}\FF^\uu_3&=\int\td\mu \[\pd_{x_1}\(\frac{x_1}{X_1-Y_1-2Y_2}\bOme_{\FF^\uu_3}\)+\pd_{x_2}\(\frac{x_2+X_1+X_2}{X_1-Y_1-2Y_2}\bOme_{\FF^\uu_3}\)\] 
\nn\\
&=\vep \[\(\frac{1}{X_2+Y_1+2Y_2}-\frac{1}{X_1-Y_1-2Y_2}\)\bOme_{\FF^\uu_3}-\frac{1}{X_2+Y_1+2Y_2}\bOme_{\QQ_3}\] ,
\\[1mm]
\pd_{Y_2}\FF^\uu_3 &= 2\hs\pd_{Y_1}\FF^\uu_3 \,.
\end{align}
\eeqs
Collectively, the canonical differential equation for $\FF^\uu_3$ is given by
\begin{align}
\td\FF^\uu_3=\vep\hs\big[&\FF^\uu_3\hs\dlog(X_1\!-\!Y_1\!-\!2Y_2)+\QQ_3\hs\dlog(X_1+X_2)
\nn\\
&+(\FF^\uu_3-\QQ_3)\hs\dlog(X_2\!+\!Y_1\!+\!2Y_2)\big]\,.
\end{align}
%

\section{Tubings of Two-Site Two-Loop Bubble System}
\label{app:C}

In table\,\ref{Atab:1}, we have summarized the complete tubing graphs associated with the functions belonging to the integral family of two-site two-loop sunset diagram [cf.\,\eqrefe{eq:Sun-I}].

The differential equations are summarized in \eqrefe{eq:DE-2loop}.\ The same structure can be derived using the family tree introduced in the main text.\ For example, the differential equation for $\PP$ takes the form: 
\begin{align}
&\hspace*{-2mm}\td \PP = \vep\[
\(\PP\!-\!\sum_{i=1}^{7}\FF_i\)
\letterfig{Letter-Sun}{l1}{0.6}{0cm}\hspace*{1.3cm}
+\(\PP\!-\!\sum_{j=1}^{7}\FFt_j\)\letterfig{Letter-Sun}{l2}{0.6}{0cm}\hspace*{1.2cm}
+\FF_1~\letterfig{Letter-Sun}{l9}{0.6}{0cm}\hspace*{1.2cm}
+\FF_2~\letterfig{Letter-Sun}{l11}{0.6}{0cm}\hspace*{1.2cm}
\right.\nn\\
&\hspace*{-2mm} ~\,+\FF_3~\letterfig{Letter-Sun}{l13}{0.6}{0cm}\hspace*{1.2cm}
+\FF_4~\letterfig{Letter-Sun}{l3}{0.6}{0cm}\hspace*{1.2cm}
+\FF_5~\letterfig{Letter-Sun}{l5}{0.6}{0cm}\hspace*{1.2cm}
+\FF_6~\letterfig{Letter-Sun}{l7}{0.6}{0cm}\hspace*{1.2cm}
+\FF_7~\letterfig{Letter-Sun}{l15}{0.6}{0cm}\hspace*{1.2cm} +\FFt_1~\letterfig{Letter-Sun}{l10}{0.6}{0cm}
\nn\\
&\hspace*{-2mm}\left.\hspace*{1mm}
+\,\FFt_2~\letterfig{Letter-Sun}{l12}{0.6}{0cm}\hspace*{1.2cm}
+\FFt_3~\letterfig{Letter-Sun}{l14}{0.6}{0cm}\hspace*{1.2cm}
+\FFt_4~\letterfig{Letter-Sun}{l4}{0.6}{0cm}\hspace*{1.2cm}
+\FFt_5~\letterfig{Letter-Sun}{l6}{0.6}{0cm}\hspace*{1.2cm}
+\FFt_6~\letterfig{Letter-Sun}{l8}{0.6}{0cm}\hspace*{1.2cm}
+\FFt_7~\letterfig{Letter-Sun}{l16}{0.6}{0cm}\hspace*{1.2cm}
\]\!,
\end{align}
where the letters are summarized in table\,\ref{Atab:2}.\
The equations for the remaining functions follow analogously and are omitted.\ Additional letters can also be found in table\,\ref{Atab:2}.
\begin{table}[t]
\centering
\begin{tabular}{c|cc|c}
\hline\hline
\multicolumn{1}{c|}{Layer-0} & \multicolumn{2}{c|}{Layer-1} & \multicolumn{1}{c}{Layer-2} 
\\ \hline\xrowht{12mm}
&\quad\,$\FF_1\,$\letterfig{Function-Sun}{layer1-F4}{0.6}{1cm}\qquad~
&\!\!$\FFt_1\,$\letterfig{Function-Sun}{layer1-Ft4}{0.6}{1cm}\quad~
&~~$\QQ_1\,$\letterfig{Function-Sun}{layer2-Q4}{0.6}{1cm}\quad~
\\\cline{2-4}\xrowht{12mm}
&$\FF_2\,$\letterfig{Function-Sun}{layer1-F5}{0.6}{1cm} ~~ 
&$\FFt_2\,$\letterfig{Function-Sun}{layer1-Ft5}{0.6}{1cm}  \quad~
&\quad$\QQ_2\,$\letterfig{Function-Sun}{layer2-Q5}{0.6}{1cm}  \quad~
\\\cline{2-4}\xrowht{12mm}
&$\FF_3\,$\letterfig{Function-Sun}{layer1-F6}{0.6}{1cm}  ~~
&$\FFt_3\,$\letterfig{Function-Sun}{layer1-Ft6}{0.6}{1cm}  \quad~
&\quad$\QQ_3\,$\letterfig{Function-Sun}{layer2-Q6}{0.6}{1cm}  \quad~
\\\cline{2-4}\xrowht{12mm}
~$\PP\,$\letterfig{Function-Sun}{layer0-P}{0.6}{1cm}\quad\,  
& $\FF_4\,$\letterfig{Function-Sun}{layer1-F1}{0.6}{1cm}  ~~
&$\FFt_4\,$\letterfig{Function-Sun}{layer1-Ft1}{0.6}{1cm}  \quad~
&\quad$\QQ_4\,$\letterfig{Function-Sun}{layer2-Q1}{0.6}{1cm} \quad~
\\\cline{2-4}\xrowht{12mm}
&$\FF_5\,$\letterfig{Function-Sun}{layer1-F2}{0.6}{1cm}  ~~
&$\FFt_5\,$\letterfig{Function-Sun}{layer1-Ft2}{0.6}{1cm}  \quad~
&\quad$\QQ_5\,$\letterfig{Function-Sun}{layer2-Q2}{0.6}{1cm} \quad~
\\\cline{2-4}\xrowht{12mm}
&$\FF_6\,$\letterfig{Function-Sun}{layer1-F3}{0.6}{1cm}  ~~
&$\FFt_6\,$\letterfig{Function-Sun}{layer1-Ft3}{0.6}{1cm}  \quad~
&\quad$\QQ_6\,$\letterfig{Function-Sun}{layer2-Q3}{0.6}{1cm} \quad~
\\\cline{2-4}\xrowht{14mm}
&$\FF_7\,$\letterfig{Function-Sun}{layer1-F7}{0.6}{1cm} ~~
&$\FFt_7\,$\letterfig{Function-Sun}{layer1-Ft7}{0.6}{1cm}  \quad~
&\quad$\QQ_7\,$\letterfig{Function-Sun}{layer2-Q7}{0.6}{1cm}\quad~
\\
\hline\hline
\end{tabular}
\caption{Complete tubing configurations associated with the functions in the integral family \eqref{eq:Sun-I}.\ The functions in layer-1 and layer-2 are partitioned into seven groups according to the subsystems $\bF_1,\ldots,\bF_7$.}
\label{Atab:1}
\end{table}

\clearpage
\begin{table}[ht]
\centering
\begin{tabular}{c|c|c|c|c|c}
\hline\hline
\hspace*{-1mm}Letter\hspace*{-1mm}& Tube & $\dlog$ Form &\hspace*{-1mm}Letter\hspace*{-1mm}& Tube & $\dlog$ Form
\\\hline\xrowht{12mm}
$l_1$
&\letterfig{Letter-Sun}{l1}{0.6}{1.1cm}
&$\dlog(X_1\!+\!Y_1\!+\!Y_2\!+\!Y_3)$
&\multicolumn{1}{c|}{$l_2$}   
&\multicolumn{1}{c|}{\letterfig{Letter-Sun}{l2}{0.6}{1.1cm}}  
&$\dlog(X_2\!+\!Y_1\!+\!Y_2\!+\!Y_3)$
\\\hline\xrowht{12mm}
$l_3$
&\letterfig{Letter-Sun}{l9}{0.6}{1.1cm}  
&$\dlog(X_1\!+\!Y_1\!-\!Y_2\!-\!Y_3)$
&\multicolumn{1}{c|}{$l_4$} 
&\multicolumn{1}{c|}{\letterfig{Letter-Sun}{l10}{0.6}{1.1cm}}   
&$\dlog(X_2\!+\!Y_1\!-\!Y_2\!-\!Y_3)$ 
\\\hline\xrowht{12mm}
$l_5$
&\letterfig{Letter-Sun}{l11}{0.6}{1.1cm}  
&$\dlog(X_1\!-\!Y_1\!+\!Y_2\!-\!Y_3)$  
& \multicolumn{1}{c|}{$l_6$}
&\multicolumn{1}{c|}{\letterfig{Letter-Sun}{l12}{0.6}{1.1cm}}  
&$\dlog(X_2\!-\!Y_1\!+\!Y_2\!-\!Y_3)$ 
\\\hline\xrowht{12mm}
$l_7$
&\letterfig{Letter-Sun}{l13}{0.6}{1.1cm}  
&$\dlog(X_1\!-\!Y_1\!-\!Y_2\!+\!Y_3)$ 
& \multicolumn{1}{c|}{$l_8$}
&\multicolumn{1}{c|}{\letterfig{Letter-Sun}{l14}{0.6}{1.1cm}}  
&$\dlog(X_2\!-\!Y_1\!-\!Y_2\!+\!Y_3)$ 
\\\hline\xrowht{12mm}
$l_9$   
&\letterfig{Letter-Sun}{l3}{0.6}{1.1cm}
&$\dlog(X_1\!+\!Y_1\!+\!Y_2\!-\!Y_3)$
&\multicolumn{1}{c|}{$l_{10}$} 
&\multicolumn{1}{c|}{\letterfig{Letter-Sun}{l4}{0.6}{1.1cm}}   
&$\dlog(X_2\!+\!Y_1\!+\!Y_2\!-\!Y_3)$   
\\\hline\xrowht{12mm}
$l_{11}$ 
& \letterfig{Letter-Sun}{l5}{0.6}{1.1cm} 
& $\dlog(X_1\!+\!Y_1\!-\!Y_2\!+\!Y_3)$
&\multicolumn{1}{c|}{$l_{12}$} 
&\multicolumn{1}{c|}{ \letterfig{Letter-Sun}{l6}{0.6}{1.1cm}} 
&$\dlog(X_2\!+\!Y_1\!-\!Y_2\!+\!Y_3)$ 
\\\hline\xrowht{12mm}
$l_{13}$  
&\letterfig{Letter-Sun}{l7}{0.6}{1.1cm}   
&$\dlog(X_1\!-\!Y_1\!+\!Y_2\!+\!Y_3)$
& \multicolumn{1}{c|}{$l_{14}$}
&\multicolumn{1}{c|}{\letterfig{Letter-Sun}{l8}{0.6}{1.1cm}}  
&$\dlog(X_2\!-\!Y_1\!+\!Y_2\!+\!Y_3)$
\\\hline\xrowht{12mm}
$l_{15}$
&\letterfig{Letter-Sun}{l15}{0.6}{1.1cm} 
&$\dlog(X_1\!-\!Y_1\!-\!Y_2\!-\!Y_3)$  
& \multicolumn{1}{c|}{$l_{16}$}
&\multicolumn{1}{c|}{\letterfig{Letter-Sun}{l16}{0.6}{1.1cm}}  
&$\dlog(X_2\!-\!Y_1\!-\!Y_2\!-\!Y_3)$ 
\\\hline\xrowht{12mm}
$l_{17}$
&\letterfig{Letter-Sun}{l20}{0.6}{1.2cm}  
&$\dlog(X_1\!+\!X_2\!+\!2Y_1)$
& \multicolumn{1}{c|}{$l_{18}$}
&\multicolumn{1}{c|}{\letterfig{Letter-Sun}{l21}{0.6}{1.2cm}}  
&$\dlog(X_1\!+\!X_2\!+\!2Y_2)$
\\\hline\xrowht{12mm}
$l_{19}$
&\letterfig{Letter-Sun}{l22}{0.6}{1.2cm}  
&$\dlog(X_1\!+\!X_2\!+\!2Y_3)$
& \multicolumn{1}{c|}{$l_{20}$}
&\multicolumn{1}{c|}{\letterfig{Letter-Sun}{l17}{0.6}{1.2cm}}  
&$\dlog(X_1\!+\!X_2\!+\!2Y_1\!+\!2Y_2)$ 
\\\hline\xrowht{12mm}
$l_{21}$
&\letterfig{Letter-Sun}{l18}{0.6}{1.2cm}  
&$\dlog(X_1\!+\!X_2\!+\!2Y_1\!+\!2Y_3)$
& \multicolumn{1}{c|}{$l_{22}$}
&\multicolumn{1}{c|}{\letterfig{Letter-Sun}{l19}{0.6}{1.2cm}}  
&$\dlog(X_1\!+\!X_2\!+\!2Y_2\!+\!2Y_3)$ 
\\\hline\xrowht{12mm}
$l_{23}$
&\letterfig{Letter-Sun}{l23}{0.6}{1.2cm} 
&$\dlog(X_1\!+\!X_2)$   
&
\multicolumn{3}{l}{} 
\\ \hline\hline
\end{tabular}
\caption{The alphabet for two-site two-loop sunset wavefunction coefficient.}
\label{Atab:2}
\end{table}

\clearpage
\bibliographystyle{utphys}
\bibliography{refs.bib}

\end{document}